\documentclass[]{aastex631}
\usepackage[caption=false]{subfig}
\usepackage{tabularx}
\usepackage{capt-of}
\usepackage{hhline}
\usepackage{lineno}

\newcommand{\unit}[1]{\mathrm{#1}}
\newcommand{\GeV}{\unit{GeV}}

\newcommand{\percc}{\unit{cm}^{-3}}

\newcommand{\Msun}{M_{\odot}}

\newcommand{\sigmav}{\langle\sigma v\rangle}

\newcommand{\Lyalpha}{Ly-$\alpha$ }
\newcommand{\microm}[1]{\unit{#1}{\mu m}}
\newcommand{\Jdrop}{\mathrm{J}_{129}}
\newcommand{\Hdrop}{\mathrm{H}_{158}}

\reportnum{NORDITA-2023-027}
\reportnum{UTWI-24-2023}
\shorttitle{Observing Supermassive Dark Stars with RST}
\shortauthors{Zhang et al.}

\graphicspath{{./}{figures/}}

\begin{document}

\title{Detectability of Supermassive Dark Stars with the Roman Space Telescope} 

\author[0000-0003-1541-177X]{Saiyang Zhang}
\affiliation{Weinberg Institute for Theoretical Physics, Texas Center for Cosmology and Astroparticle Physics, \\
Department of Physics, University of Texas, Austin, TX 78712, USA}

\author[0000-0001-6262-209X]{Cosmin Ilie}
\affiliation{Department of Physics and Astronomy, Colgate University\\Hamilton, NY 13346, USA}
\affiliation{Department of Theoretical Physics, National Institute for Physics and Nuclear Engineering\\ Magurele, P.O.Box M.G. 6, Romania}

\author{Katherine Freese}
\affiliation{Weinberg Institute for Theoretical Physics, Texas Center for Cosmology and Astroparticle Physics, \\
Department of Physics, University of Texas, Austin, TX 78712, USA}
\affiliation{Department of Physics, Stockholm University, Stockholm, Sweden}
\affiliation{Nordic Institute for Theoretical Physics (NORDITA), Stockholm, Sweden}

\begin{abstract}

Supermassive dark stars (SMDS) are luminous stellar objects formed in the early Universe at redshift $z \sim 10-20$, made primarily of hydrogen and helium, yet powered by dark matter. 
We examine the capabilities of the Roman Space Telescope (RST), and find it able to identify $ \sim 10^6M_{\odot}$ SMDSs at redshifts up to $z\simeq 14$. With a gravitational lensing factor of $\mu\sim 100$, RST could identify SMDS as small as $\sim10^4\Msun$ at $z\sim 12$ with $\sim 10^6$~s exposure. 
Differentiating SMDSs from early galaxies containing zero metallicity stars at similar redshifts requires spectral, photometric, and morphological comparisons. With only RST, differentiation of SMDS, particularly those formed via adiabatic contraction with $M\gtrsim 10^5\Msun$ and lensed by $\mu\gtrsim 100$, is possible due to their distinct photometric signatures from the first galaxies. Those formed via dark matter capture can be differentiated only by image morphology: i.e. point object (SMDSs) vs. extended object (sufficiently magnified galaxies).
By additionally employing James Webb Space Telescope (JWST) spectroscopy, we can identify the HeII $\lambda$1640 absorption line, a "smoking gun" for SMDS detection. Although RST doesn't cover the required wavelength band (for $z_{\rm emi}\gtrsim 10$), JWST does, hence the two can be used in tandem to identify SMDS. 
The detection of SMDS would confirm a new type of star powered by dark matter and may shed light on the origins of the supermassive black holes powering bright quasars observed at $z\gtrsim 6$.

\end{abstract}

\section{Introduction} \label{sec:intro}
The  ``dark ages'' of the universe end with the formation of the first stars. This event  -the cosmic dawn - happens roughly 200-400 Myrs after the Big Bang, when pristine, zero metallicity, molecular hydrogen  clouds at the center of $10^6-10^8 ~\rm M_\odot$ minihaloes  start to undergo gravitational collapse, a necessary first step towards the formation of stars.  In the standard picture of the formation of Population III stars~\citep{Abel:2001,Barkana:2000,Bromm:2003,Yoshida:2006,OShea:2007,Yoshida:2008,Bromm:2009}, the collapsing hydrogen clouds eventually become protostars $\sim 10^{-3} ~\rm M_\odot$, which grow by accretion as long as radiative feedback doesn't halt accretion.  They are thought to grow to $100-1000 ~\rm M_\odot$.\footnote{The question as to the numbers of these stars forming in a given minihalo is as yet unsettled, due to the possibility of fragmentation of the hydrogen cloud. In simulations, the fragmentation has been shown to be suppressed due to the effects of dark matter heating~\citep{Smith:2012,Stacy:2012,Stacy:2014}.}
On the other hand, the large dark matter abundance at the centers of the minihalos may alter these conclusions.  Dark matter, whether it be Weakly Interacting Massive Particles, Self-Interacting Dark Matter, or other candidates, may provide a heat source that halts the collapse of the hydrogen clouds (\cite{Spolyar:2008dark}) and leads to a Dark Star, a star made almost entirely of hydrogen and helium but powered by dark matter heating (for a review, see \cite{Freese:2016dark}).
 These Dark Stars may grow to becomes very massive and bright Supermassive Dark Stars (SMDS) (\cite{Freese:2010smds}), up to $10^6 ~\rm M_\odot$ and $10^{10} L_\odot$, and be visible in current and upcoming telescopes, including the   James Webb Space Telescope (JWST) ~\citep{Ilie:2012} and, as is the subject of the current paper, the upcoming Roman Space Telescope (RST). The goal of the present paper is to make predictions for the observability of Dark Stars in RST and suggest how RST and JWST can be used in tandem to discover SMDS candidates and confirm them spectroscopically. 

Before the era ushered in by the James Webb Space Telescope (JWST), our understanding of the formation and properties of first luminous objects in the universe came only from a combination of theoretical and numerical models. Already, only a few months after becoming operational, photometric data from JWST are finding the most distant galaxy candidates to date
~\citep[e.g.][]{GLASSz13,Maisies:2022,z16.CEERS93316:2022,z17.Schrodinger:2022,Labbe:2022,JADES:2022a,JADES:2022b} and stand to discover SMDSs. Those findings significantly strengthen the tension that begun to emerge during the HST era between the picture one gets via numerical simulations of the formation of the first stars and assembly of the first galaxies in the universe~\citep[e.g.][]{Gnedin:2016,Dayal:2018,Yung:2019,Behroozi:2020} and what nature offers us~\citep[e.g.][]{Stefanon:2019,Bowler:2019,HD1,Bagley:2022,Labbe:2022}. Simply put, we are observing too many extremely bright galaxy candidates too early in the  universe.  If some of those objects were supermassive dark stars ($M\sim 10^6\Msun$) instead of galaxies, this tension would be alleviated.  Important theoretical uncertainties, such as how massive the first generation of stars can be, or if they can be powered by dark matter annihilations (Dark Stars) or only by nuclear fusion (Population III stars), still remain open questions, that are now more relevant than ever. 

In a recent paper~\citep{Ilie:2023JADES}, two of us showed that JWST may have already discovered Dark Stars, although differentiation from early galaxies is not yet possible, given the low S/N of the available spectral data for the SMDS candidates we identified based on photometric data, as well as the lack of resolution to tell if the detected objects are point or extended in nature.
 As yet only $\sim 10$ high redshift objects found by JWST have spectra, which is required to prove that the objects are indeed at $z>10$ based on clean observation of the Lyman break.  We examined the four high redshift objects in the JADES (JWST Advanced Extragalactic Survey) data~\citep{JADES:2022a,JADES:2022b}, and showed that three of them (JADES-GS-z13-0, JADES-GS-z12-0, JADES-GS-z11-0) are consistent with being Dark Stars.  In addition to JWST,  the upcoming RST will also observe some of the first galaxies, and potentially even the first stars. In this paper we address the following question: how can one disambiguate between supermassive dark stars (SMDS) and the first galaxies made entirely of standard Population III stars with RST.      

The standard picture of the formation of the first stars may be drastically changed due to the role of the large dark matter abundance at the center of the host  minihalo, leading to a heat source for the collapsing molecular cloud. 
\cite{Spolyar:2008dark} first considered the possibility that, if dark matter consists of Weakly Interacting Massive Particles (WIMPs), their annihilation products would be trapped inside the collapsing molecular cloud, thermalize with the cloud, and heat it up. This dark matter heating can overcome the dominant hydrogen cooling mechanisms and thereby halt the collapse of the protostellar gas cloud.  As such, DM heating could lead to the formation of a new phase in the stellar evolution, a Dark Star. 
These are actual stars, made almost entirely of hydrogen and helium, with dark matter providing only 0.1\% of the mass of the star.  The stars are in hydrostatic and thermal equilibrium, obeying all the equations of stellar structure.  Initially $\sim 1 ~\rm M_\odot$, dark stars grow by accretion and some can become very massive ($> 10^6 ~\rm M_\odot$) and very bright ($>10^9 L_\odot$). They are puffy diffuse objects, $\sim 10$ A.U. in size, with dark matter annihilation power spread uniformly throughout the star.  They have low surface temperatures ($T_{\rm eff} \sim 10^4$K), leading to very little ionizing radiation and thus very little feedback preventing their further accretion.  Thus, some of the dark stars can grow to be supermassive SMDSs.

In a recent paper~\citep{Wu:2022SIDMDS}, one of us considered the possibility that dark matter, rather than being  comprised entirely of WIMPs, consists instead of a different type of particle, namely Self Interacting Dark Matter (SIDM).  In the case of SIDM, it was found that the deepening gravitational potential can speed up gravothermal evolution of the SIDM halo, yielding sufficiently high dark matter densities for Dark Stars to form. The SIDM-powered Dark Stars can have similar properties, such as their luminosity and size, as Dark Stars predicted in WIMP dark matter models. 
Regardless of the nature of the DM particle powering a SMDS (WIMPs vs SIMPs vs any other potential candidates that could form a DS) the final fate of SMDSs is the same: once the DM fuel runs out the object eventually collapses to a supermassive black hole (SMBH), leading to an interesting possible explanation of the many puzzling SMBHs in the Universe found even at high redshifts.

In this work we consider the possible detection of Supermassive Dark Stars (SMDSs) with the upcoming Roman Space Telescope (RST), which is set to launch in the mid-2020's.  
Some, but not all, supermassive dark stars vs  early galaxies comprised only of standard Pop~III stars  have different photometric signatures in JWST~\citep[see][]{Zackrisson:2011,Ilie:2012}. 
In order to familiarize the reader with the subject of SMDS detection and their observable signatures, in the remainder of this paragraph we summarize  the results obtained by two of us~\citep{Ilie:2012} regarding the predictions of SMDSs for JWST;  many of the remarks here will apply also to RST. One of the main results of~\cite{Ilie:2012} was that at reshifts $z_{\rm emi}\sim10$ SMDSs of mass $10^6\Msun$ or higher are bright enough to be detectable in all the bands of the NIRCam instrument on JWST, even without gravitational lensing. With photometry distant objects ($z\gtrsim 6$) are detectable as dropouts, i.e. the presence of an image in a band and its absence in the immediately adjacent band probing shorter wavelengths. This is due to the efficient attenuation  of Lyman-$\alpha$ photons by the neutral H present in abundance at redshifts higher than $z\sim 6$. 
For instance, with JWST, SMDS will show up as either J-band ($z_{\rm SMDS} \sim 10$), H-band ($z_{\rm SMDS} \sim 12$), or K-band dropouts ($z_{\rm SMDS} \sim 14$). Moreover, based on null detection results with HST (therefore assuming SMDS survive until $z\sim 10$ where HST bounds apply) in~\cite{Ilie:2012} we estimated that a multiyear deep parallel survey with JWST covering an area of 150 arcmin$^2$ can find anywhere between one and thirty unlensed $10^6\Msun$ SMDSs 
as either K band or H band dropouts. Of course this number can be even higher~\footnote{as high as $10^5 f_{\rm SMDS} $, where $f_{\rm SMDS} \ll1$ is the fraction of early DM halos hosting SMDS.} if SMDSs do not survive until $z\sim 10$, and therefore our bounds based on HST null detection do not apply. Compared to JWST, the  wider effective field of view (FOV) of the RST will increase the probability of detection so that the predicted number of objects will be larger.

In principle there are three techniques to differentiate between SMDSs and early galaxies dominated by Pop~III stars:
(a) SMDSs are essentially point sources vs. early galaxies are extended objects, (b) different locations in color-color plots, and (c) spectra including specific spectral lines. In somewhat more detail: 
First, SMDSs are point objects, whereas Pop~III galaxies are not. This can show as a  point spread function (PSF) difference, if the telescope has high enough resolution, and/or if the objects are gravitationally lensed. For RST we apply this technique in Sections~\ref{sec:compare} and ~\ref{sec:lensing}. Secondly the photometric signatures in color-color plots can be quite different for SMDS for which nebular emission is negligible vs. Pop~III galaxies. This was shown in~\cite{Zackrisson:2011,Ilie:2012} for the case of JWST, and for RST we do this analysis in Sections~\ref{sec:compare} (see fig.~\ref{fig:Colorcolor}) and~\ref{sec:lensing} (see figs.~\ref{fig:Colorcolor100} and \ref{fig:comparecolor100}). The differences are due to different spectral features, that, in turn, are converted into different colors in various bands. Third, the most precise way to disambiguate is to have a full spectral analysis. We use the \textsc{TLUSTY} code~\citep{hubeny1988computer} to obtain spectra for the SMDSs, as presented in Sec.~\ref{sec:Spectra} (see also \citet{Ilie:2012} for previous work).
Since SMDSs are stars, their spectra can roughly be modeled as black bodies.  The more detailed spectra obtained using \textsc{TLUSTY} also include the effects of the stellar atmospheres, while ignoring any possible effects of nebular emission. For comparison with spectra of Pop~III galaxies, we use simulations from the Yggdrasil model grids of early galaxies~\citep{Zackrisson2011}.\footnote{see also ~\url{https://www.astro.uu.se/~ez/}} SMDS are usually too cool to produce any nebular emission lines, whereas the spectrum of a Pop~III galaxy is usually dominated by nebular emission.\footnote{For completeness in this paper we also consider the case of Pop~III/II galaxies with no nebular emission} Particularly relevant for the JWST are the He-II emission line at $\lambda \sim 1.6 \mu$m (also called the He~II $\lambda1640$ line) and the H-$\alpha$ emission. Those would be telltale  signatures of a  Pop~III galaxy. By contrast, for SMDSs, at the same wavelength we would have an absorption feature. Other prominent SMDS spectral features (that get redshifted into the region RST is sensitive to) are He-I absorption and the Lyman-$\alpha$ break for the case of the coolest SMDS ($T_{\rm eff} \lesssim 2\times 10^4$K), whereas for the hottest SMDS ($T_{\rm eff} \sim 5\times 10^4$K) the main spectral features will be He~II absorption lines. A detailed discussion of the SMDS spectra can be found in Sec.~\ref{sec:Spectra}. 

Regarding the spectra of Dark Stars: we have in this paper not included the effects of nebular emission, which could make differentiation from early galaxies more difficult. We expect negligible nebular emission for the cooler dark stars, those in which the DM is obtained only gravitationally via AC.  For the hotter DS formed via capture, we make the assumption of no nebular emission in the studies in this paper but recognize the need to do better in future work. In more detail:
SMDS with $T_{\rm eff} \lesssim 2\times 10^4$~K, such as all SMDSs formed via Adiabatic Contraction and powered by WIMPs more massive than $100$ GeV (see Fig.~\ref{fig:SMDSHR}), are too cool to produce any nebular emission. Moreover, the neutral H present in abundance in their atmospheres absorbs very efficiently the vast majority of their photoionizing flux, as one can see in Fig.~\ref{fig:spectra}. For the hotter ones (such as those formed via DM Capture, or via Adiabatic Contraction and powered by WIMPs lighter than $100$ GeV) there is a possibility that they might create an ionization bounded nebula, and therefore exhibit nebular emission~\citep{Zackrisson:2011}. The main uncertainty in this scenario is how much gas is available around a SMDSs with $T_{\rm eff} \gtrsim 2\times 10^4$~K. 
In this paper we consider only SMDSs of purely stellar spectra, i.e., the photometric differences and spectral differences we work out in this paper are only valid under the assumption of SMDSs without nebular emission.  We emphasize the caveat that the neglect of nebular emission may alter the results for the hotter SMDS formed via capture considerably, as must be studied in future work. On the other hand, the spectrum of a Pop~III/II galaxy is usually dominated by nebular emission.

The literature on the possibility that DM annihilation might have effects on  stars dates back   to the $'80$s and early $'90$s, with the initial work studying the effects on current day stars~\citep[e.g.][to name a few]{Krauss:1985,Press:1985,Spergel:1985,Gould:1987,Salati:1987,Gould:1990WimpConduction,Gould:1992ApJ}. Regarding the DM heating effects on the {\it{first stars}} we mention~\cite{Spolyar:2008dark,Spolyar:2009,Freese:2008cap,Freese:2008ds,Freese:2010smds,Taoso:2008PhRvD,Yoon:2008,Iocco:2008,Casanellas:2009dp,Ripamonti:2009xw, Ripamonti:2010ab,Gondolo:2010dmds,Hirano:2011ig, Sivertsson:2011,Ilie:2012,Gondolo2022ApJ...935...11G}. For reviews see Ch. 29 (``Dark Matter and Stars'') of~\citet{bertone2010}, ~\cite{Tinyakov:2021lnt} and~\cite{Freese:2016dark}.  

We organize this paper as follows. We start, in Sec.~\ref{sec:DS} with a brief review of dark stars and their properties. A discussion of the simulated spectra of supermassive dark stars (SMDSs) is presented in Sec.~\ref{sec:Spectra}. Then, in Sec.~\ref{sec:Detect}, we study the detectability of SMDSs with the Roman Space Telescope and make predictions on observational results. In Sec.~\ref{sec:compare}, we consider different possibilities 
for the observed high redshift ($z\gtrsim 10$) objects, such as  Pop~III galaxies, and compare their observable properties to those of SMDSs. The effect of gravitational lensing on our results is discussed in Sec.~\ref{sec:lensing}. In Sec.~\ref{sec:conclusion}, we present  conclusions and summarize our study. We end three appendices: Appendix~\ref{Pandeia} where we present key parameters used for our simulated images,  Appendix~\ref{ap:Nebular} where we estimate the maximum mass for a SMDS formed via DM capture that would have sufficient gas surrounding it to form a ionization bounded nebula, and Appendix~\ref{ap:SMDSvsSMSs}, where we discuss the comparison between SMDSs and other types of primordial Supermassive Stars commonly considered in the literature. Throughout this paper, we will assume $\Lambda \mathrm{CDM}$ Cosmology and use the following cosmological parameters: $\Omega_{\Lambda}=0.73, \Omega_{\rm M} =0.27$, and $H_{0}=72 \mathrm{~km} \mathrm{~s}^{-1} \mathrm{Mpc}^{-1}$.

\bigskip
\section{Dark Stars}\label{sec:DS}

The first stars form as clouds of molecular hydrogen start to collapse inside the dark matter rich centers of $10^6 ~\rm M_\odot$ minihalos at $z \sim 10-20$.
As first showed by~\cite{Spolyar:2008dark}, the collapsing baryons gravitationally  pull DM in with them, increasing even further the DM abundance.  
If the DM is made of Weakly Interacting Massive Particles (WIMPs), they annihilate among themselves.
The WIMP
annihilation rate is $n_\chi^2 \langle \sigma v \rangle$ where
$n_\chi$ is the WIMP number density, the standard annihilation cross
section is


\begin{equation}
\label{eq:sigmav}
\langle \sigma v \rangle \simeq 
3 \times 10^{-26} {\rm cm}^3/{\rm s},
\end{equation}
and WIMP masses are in the range 1 GeV-10 TeV.  
WIMP annihilation produces energy at a rate per unit volume 
\begin{equation}
\hat Q_{\rm DM}  = n_\chi^2 \langle \sigma v \rangle m_\chi c^2 =
\langle \sigma v \rangle \rho_\chi^2 c^2 /(m_\chi) ,
\label{eq:Q}
\end{equation}
 $m_\chi$ is the WIMP mass,\footnote{ From here on we adopt the natural system of units, where $c=1$ and therefore mass has the same dimensions as energy.} 
and $\rho_\chi$ is the WIMP energy density. 
The cross section in Eq.(\ref{eq:sigmav}) is the canonical value used for WIMP annihilation by the dark matter community: WIMPS annihilating with this (weakly interacting) value in the early Universe automatically lead to the correct relic dark matter abundance today to explain the dark matter. This value does have model-dependence, with detailed models predicting slightly different values of the annihilation cross-section. Hence 
it is important to emphasize that Dark Stars are produced for a wide variety of WIMP masses and annihilation cross sections.  Indeed the cross section may be several orders of magnitude larger or smaller than the canonical value in Eq.(\ref{eq:sigmav}), and yet (roughly) the same DSs result.  Further, we note that the canonical cross-section used in the paper is in agreement with all experimental bounds including those from indirect detection.
In previous work we 
studied WIMP masses $m_\chi \sim 1{\rm GeV} - 10~\rm TeV$ (as we will show below, see Fig. 1, the lighter WIMPs provide 
more heating and therefore slightly different DSs since $Q \propto 1/m_\chi$).  Since the heating rate scales as $Q \propto \rho_\chi^2/m_\chi$,
considering a variety of WIMP masses is equivalent to studying a variety of annihilation cross sections.  

The annihilation products
typically are electrons, photons, and neutrinos. The neutrinos escape
the star, while the other annihilation products are trapped in the
dark star, thermalize with the star, and heat it up.  The luminosity
from the DM heating is
\begin{equation}
\label{eq:DMHeating}
L_{\rm DM}  \sim f_Q \int \hat Q_{\rm DM}  dV 
\end{equation}
where $f_Q \sim 1$ is the fraction of the annihilation energy deposited in
the star and $dV$ is the volume element.

Dark stars are born with masses $\sim 1 ~\rm M_\odot$.  They are giant puffy
($\sim 10$ AU), cool (surface temperatures $<10,000$K), yet bright
objects~\citep{Freese:2008ds}.  They reside in a large
reservoir of baryons, i.e., $\sim 15$\% of the
total minihalo mass $10^6 - 10^8 ~\rm M_\odot$.  These baryons can start to accrete onto the dark
stars. Dark stars can continue to grow in mass as long as
there is a supply of DM fuel.  Two mechanisms--discussed below--can provide enough DM fuel to potentially 
allow the DS to become supermassive 
 ($M_{\rm SMDS}  >10^5 \Msun$) and very luminous ($L>10^9 L_\odot$).  \hfil\break
  1) {\it Extended Adiabatic Contraction (AC):}   The infall of baryons into the center of the minihalo provides a deeper potential well that increases the DM density.
  A simple approach towards this gravitational
  effect is the Blumenthal method for adiabatic contraction. We~\citep{Freese:2008dmdens} and others~\citep{Iocco:2008,Natarajan:2009}
  have previously confirmed that this simple method provides
  dark matter densities accurate to within a factor of two, sufficient for these studies.
  There remains the question of how long this process can continue.
   In the central cusps of triaxial DM halos DM particles follow a variety of centrophilic orbits \citep[box orbits and chaotic orbits][]{Valluri:2010Triaxial} whose population is continuously replenished, allowing DM annihilation to continue for a very long time (hundreds of
  millions of years or more).  \cite{Freese:2010smds} then followed the
 growth to supermassive dark stars (SMDSs) of mass
 $M_{DS} >10^5 \Msun$.   \hfil\break 
2) {\it Capture:} At a later stage, there is  an additional mechanism that provides DM fuel.

  Once the initial DM reservoir is exhausted (or the DS is kicked away from the DM rich region), the star shrinks, its density
  increases, and subsequently further DM is captured from the surroundings~\citep{Freese:2008cap,Iocco:2008cap,Sivertsson:2011} as it scatters elastically off of nuclei in the star.  In this
  case, the additional particle physics ingredient of WIMP scattering
  is required.  This elastic scattering is the same process that
  direct detection experiments (e.g. LUX/XENON, SuperCDMS, DAMA, CRESST) rely upon in WIMP direct detection searches.  The DM capture rate, and, in turn, the properties of the ensuing Dark Star, will depend on two key parameters: the ambient DM density ($\bar{\rho}_\chi$) and the elastic scattering cross section $\sigma$. In what follows we assume $\sigma \bar\rho_\chi = 10^{-40} {\rm cm}^2 \times 10^{14}$GeV/cm$^3$. The DM density assumed is consistent with estimates based on the adiabatic contraction prescription~\citep{Spolyar:2008dark,Freese:2008dmdens} and consistent with numerical simulations~\citep{Abel:2001}, while the scattering cross section is within the allowed region of parameter space for Spin Dependent interactions~\citep{Amole:2019fdf,Aprile:2019}.

\bigskip
\subsection{Dark Star Properties}
The properties of dark stars have been studied in a series of papers, first using polytropic models~\citep{Freese:2008ds,Spolyar:2009},\footnote{A polytrope of index $n$ is defined by the following relationship betwen pressure and density in a star: $P(r)\sim\rho(r)^{1+1/n}$. Whenever such a relationship exist, the mechanical structure of the star, i.e. $P(r)$ and $\rho(r)$, can be found without any knowledge of the energy source powering the star.} and then using the MESA stellar evolution code. For the case of polytropic models for supermassive dark stars see~\cite{Freese:2010smds}. Dark Stars start as $\sim 1~\Msun$ convective stars in thermal and hydrostatic equilibrium, powered exclusively by adiabatically contracted DM annihilations. As such they can be well modeled by $n=3/2$ polytropes. Accretion leads to their growth, and by the time they reach roughly $100-500 \Msun$ they become radiation pressure dominated and can be well modeled by $n=3$ polytropes. As any radiation pressure dominated star, dark stars more massive than $\sim 500 \Msun$ will have a luminosity approximated well by the Eddington limit:
\begin{equation}
    L_{\rm Edd} = \frac{4\pi c G M_\star}{\kappa_\star},
\end{equation}
where $G$ is the Universal gravitational constant, $c$ is the speed of light, $M_\star$ is the mass of the star in question, and $\kappa_\star$ is the stellar atmospheric opacity.
To get an order of magnitude estimate we can assume that the dominant opacity source in metal-free atmospheres of supermassive dark stars is due to Thompson electron scattering. This is a function of the hydrogen fraction ($X$) of the star: $\kappa_\star \simeq \kappa_{es} = 0.2(1+X) \unit{cm}^2 \unit{g}^{-1}$.  Further assuming a big bang nucleosynthesis (BBN) composition of the stellar atmospheres allows one to recast the Eddington limit as:
\begin{equation}\label{eq:Ledd}
    L_{\rm Edd} \simeq 3.7 \times 10^{4} (M_\star/M_\odot)L_\odot
\end{equation}
For the case of adiabatically contracted DM powered Dark Stars
the polytropic approximation has been tested and confirmed by~\cite{Rindler-Daller:2015SMDS} using numerical solutions of the full stellar structure equations using the Modules for Experiments in Stellar Astrophysics (\textsc{MESA}) 1D stellar evolution code. By the time their mass becomes $M_\star\sim 10^5-10^7\Msun$, for SMDS formed via either the extended AC (without capture) mechanism or via DM capture, their luminosity is as high as $L_{\rm DS}\sim 10^9-10^{10}L_{\odot}$. 
\begin{figure}[!tb]
    \centering
    \includegraphics[width=.6\linewidth]{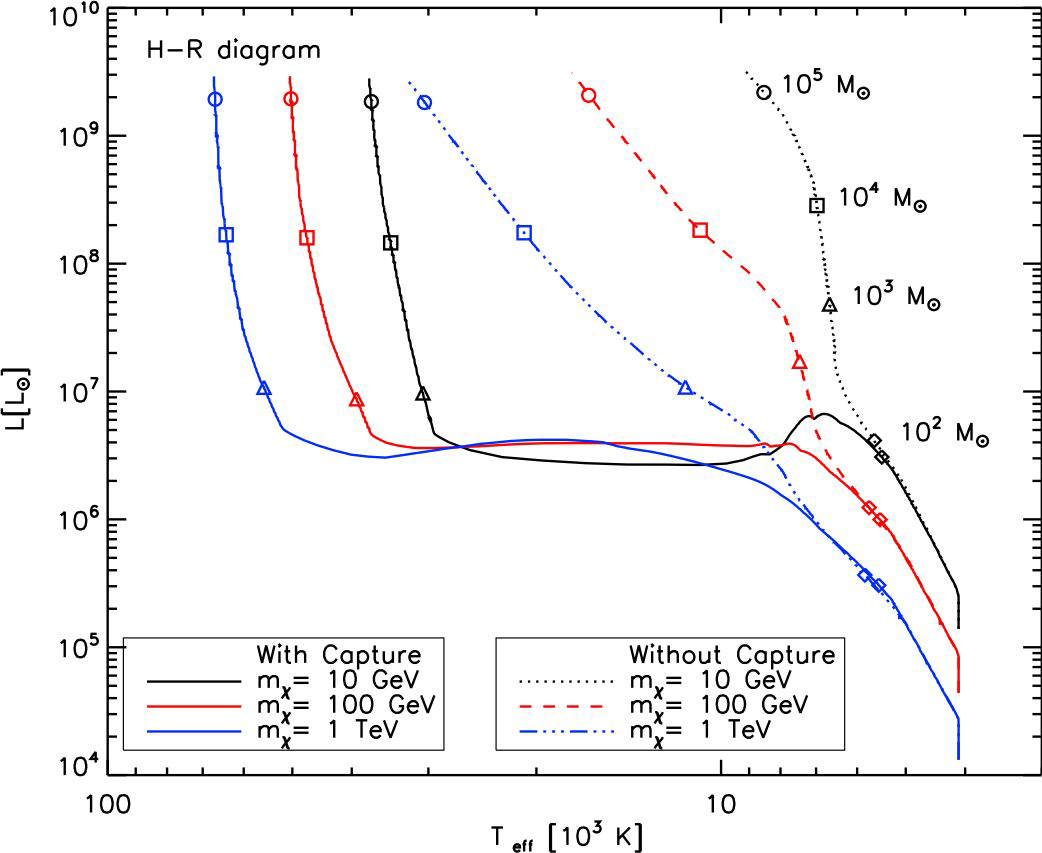}
    \caption{Hertzsprung-Russell (HR) diagram for dark stars for accretion rate $\dot M = 10^{-3} \Msun$/yr  and a variety of WIMP masses as labeled for the two cases: (i) ``without capture'' but with extended adiabatic contraction (dotted lines) and (ii) ``with capture''(solid lines).  The case with capture is for product of scattering cross section times ambient WIMP density $\sigma \bar\rho_\chi = 10^{-40} {\rm cm}^2 \times 10^{14}$GeV/cm$^3$.  Also labeled are stellar masses reached by the DS on its way to becoming supermassive. The final DS mass was taken to be $1.5\times 10^5 \Msun$ (the baryonic mass inside an assumed $10^6\Msun$ DM host halo), but it could be larger, depending on the mass of the host halo. (Figure reproduced from~\cite{Freese:2010smds}).}
  \label{fig:SMDSHR}
\end{figure}
In Fig.~\ref{fig:SMDSHR} (reproduced from ~\cite{Freese:2010smds}) we plot the evolutionary tracks in an HR diagram for dark stars that grow via accretion and reach a supermassive phase (SMDS) via either dark matter capture (labeled ``with capture'') or extended AC (labeled ``without capture''). Those evolutionary tracks are obtained by solving numerically the equations of stellar structure under the simplifying assumption of a polytrope of variable index. From Eq.~(\ref{eq:Ledd}) we can see that for even more massive dark stars, the luminosity scales linearly with the stellar mass. Note that lighter DM particles are more efficient in heating up a Dark Star, in both the Extended AC and the ``with capture'' cases, as $L_{\rm DM} \sim 1/m_\chi$ (this fact is easy to see for the extended AC case by combining equations~\ref{eq:Q} and~\ref{eq:DMHeating}). This effect is manifest in Fig.~\ref{fig:SMDSHR} in two ways. First, at the same effective temperature ($T_{\rm eff} $) a SMDS powered by a lighter WIMP will be brighter and puffier for both formation mechanisms considered. For a given luminosity a SMDS powered by a lighter DM particle (in either scenario) will be cooler and puffier than those powered by heavier DM particles. We emphasize that the evolutionary tracks of Fig.~\ref{fig:SMDSHR} are largely insensitive to the assumed accretion rate, which only dictates how fast a DS can grow from the moment it was formed ($z_{\rm form}$) until the moment it emits light to be observed by a telescope ($z_{\rm emi}$). As such, the  observable properties of SMDSs are largely independent of the assumed formation redshift or accretion rate, and are instead controlled by the stellar mass, stellar composition, surface temperature, and radius. In turn, the later two are fixed by DM parameters such as its mass ($m_\chi$), annihilation cross section ($\sigmav$), and, for the case ``with capture'' ambient DM density ($\rho_\chi$).
For future reference we list in Table~\ref{tab:SMDSpara} relevant parameters for  SMDSs obtained via the polytropic approximation
for the case of 100~GeV WIMPs and SMDS formed in $10^8 M_{\odot}$ DM halos at redshift $z_{\rm form}=15$.
\begin{table}[h!]
\centering
 \begin{tabular}{lccccc}
 \hline
 Formation Mechanism &$M_{*}$  & $L_{*}$ & $R_{*}$  & $T_{\text {eff }}$ & g \\ &$\left(M_{\odot}\right)$ & $\left(10^{6} L_{\odot}\right)$ & $(\mathrm{AU})$ & $\left(10^{3} \mathrm{~K}\right)$ &$m/s^2$  \\ [0.5ex] 
 \hline\hline
Extended AC & $2.04\times 10^{4}$ & 407 & 31 & 10 &0.126  \\
Extended AC & $10^{5}$ & $2.42\times 10^3$ & 39 & 14 & 0.390 \\
Extended AC & $10^{6}$ & $2.01 \times 10^{4}$ & 61 & 19 & 1.59 \\
Capture & $4.1 \times 10^{4}$ & 774 & 1.8 & 49 & 75.1\\
 Capture & $10^{5}$ & $1.75 \times 10^{3}$ & 2.7 & 51 &81.4 \\
 Capture & $10^{6}$ & $2.03 \times 10^{4}$ & 8.5 & 51 & 82.1 \\ \hline
 \end{tabular}
 \caption{Parameters of SMDSs used in this paper. The values listed here are adopted from~\cite{Freese:2010smds} (see their Tables~3 and 4). We assume both type SMDSs are powered by annihilations of 100~GeV WIMPs and formed in $10^8 M_{\odot}$ DM halos at redshift $z_{\rm form}=15$ and grow via accretion, at a rate of $\dot{M}=10^{-1} M_{\odot} \mathrm{yr}^{-1}$. For the case of a SMDS formed via DM capture, we further assume that the product between the ambient DM density and the DM-proton scattering cross section is: $\rho_\chi\sigma=10^{14}~\GeV\percc\times 10^{-40}\unit{cm}^2$}\label{tab:SMDSpara}
\end{table}

\subsection{Death of Dark Stars and the Formation of Supermassive Black Holes}
Dark stars will continue to shine and grow as long as there are situated in a dark matter rich environment that provides the fuel for their existence.  The minihaloes they formed in will merge with other minihaloes, potentially providing a new fuel source for the DS.  Further DM may be captured from the surroundings.  Eventually the DM fuel will run out. At that point, 
in smaller DSs, the ignition of nuclear fusion in the star after the DM fuel is gone may lead to a period on the Zero Age Main Sequence (ZAMS) lasting $\sim 10^6$ years before collapse to a black hole.  Supermassive Dark Stars
 heavier than $\sim 10^4 M_\odot$, on the other hand, will collapse directly to SMBHs. 
 
 As such, supermassive dark stars offer an interesting explanation for the more than 50 SMBHs powering the high redshift ($z\gtrsim6.5$) quasars observed over the past decade~\citep[for statistical studies see][]{Wang:2019}. For instance, the most distant quasar observed, as of this writing, is UHZ1, a recently discovered galaxy at $z\sim 10$, harboring a very bright x-ray luminous quasar ($L_{\rm bolo}\sim 5\times 10^{45}~\unit{erg\, s^{-1}}$) with mass $10^7 - 10^8 M_\odot$ (inferred assuming Eddington accretion), as shown by~\cite{Bogdan:2023UHZ1}.
If one insists on explaining the UHZ1 data with this ``low'' mass seeds, such as those generated by the BHs formed by the collapse of $10-1000~\Msun$ Population~III unrealistic sustained super-Eddington accretion rates are necessary~\citep[see for e.g. Fig.~4 of][where a radiative efficiency of $\eta\simeq 0.1$ is assumed]{Bogdan:2023UHZ1}\footnote{A lower radiative efficiency would lead to larger mass of the BH, given the same amount of time and starting with the same seed; we note that $\eta\simeq 0.1$ is the standard value in the literature.}. Therefore the need for heavy, or even supermassive Black Hole seeds becomes evident when attempting to explain the mass of the SMBHs powering the most distant quasars~\citep{Natarajan:2023UHZ1}. Supermassive dark stars are an ideal candidates for such massive BH seeds.\footnote{We mention here an alternative supermassive BH seed: direct collapse to black holes of very metal poor low angular momentum gas clouds via dynamical instabilities~\citep{Loeb:1994wv,Belgman:2006,Lodato:2006hw,Natarajan:2023UHZ1}. For a recent review see~\citet{Inayoshi:2020}} As soon as the DM fuel runs out the SMDS will convert almost instantaneously into a SMBH, and this process could happen at redshifts as high as $z\sim15$ or even higher. 
\bigskip
\section{Dark Star Spectra\label{sec:Spectra}} 
\bigskip
Since the Dark Star spectra control most observable properties (such as color in various bands, signal to noise ratio, etc.) we discuss here the most prominent properties of the spectra, and the various differences in the two formation scenarios considered: with capture and extended AC. The spectra of SMDSs were first presented and discussed by two of us in~\cite{Freese:2010smds,Ilie:2012}.  

To zeroth order the spectrum of a DS (like any other star) is a blackbody.  
The role of the SMDSs stellar atmospheres in reprocessing of  photons and thus reshaping the  black body spectrum approximation can be investigated  by using the \textsc{TLUSTY} code~\citep{hubeny1988computer}. To simplify our discussion here, we only consider SMDSs of three masses: $\sim 10^4, 10^5$, and $10^6\Msun$. Lower mass SMDS have very little chance of detection, unless they are significantly gravitationally lensed, as we will show in Fig.~\ref{fig:DSvsMu}. 

\begin{figure}
\centering 
\includegraphics[width=0.99\columnwidth]{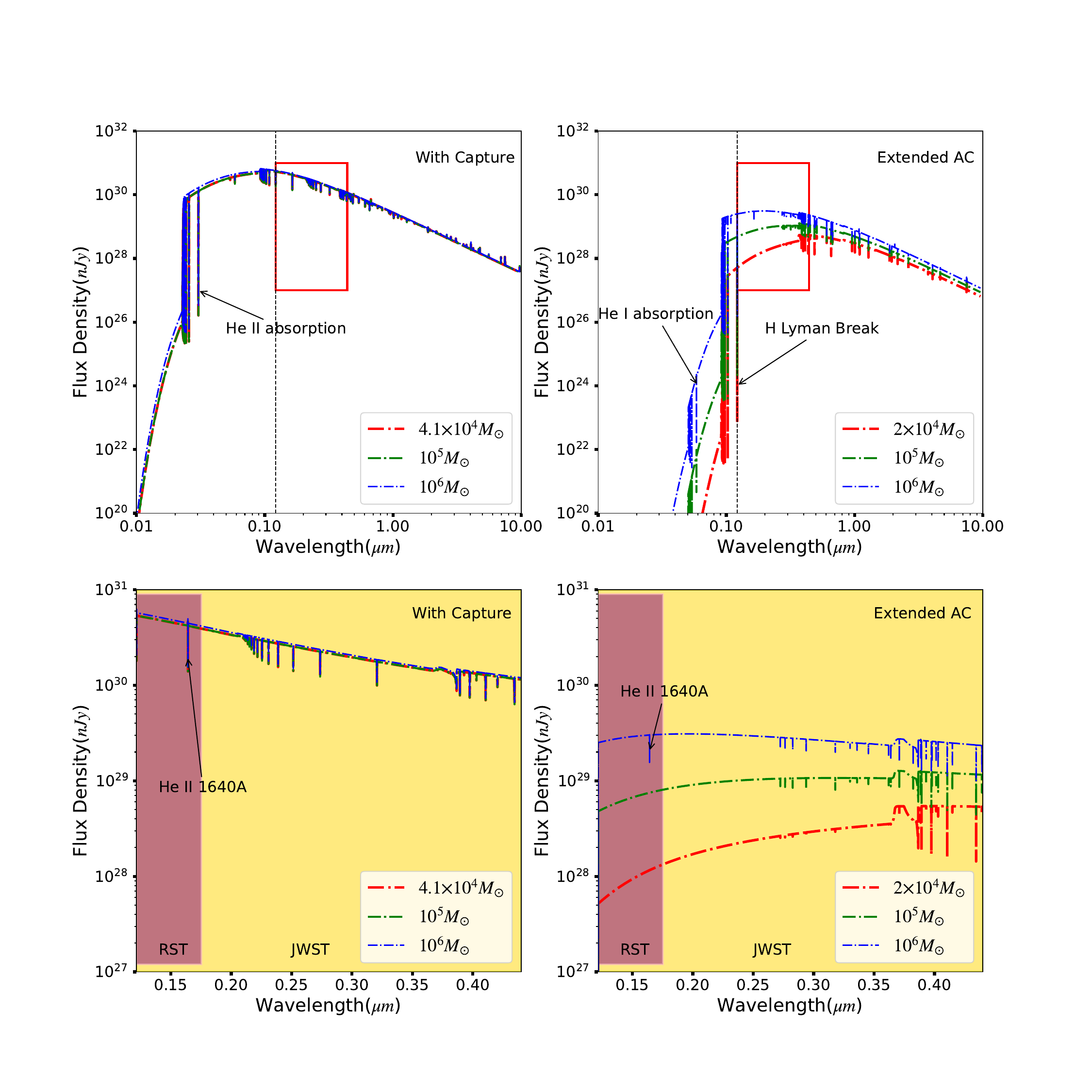}
\caption{\textsc{TLUSTY} simulated restframe SEDs of supermassive dark stars of three different masses, as labeled. Following~\cite{Ilie:2012} we assume a 100 GeV WIMP DM particle. 
Left panels: SMDS formed via capture with $M_{*} = 4.1\times 10^4, 10^{5}, 10^{6} M_{\odot}$, effective temperatures $T_{\rm eff}=4.9, 5.1, 5.1 \times 10^4 \mathrm{K}$, and  radii of $R_{*}= 1.8, 2.7, 8.5 \mathrm{AU}$ respectively. Right panels: SMDS formed via Extended AC with $M_{*} = 2\times 10^4, 10^{5}, 10^{6} M_{\odot}$, effective temperatures $T_{\rm eff}=1.0, 1.4, 1.9 \times 10^4\mathrm{K}$, and radii of $R_{*}= 31, 39, 61 \mathrm{AU}$ respectively. Top panels are fluxes for rest frame wavelengths up to $\microm{10}$, whereas in the lower panels we zoom in the red boxed regions of the spectra, i.e. from the \Lyalpha line up to $\microm{0.5}$. The zoom in region corresponds to rest frame wavelengths that can be probed with NIRSpec on JWST (shaded yellow), or the Grism spectrometer on RST (shaded with clay color), assuming $z_{\rm emi}\simeq 10$, and only considering wavelengths higher than the \Lyalpha line. To enable comparison of predicted spectra of SMDS and galaxies with RST and JWST,
we took the range of wavelengths of sensitivity for each instrument (Grism and NIRSpec), and then scaled these wavelengths back
as $1/(1+z_{\rm emi}) = 1/11$ to the epoch of the restframe SMDS and galactic SEDs, assuming $z_{\rm emi}\simeq 10$.
Note that the JWST region extends over the all the wavelengths shown in the plot (the yellow JWST region merely looks obscured by the darker clay color of RST).
The vertical line in the top panels represents the wavelength of the Lyman-alpha line, which is relevant since for objects at redshifts higher than $z\sim 6$ the flux at shorter wavelengths than the \Lyalpha line is highly suppressed via the Gunn-Peterson trough effect due to the neutral H gas abundant in the Intergalactic Medium at $z\gtrsim 6$. For objects emitting at $z\gtrsim 10$, the Lyman break is within the wavelengths probed by both Grism (on RST) and NIRSpec (on JWST). For RST, the most significant spectral features from SMDSs are the He~II absorption lines at $\microm{0.1640}$, marked in the standard notation as He~II 1640A in the lower two panels.}
\label{fig:spectra}
\end{figure}

In Fig.~\ref{fig:spectra} we present the rest-frame spectra of the three different mass SMDSs considered for both formation scenarios.  The left panel illustrates the spectra for the case of ``With Capture", i.e. when WIMPs captured due to elastic scattering are included.   The right panel illustrates the DS spectra for SMDSs in the case ``Extended AC", where only the WIMPs gravitationally pulled into the center of the minihalo provide the heat source.  In this figure we have taken the WIMP mass to be 100 GeV.  In the remainder of this section we discuss the wavelength, flux density, and most prominent spectral features (due to absorption by neutral H and He in the stellar atmospheres), since these will be important features in detecting SMDSs with Roman Space Telescope as well as differentiating them from high redshift galaxies containing more standard Pop III stars.

{\it Peak Wavelength:} As discussed above, the DS formed ``with capture" are hotter than those formed via ``extended AC" (as a reminder, DS ``with capture" form at a later stage of stellar evolution, when the original gravitational DM has run out, and the star has collapsed via Kelvin-Helmholtz contraction to a denser object capable of capturing further WIMPs.) As a consequence the DSs ``with capture" have spectra peaking at shorter wavelengths
 in the UV range at roughly $\microm{0.1}$.    The flux does not decrease substantially (with decreasing wavelength) until after the HeII (singly ionlized He) absorption break (due to absorption by the stellar atmosphere, see discussion below) at wavelengths ranging between $[\microm{0.023}-\microm{0.03}]$ (see top left panel of Fig.~\ref{fig:spectra}). In contrast, for SMDSs formed via the Extended AC mechanism, absorption by neutral H or He cuts off most of the UV flux at wavelengths shorter than the \Lyalpha line (see top right panel of Fig.~\ref{fig:spectra}).  Yet SMDSs ``with capture" have a steeper UV continuum slope $\beta$ ($f_{\lambda}\propto \lambda^{\beta}$), which is one of the factors leading to those objects having a larger magnitude difference between neighboring IR bands relevant for both JWST and RST (and used to search for dropouts as described below).

{\it Rest Frame Flux Density:} First, we note that  the flux density for SMDS formed via capture is roughly insensitive to the stellar mass since they have nearly equal surface temperatures.  Secondly,
by comparing the top two panels of Fig.~\ref{fig:spectra} one can see that the flux density around the peak for SMDS of the same mass formed via DM capture is about $100\times$ larger than that formed via Extended AC. This is due to the higher effective temperature ($T_{\rm eff} $) in the case ``with capture.'' However, as they reach supermassive status both type SMDS will be Eddington limited, and therefore their brightness is determined largely by the stellar mass, and to a lesser degree by their composition via the opacity of the stellar atmosphere ($\kappa_\star$), as per Eq.~(\ref{eq:Ledd}). The fact that different flux densities (extended AC vs capture) lead to the same total luminosity (Eddington limit) for both cases can be traced to their different radii: extended AC stars are more puffy since they never undergo a contraction phase. 

We also note that the question of mass loss for Dark Stars, in view of possible super-Eddington winds or pulsations was studied by~\cite{Rindler_Daller_2021}. That work  found that Dark Stars are not significantly affected by those effects. Hence wind  mass loss does not affect the spectral features of dark stars.

{\it Spectral Features:} The most prominent spectral features of SMDSs,  whenever nebular emission is negligible, are due to absorption by neutral H and He in the SMDSs stellar atmospheres (see Fig.~\ref{fig:spectra}). This is due to their relatively cool surface temperatures ($T_{\rm eff} \sim 10^4$~K). In contrast, the fluxes from Pop~III galaxies are, in many scenarios, dominated by nebular emission~\citep{Zackrisson2011}.
Going back to the features in the SMDSs spectra, below we contrast them for the two formation scenarios considered. The relatively low surface temperature ($T_{\rm eff} \lesssim 2\times 10^4$~K) for the SMDS formed via the extended AC scenario (right panels of Fig.~\ref{fig:spectra}) leads to a larger fraction of neutral H and He in their stellar atmospheres, when compared to a SMDS of the same mass but formed via the DM capture mechanism. This explains the strong absorption lines at wavelengths corresponding to the Lyman series~($\microm{0.1216}-\microm{0.0912}$), caused by neutral H, and, at shorter wavelengths~($\sim\microm{0.05}-\microm{0.06}$) the HeI break. Additionally, the presence of neutral H in the cooler SMDSs formed via AC leads to a prominent Balmer break at $\lambda\simeq\microm{0.36}$, followed by a sequence of strong absorption lines in the Balmer series, as can be seen in the lower right panel of Fig.~\ref{fig:spectra}. Conversely, the higher surface temperature (T$_{\rm eff} \sim 5\times 10^4$~K) of SMDS formed via DM capture implies a large ionization fraction for H, hence the  Lyman or Balmer  absorption lines are weaker. The most prominent SED feature of SMDS formed via capture is the HeII (singly ionized He) break at wavelengths ranging between $\microm{0.023}$ and $\microm{0.030}$. Common between the two cases are  HeI lines  at wavelengths $\sim~[\microm{0.3}-\microm{0.45}]$, the isolated HeII $\lambda1640$ absorption line at $\microm{0.1640}$, a sequence of HeII lines at
wavelengths $\sim\microm{0.46}$, and more HeI lines at $\sim~[\microm{0.47}-\microm{1.0}]$. Note that JWST and RST will be sensitive to a relatively narrow bin of restframe wavelengths, as shown in the lower two panels of Fig.~\ref{fig:spectra}. Thus, many of the spectral features discussed above will not be observable with either instrument. Since we expect SMDSs to be at $z_{\rm emi}\gtrsim 10$, this implies that all the features to the left of the restframe \Lyalpha line (i.e. $\microm{0.1216}$) are going to be erased by the Gunn-Peterson trough. At the other end, the highest value of the restframe wavelength probed is just $\lambda_{\rm max}/(1+z_{\rm emi})$, with $\lambda_{\rm max}$ being the maximum wavelength to which each instrument is sensitive (i.e. $\microm{1.93}$ for the Grism spectrometer of RST and $\microm{5}$ for NIRSpec onboard JWST). In turn, this implies that for objects at $z_{\rm emi}\gtrsim 10$ (such as SMDSs) RST will only probe the SEDs spectroscopically up to $\lambda_{\rm rest}\lesssim \microm{0.1754}$, whereas JWST will be sensitive to features in the SEDs up to $\lambda_{\rm rest}\lesssim \microm{0.44}$. The single most intriguing spectral feature that can be potentially detected in the region of overlap of both instruments is the HeII$ \lambda1640$ absorption line at restframe wavelength $\microm{0.1640}$. The observation of this feature would be a smoking gun for a SMDS, since at the same wavelength galaxies would typically exhibit a nebular emission line instead. 

{\it SMDSs vs. compact clusters of Pop~III stars:}  Pop~III stars, or clusters thereof, are unlikely to have similar spectra with SMDSs, due to the differences in combinations of their surface gravity ($g$) and $T_{\rm eff}$. Using the Pop~III parameters from \cite{Ilie:2020PopIII}, we compare $g$ and $T_{\rm eff}$ to the values listed in Table~\ref{tab:SMDSpara} (for SMDSs). Pop~III stars more massive than $~10 \Msun$ will be hotter than the hottest SMDSs considered in our work. In fact, by the time they reach $100\Msun$ Pop~III stars are as hot as $10^5$K. As such Pop~III stars occupy different locations in a HR diagram, and have significantly different continuum of their SEDs. Moreover, the surface gravity $g$ of SMDSs are less than $\sim 100 \, {\rm m/s}^2$ since they are puffy; whereas for all Pop III stars,  $g \gg 100 \, {\rm m/s}^2$. The differences in $T_{\rm eff}$ and $g$ would lead to very different spectral features.

\section{Detectability of Supermassive Dark Stars with the  Roman Space Telescope}\label{sec:Detect}

Here we examine the detectability of SMDSs with the Roman Space Telescope (RST). In this section we assume that the objects are unlensed\footnote{In future sections we will consider the case of SMDS with magnified imagaes due to lensing by foreground objects.}. In  Sec.~\ref{ssec:Detector}, we estimate in which RST wavelength bands SMDSs of various masses will be observable with exposure times ranging from $10^4-10^6$~s. Then in Section 4.2 
we examine the use of photometric dropout criteria for SMDSs in RST. We will show that SMDSs of mass $\sim10^6\Msun$ can be detected by RST with exposure times of $10^6$~s as J, H and H/K band dropouts, corresponding to emission redshifts of $z\sim 11$, $z\sim 13$, and $z\sim 14$, respectively.  

\subsection{Detector Capability and SMDS redshifted spectra}\label{ssec:Detector} 

\begin{figure}[!htb]
\includegraphics[width=\linewidth, height=8cm]{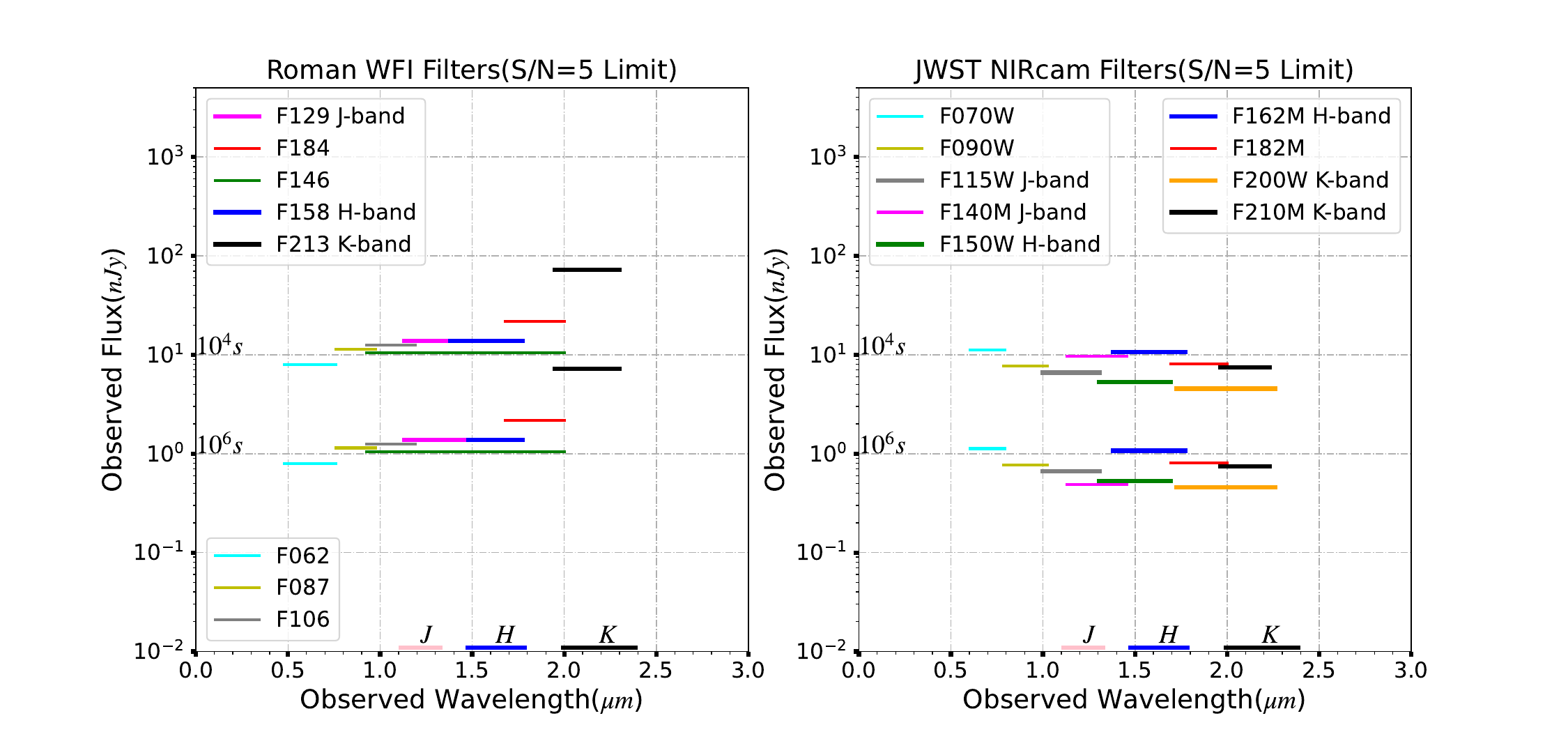}
\caption{Left panel: Detection limits (i.e. flux necessary to achieve a signal to Noise ratio (S/N) of 5) for all the Roman WFI filters assuming $10^4$ and $10^6$s exposure time. Right panel: same as left panel for JWST NIRcam SW (short wavelength where $\lambda_{\rm obs}< 2.5 \mu m$) filters. In general, there is no significant difference between the performance of JWST NIRCam and Roman WIFI in terms of sensitivity for wavelengths up to $\sim\microm{1.8}$. For filters with larger central wavelength NIRCam is more sensitive, as can be seen by contrasting the limits for the F213 RST WFI band (left panel) to those of the F200W or F210M  of NIRCam (right panel). However, with a wider field of view, RST has better odds than JWST to detect SMDSs as photometric dropouts in deep field surveys.}
\label{fig:telescope}
\end{figure}

Roman’s wavelength coverage of visible and infrared light will span 0.5 to 2.3 microns.  In Fig.~\ref{fig:telescope} we plot the projected sensitivity limits of the Roman Space Telescope in its Wide Field Instrument (WFI)  filters\footnote{\url{https://roman.ipac.caltech.edu/}}~(left panel) and contrast them to those of the NIRCam filters covering the same wavelength range ($\microm{0.5}-\microm{2.3}$) onboard the James Webb Space Telescope~\footnote{\url{https://jwst-docs.stsci.edu/jwst-near-infrared-camera/nircam-performance/nircam-sensitivity}}~(right panel). 
For both detectors we show the flux necessary to achieve a signal to noise ratio S/N $\simeq5$ for two exposure times considered: $10^4$ and $10^6$s.
The two telescopes have comparable sensitivity below $\sim\microm{1.8}$ 
whereas at higher wavelengths RST loses sensitivity, as seen by the gradual increase in the necessary flux to achieve a detection (S/N=5) in the F184 and F213 WFI bands of RST. This is in contrast with the  NIRCam instrument, for which the sensitivity actually mildly improves for filters probing those wavelengths. For instance the projected detection limits of the F213 WFI RST filter are about an order of magnitude weaker than the corresponding ones for the F200W NIRCam filter. This, in turn, corresponds to JWST being able to detect, in the F200W filter, objects that are ten times dimmer (i.e. 2.5 larger magnitude) than RST with the F213 WFI filter, assuming the same exposure time. For most other overlapping bands the difference in their detector capability is minimal. 

JWST is able to probe the universe deeper than RST, i.e. to higher redshifts, due to it being sensitive to larger wavelengths (the Mid-Infrared Instrument (MIRI) is sensitive up to 28 microns). However, RST excels in its Field of View (FOV) capability. For instance, each of the RST deep field images will cover an area of the sky roughly equivalent to the apparent size of a full moon. Conversely, Webb's First Deep Field image covers an area of the sky smaller than the apparent size of a grain of sand as viewed at armlength. As such, the probability of detecting SMDSs at redshifts as high as $z_{\rm emi}\sim 14$ using wide field surveys is much larger for RST\footnote{As we will see in Secs.~\ref{ssec:Dropouts} RST will be able to detect SMDS to redshifts as high as $z_{\rm emi}\sim 14$ as photometric dropouts.} than JWST, simply because each survey will cover a much larger area.

\begin{figure}[!htb]
\includegraphics[width=\linewidth, height=7cm]{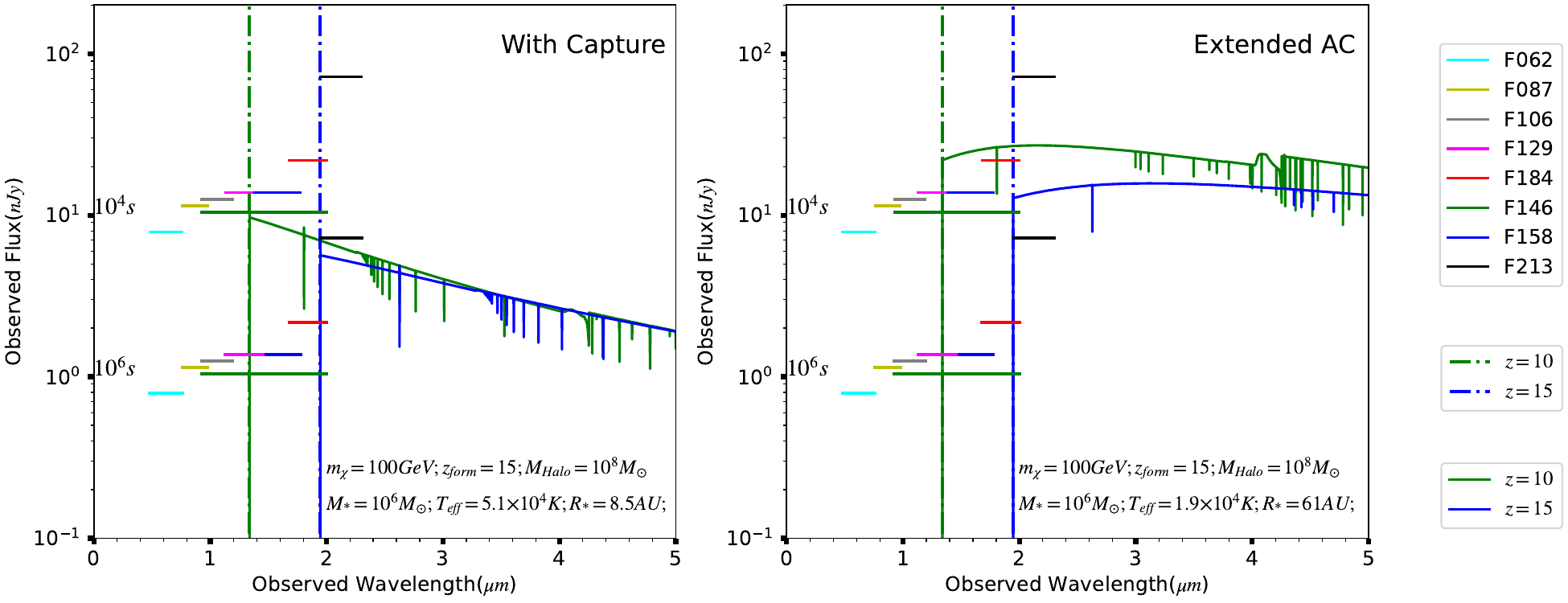}
\caption{Detectability of SMDSs with Roman Space Telescope, for $10^6 M_{\odot}$ SMDS powered by annihilations of 100 GeV WIMPs.  In both panels, green and blue solid lines are \textsc{TLUSTY} simulated SMDSs spectra at emission redshifts of 10 and 15, respectively. Left panel represents the spectra of SMDSs formed via Capture whereas the right panel are the SMDSs formed via Extended AC. The short horizontal lines are detection limits of all the Roman WFI filters obtained requiring a S/N=5 achieved in  $10^4$ (top lines) or $10^6$s (bottom lines) of exposure time. The vertical lines represent the redshifted \Lyalpha lines. For sources at $z\gtrsim 6$ the flux shortward of the redshifted \Lyalpha lines will be significantly reduced via attenuation  due to neutral H in the IGM. One can see that SMDSs formed via AC (right panel) will appear brighter in all bands covered by RST, when compared to SMDSs of the same mass formed via DM capture (left panel)}
\label{fig:spectra_z}
\end{figure}

Given the sensitivity limits of RST  and JWST, we can then overlay the relevant DS spectra to determine their detectability of a individual Dark Star in these instruments. We take the spectra in the rest frame of the star obtained in Section~\ref{sec:Spectra} and redshift them appropriately:  
\begin{equation}\label{eq:flux}
    F_{v}(\lambda_{\rm obs} ; z_{\rm emi})=\frac{(1+z_{\rm emi}) 4 \pi R_{*}^2 F_{v}\left(\lambda_{\rm emi}\right)}{4 \pi D_{\mathrm{L}}^{2}(z_{\rm emi})}, \ \lambda_{\rm obs}=(1+z_{\rm emi})\lambda_{\rm emi},
\end{equation}
where $\lambda_{\rm obs}$ and $\lambda_{\rm emi}$ represents the observed and emitted wavelength, $R_{*}$ is the radius of SMDSs, $D_{\mathrm{L}}$ is the luminosity distance, and $ F_{v}\left(\lambda_{emi}\right)$ and $F_{v}(\lambda_{\rm obs} ; z_{\rm emi})$ represents the original and shifted spectra. To be concrete, we assume a $10^6\Msun$ SMDS formed via either mechanism survives to various redshifts, as shown in Fig.~\ref{fig:spectra_z}. Moreover, we add the sensitivity limits discussed above, in order to assess the potential observablility of SMDSs with RST. Note that the SMDSs formed via AC (right panel) will appear brighter in all bands covered by RST, when compared to SMDSs of the same mass formed via DM capture (left panel). This is a combination of two factors. First, the SMDSs formed via DM capture are typically more compact, having undergone a Kelvin-Helmholtz contraction phase. Therefore, the $R_\star^2$ enhancement factor (see Eq.~(\ref{eq:flux})) is milder. Moreover, the slope of the UV continuum for the restframe fluxes of SMDSs formed via DM capture is steeper, as noted in our discussion of Fig.~\ref{fig:spectra}. This, in turn, will lead to a faster decrease with wavelength of their redshifted fluxes, as one can see from comparing the left and right panels of Fig.~\ref{fig:spectra_z}. 
Unlensed $10^6\Msun$ SMDSs formed via extended AC could be observed if emitting at $z_{\rm emi}\sim10$ (green line in right panel of Fig.~\ref{fig:spectra_z}) even with $10^4$~s exposure time with RST in the following bands: F129, F146, F184, F158, F184. For the same objects, a greater exposure time is needed for observation in the F213 band, as the detector loses sensitivity in that higher wavelength filter. On the other hand, SMDSs of the same mass formed via DM capture (left panel of Fig.~\ref{fig:spectra_z}) require an exposure time longer than $10^4$s (varying in the range $10^4 - 10^6$ depending on the band) in order to be detected at the level of S/N=5.

One common feature between all spectra presented in Fig.~\ref{fig:spectra_z} is the sharp cutoff of the flux at wavelengths shortward of the redshfited \Lyalpha line, which, in turn, will allow for the use of the photometric dropout detection technique for SMDSs or high redshift galaxies. Photons emitted from distant sources can be efficiently scattered by the neutral H in the intergalactic medium (IGM). For sources at redshifts $z\lesssim 6$ this effect is minimal, since it is roughly around that redshift that the universe has become fully reionized. In contrast, for more distant objects ($z\gtrsim 6$) redshifted photons that reach a cold gas cloud with neutral H  can get resonant-scattered out of the line of sight  very efficiently if their redshifted wavelengths as they reach the neutral H have a value of 1216\AA ~or, equivalently, $\microm{0.1216}$ (the \Lyalpha line). As a result, for any object emitting from $z_{\rm emi}\gtrsim 6$, the observed fluxes shortward of the redshifted \Lyalpha line, i.e. for $\lambda_{\rm obs}\lesssim (1+z_{\rm emi})1216$~\AA,
are highly suppressed.\footnote{This is the so called Gunn-Peterson trough~\citep{Gunn-Peterson:1965}, and it was observed for a large number of quasars since 2001~\citep{Becker:2001}.} Depending on the model assumed for reionization, the effect of the attenuation  of \Lyalpha photons by neutral H in the intergalactic medium (IGM) will be slightly different. However, for quasars emitting at $z_{\rm emi}\gtrsim 6$ observations show that most of the flux shortward of the redshifted \Lyalpha line is completely attenuated. In turn, this implies that the redshifted SEDs of SMDSs are cut off to the left of the horizontal lines plotted in Fig.~\ref{fig:spectra_z}.  In summary, the SMDSs formed via AC  will appear brighter in all bands covered by RST, when compared to SMDSs of the same mass formed via DM capture. Unlensed SMDSs at $z\sim 10$, of $10^6 M_\odot$ formed via AC can be detected at the level of S/N=5 with $10^4$s of exposure time in RST.

\bigskip
\subsection{Photometric dropout  criteria for Roman Space Telescope}\label{ssec:Dropouts}

In this subsection we turn the information regarding the redshifted fluxes of SMDSs into dropout criteria of their potential observation with RST. For luminous objects at $z\gtrsim 6$ photometry alone, i.e. color magnitudes in various bands, is sufficient to give a rough estimate of the emission redshift via the so called ``dropout technique" pioneered by \citet{Steidel:1996ym}. 
This photometric redshift determination method  requires a 5-sigma detection of an object  in one band but a non-detection in a adjacent band of lower wavelength. 
The absence of  emission in the latter bands  is assumed to occur due to Ly$-\alpha$ attenuation  by hydrogen clouds in
between the source and us, allowing  for an approximate estimate of the redshift of the object.  More specifically we take as our dropout criterion

\begin{equation}\label{eq:dropout}
\Delta m_{\rm AB} \geq 1.2
\end{equation}
where $\Delta m_{\rm AB}$ is the difference in apparent magnitude between
adjacent bands of observation.
 
For instance, for the $10^6~\Msun$ SMDS at $z\sim 10$ and formed via the extended AC mechanism considered in the right panel (green line) of Fig.~\ref{fig:spectra_z}, there will be a sharp increase in magnitudes (decrease in flux) as observed in any filter with central wavelength shorter than $\sim\microm{1.34}$. This, in turn, will lead to the SMDSs at $z\sim 10$ appearing as J band dropouts. 
 
 The ``dropout technique'' has been  applied extensively to J and H band observations of the Hubble Ultra Deep Field (HUDF). For example,~\cite{Bouwens:2009qs} used it to detect the first galaxy at $z\sim10$ as a "J-band dropout." This object was observed in the 1.60 $\mu m$  (H-band)  but was {\it not} seen in the 1.15$\mu m$  (Y -band) or 1.25 $\mu m$ (J- band).  Since then this technique has been used successfully to identify many high redshift luminous  objects. For instance, the pre-JWST record holder as the most dinstant galaxy candidate was HD1, at a whopping $z\sim 13$, i.e. only 300 Myrs after the Big Bang. This object (HD1) has been recently identified as a H band dropout by~\cite{HD1}. JWST data already broke this record multiple times~\citep[e.g.][]{GHz2,GLASSz13,JADES:2022a,z16.CEERS93316:2022,z17.Schrodinger:2022}, and all those new candidates for the title of ``most distant galaxy ever observed'' were first detected as photometric dropouts. It is worth emphasising that spectroscopy is the only available tool to confirm such high $z$ candidates.\footnote{The current record holder for the most distant spectroscopically confirmed Lyman break object is JADES-GS-z13-0~\citep{JADES:2022b}, an object also consistent with a SMDS interpretation, as we have shown in~\cite{Ilie:2023JADES}} If confirmed as galaxies those objects could pose  challenges to our current standard models of the formation of the first stars and galaxies, as they imply a much faster star formation rate and growth of galaxies in the cosmic dawn era than any of the available numerical simulations predict.

\begin{figure}[!htb]
\includegraphics[width=\linewidth, height=10cm]{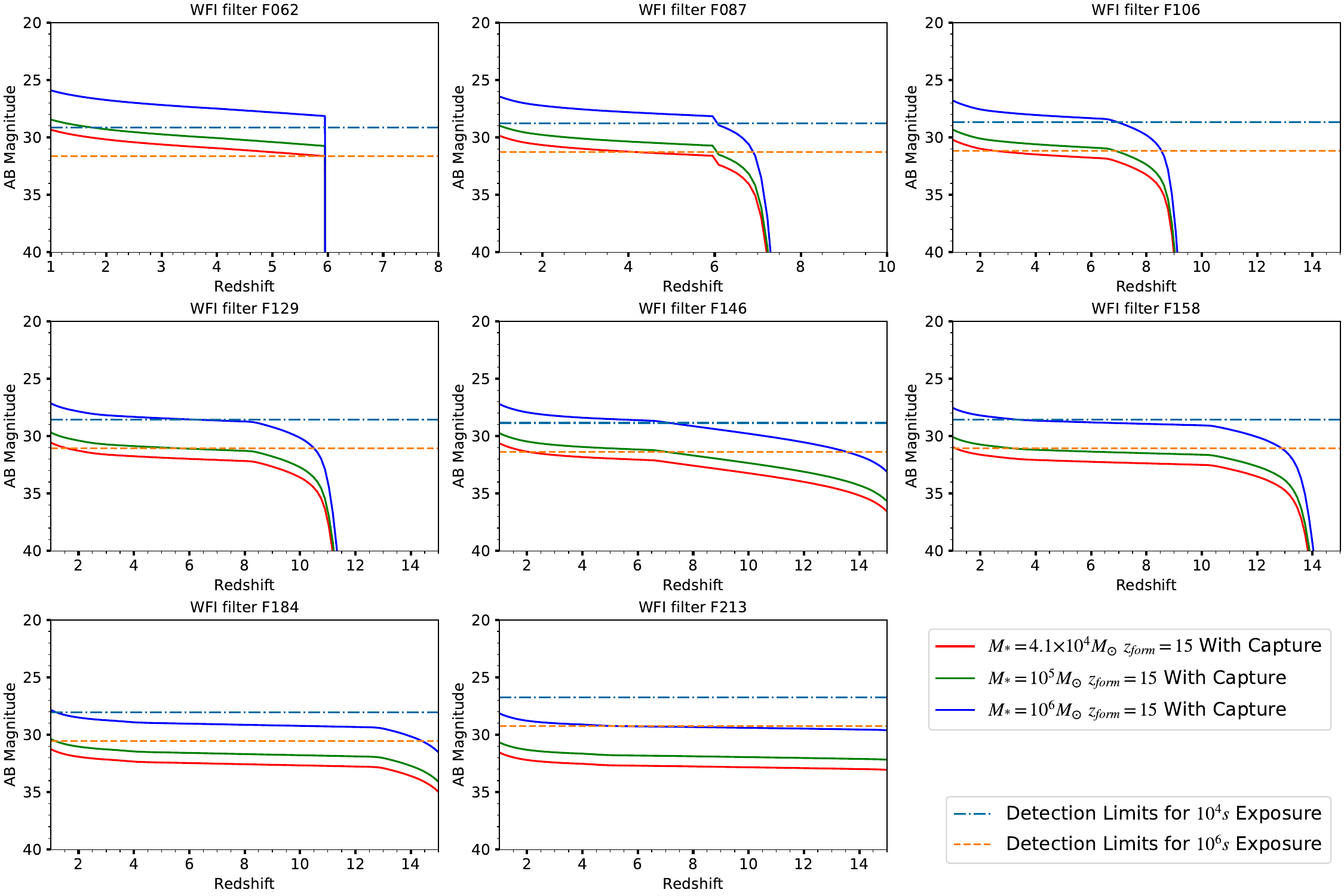}
\caption{AB magnitudes of SMDSs in Roman Space Telescope as a function of emission redshift, for SMDSs of various masses labeled in the legend, formed via the DM capture mechanism. (We note that magnitudes increase in the downward direction, i.e. the stellar luminosity increases in the upward direction for the y-axis). Each panel corresponds to a different filter, ordered in increasing order of central wavelength. The horizontal lines correspond to the detection limit (S/N=5) for each filter assuming exposure times of $10^4$s (dash-dotted line) or $10^6$s (dashed line). At $z\lesssim 6$ even the dimmer $2\times 10^4\Msun$ SMDS (red lines) formed via capture would be observed in the WFI F062 band of RST (see top left panel). The most massive SMDS considered here ($10^6\Msun$; blue lines) can be observed at redshifts as high as $z\sim 14$ in the WFI F184 and F213 filters (bottom panels). The sharp increase in AB magnitudes (i.e. dimmer object) that can be observed in all filters (except F213) at redshifts ranging from 6 to 14 is due to the \Lyalpha line (and thus the Gunn-Peterson trough) entering each filter at a different redshift.}
\label{fig:ABmagCap}
\end{figure}

\begin{figure}[!htb]
\includegraphics[width=\linewidth, height=10cm]{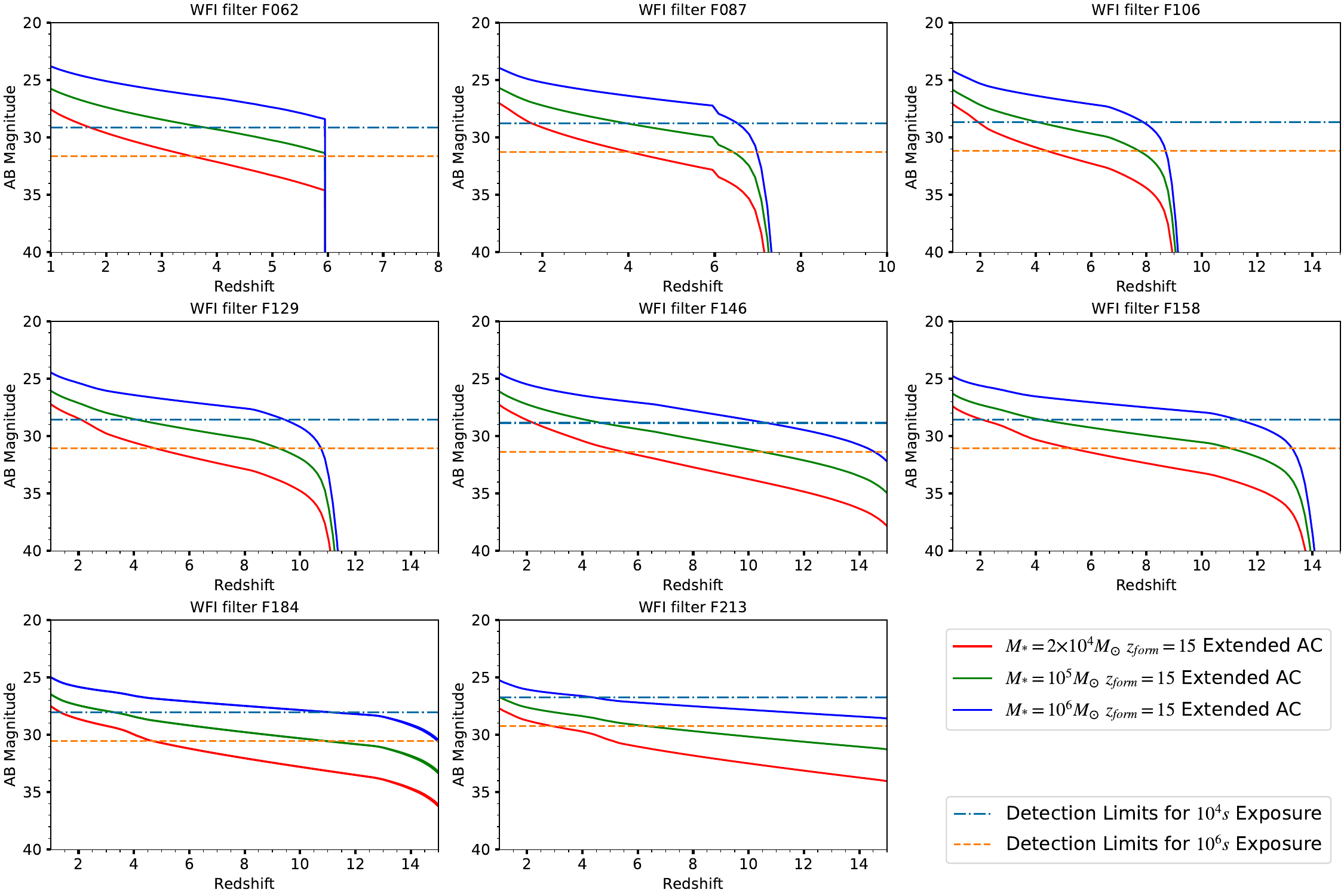}
\caption{Same as Fig.~\ref{fig:ABmagCap}, but now for SMDSs formed via the Extended AC mechanism. One general trend is that SMDSs formed via this mechanism will appear brighter, and thus easier to detect, than their counterparts formed via DM capture (Fig.~\ref{fig:ABmagCap}). The rest of the features of the magnitude vs redshift curves for SMDSs formed via the extended AC mechanism (presented in this figure) are very similar to those of the SMDSs formed via DM Capture (presented in Fig.~\ref{fig:ABmagCap})}. 
\label{fig:ABmagAC}
\end{figure}

In Fig.~\ref{fig:ABmagCap} and~\ref{fig:ABmagAC} we plot the predicted AB magnitudes in different Roman WFI filters as a function of the emission redshift for SMDSs of three various masses: $\sim 10^4\Msun$, $10^5\Msun$, and $10^6\Msun$. 
Those stars are powered by annihilations of $100~\GeV$ WIMPs and are assumed to have formed 
via either the Extended AC (Fig.~\ref{fig:ABmagAC}) or the DM capture mechanism (Fig.~\ref{fig:ABmagCap}). The apparent AB magnitude ($m_{\mathrm{AB}}$) in any band (filter) is calculated by the following prescription~\citep{Oke:1983ABmag}:
\begin{equation}\label{eq:ABmag}
    m_{\mathrm{AB}}=-2.5 \log \left[\frac{\int_{\lambda_{\rm min}}^{\lambda_{\rm max}} T(\lambda_{\rm obs}) F_{\nu}(\lambda_{\rm obs} ; z_{\rm emi}) \mathrm{d} \lambda_{\rm obs} /  \lambda_{\rm obs}}{\int_{\lambda_{\rm min}}^{\lambda_{\rm max}} T(\lambda_{\rm obs}) \mathrm{d} \lambda_{\rm obs} /  \lambda_{\rm obs} }\right]+31.4, \end{equation}
where $ T(\lambda_{\rm obs})$ is the throughput curve of the filter considered\footnote{We used values obtained from the following website: \url{https://roman.gsfc.nasa.gov/science/Roman_Reference_Information.html}. Here, the effective area is equivalent to the throughput curve in calculating the magnitude.} and $F_{\nu}(\lambda_{\rm obs} ; z_{\rm emi})$ is the redshifted flux density, obtained via Eq.~(\ref{eq:flux}). By comparing Figs.~\ref{fig:ABmagCap} and~\ref{fig:ABmagAC} we find, just as we did in Fig.~\ref{fig:spectra_z}, that at the same mass a SMDS formed via the extended AC mechanism has a larger observed flux, and therefore a smaller AB magnitude, rendering it easier to detect.  This is mainly due to their larger radii, and therefore larger observed fluxes (see Eq.(\ref{eq:flux})). However, at around $z_{\rm emi}\sim 12$, for SMDSs of $10^5 M_{\odot}$ and below, their magnitudes fall below the S/N=5 detection limit even with $10^6 s$ of exposure. Therefore, for the case of unlensed SMDSs, in the subsequent sections we will limit our discussion to the case of a hypothetical $10^6 M_{\odot}$ SMDS.

A trend clearly seen in Figs.~\ref{fig:ABmagCap} and~\ref{fig:ABmagAC} is the increase of AB magnitudes with the redshift of emission. There are two reasons for this behavior. First, and foremost, the same object viewed from farther away will be dimmer. This very intuitive fact can be easily explained by increase of the luminosity distance $d_L(z)$ with redshift, and therefore a larger suppression of the observed flux as per Eq.~(\ref{eq:flux}). Further the sharp increase in the AB magnitudes at redshifts $z\gtrsim 6$ corresponds to the Gunn-Peterson trough affecting different filters at different wavelengths. For instance for the WFI F106 filter, with a center wavelength of $\microm{1.06}$ and covering the $[\microm{0.927}-\microm{1.192}]$ band will start to be affected by the Gunn-Peterson trough at $z=0.927/0.1216 -1\sim 6.6$ and by $z=1.192/0.1216-1\sim 8.8$ the entire flux in the F106 band is completely suppressed by attenuation due to neutral H along the line of sight. Those estimated values of the redshift where the F106 band would be affected by the Gunn-Peterson trough can be confirmed by looking at the third panel on the top row of either Fig.~\ref{fig:ABmagCap} or Fig.~\ref{fig:ABmagAC}.

In the next three subsections we apply the dropout criterion in Eq.~(\ref{eq:dropout}) to show that SMDSs at $z\sim 11$ could be observed with RST as $\Jdrop$ dropouts, whereas those at $z\sim 13$ would appear as $\Hdrop$ dropouts, and those at $z\sim 14$ as $F_{184}$ (H/K) droputs.\footnote{For the relation between the colors of JHKLM photometric system and other standard systems such as SAAO, CIT/CITIO, etc. the interested reader should consult~\cite{Bessell:1988JHKLM}.}

\subsubsection{Detection of SMDSs with RST at $z\sim [10,12]$ as $J_{129}$ band dropouts}\label{sec:J-band}

\begin{figure}[!htb]
\includegraphics[width=\linewidth, height=7cm]{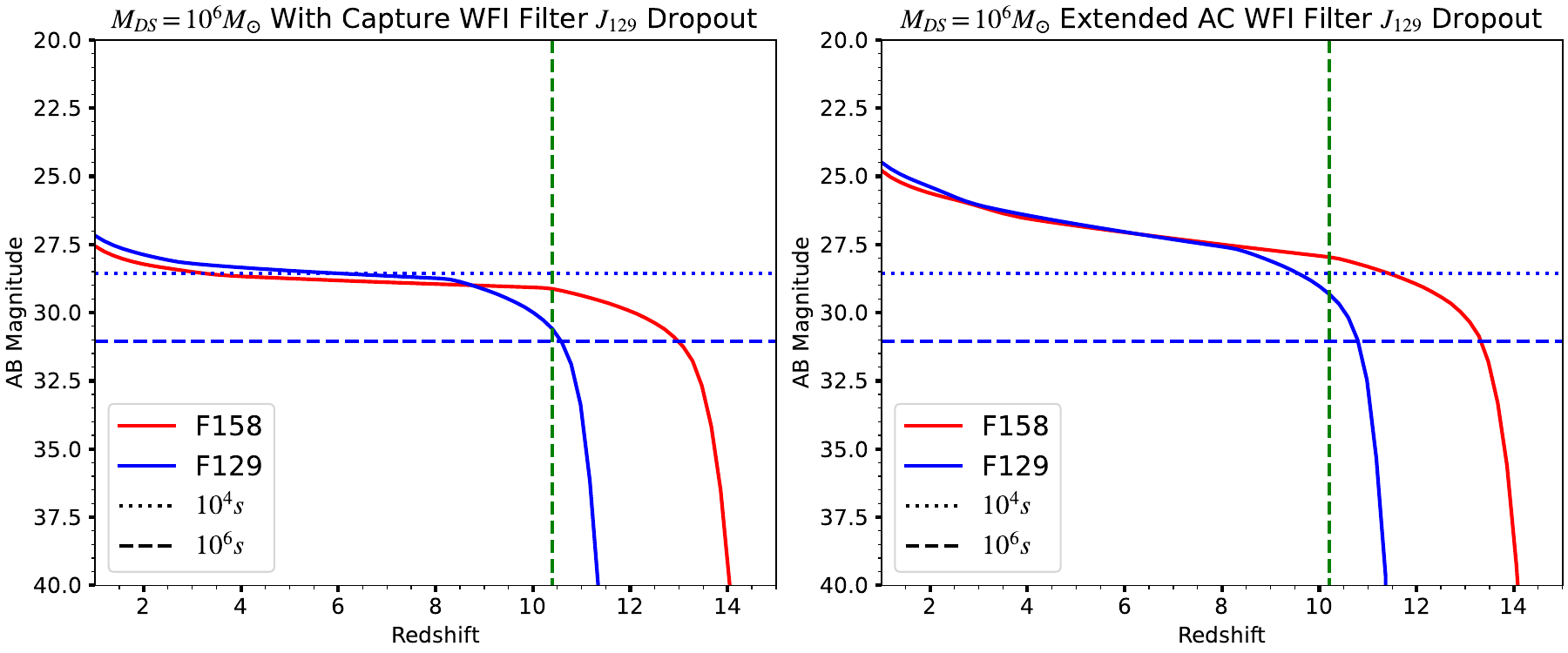}
\caption{J-band dropouts: Predicted SDMS magnitude vs. redshift for a $10^6\Msun$ SMDS in RST. The two panels differ in the different SMDS formation mechanism: DM Capture~(left) and via extended AC~(right). The horizontal dotted lines corresponds to the S/N=5 detection limits with $10^{4}$~s and dashed lines to $10^6$~s of total exposure time.  Here, the lines with same color (red for F158 and blue for F129) represent the AB magnitude (solid) and detection limit (dotted) of the same filter. In this case, both filters have essentially the same sensitivity limits (shown in blue in this figure but relevant for both filters); however for higher center wavelength bands sensitivity decreases significantly even for adjacent filters (see Figs.~\ref{fig:H158} and~\ref{fig:F184}). For this reason we color code both the sensitivity limits and the AB magnitudes. The vertical green dashed corresponds to the redshift where the magnitudes differ by 1.2, i.e. the dropout criterion would be satisfied. Hence SDMS of both types emitting at $z_{\rm emi}\sim 11$ could be identified as J-band dropouts.}
\label{fig:J129}
\end{figure}
Here we show that $M\sim 10^6\Msun$ SMDSs at $z \sim 11$ can be found as $J_{129}$-band dropouts in RST data.
Fig.~\ref{fig:J129} illustrates this result, in the left panel for SDMSs formed via AC and in the right panel for those formed with capture.  For the two filters F158 and F129, the figure shows the predicted SMDS magnitudes vs. redshift as solid lines, and the S/N=5 detection limits of RST assuming $10^4$ and $10^6$s exposures. 
For a $10^6\Msun$ SMDS the difference in the magnitudes in the two bands will satisfy the dropout criterion at $z\gtrsim 10$, independent of the formation mechanism, as marked by the vertical green dashed line that corresponds to the redshift where the magnitudes differ by 1.2. 
The Ly-$\alpha$ attenuation  will completely cutoff the flux in the RST $J_{129}$ band at $z\gtrsim 11$, which is slightly larger than the corresponding value ($z\gtrsim 9.5$) for JWST at $J_{115}$ band . This difference can be accounted for the fact that the Roman J-band filter ($J_{129}$) covers wavelengths up to $\microm{1.454}$ whereas the corresponding JWST J-band filter ($J_{115}$) only covers wavelengths up to $\microm{1.282}$. In turn, this implies that J-band dropouts with RST have a slightly larger estimated photometric redshift than those detected with JWST. Of course, the actual redshift of any object can only be accurately determined via spectroscopy. We point out that for a $\sim 10^6\Msun$ SMDS formed via capture, an exposure time longer than $\sim10^4$~s is necessary in order for them to show as $\Jdrop$ band dropouts in RST. For the same mass SMDSs formed via extended AC even slightly lesser exposure times would suffice, as seen from contrasting the left and right panels of Fig.~\ref{fig:J129}.

\subsubsection{Detection of SMDSs with RST at $z\sim [12,14]$ as $H_{158}$ band dropouts} \label{sec:H-band}

\begin{figure}[!htb]
\includegraphics[width=\linewidth, height=7cm]{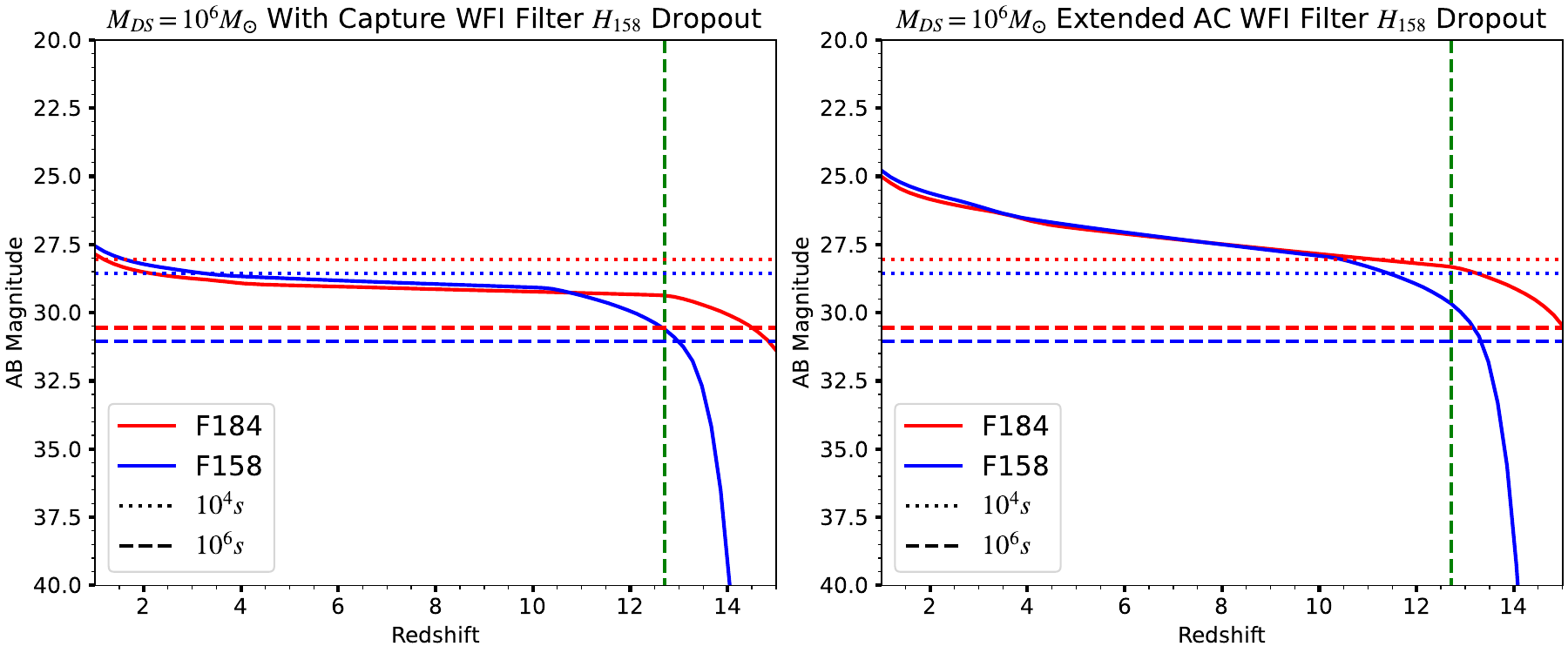}
\caption{Similar to Fig.~\ref{fig:J129} but for H-band Dropouts: SMDSs of $M\sim10^6~\Msun$, formed via either of the two mechanisms (``with Capture'' or Extended AC), would show as $H_{158}$ band dropouts in RST, at redshifts of  $z_{\rm emi}\sim 13$. The dropout criterion starts to be satisfied at $z_{\rm emi}\sim 12.5$, in both cases, as evidenced by the location of the green vertical lines that marks where the difference in the AB magnitudes in the two bands  ($H_{158}$(blue) vs $K_{184}$(red)) becomes larger than 1.2. For the SMDSs formed via extended AC the H band dropout is observable with exposure times as low as $10^4$~s, whereas the slightly dimmer SMDSs formed via DM capture  would require slightly more than $10^4$~s exposure time to present as a H band dropout at $z\sim 13$.}
\label{fig:H158}
\end{figure}

Objects of redshift higher than $z\sim 11$ would show as photometric dropouts in filters of increasingly higher central wavelength. For instance an H-band dropout usually implies a photometric redshift of $z\sim 13$, a K-band dropout a photometric redshift of $z\sim 15$, and so on. In this subsection we demonstrate that SMDSs of mass $M\sim 10^6\Msun$ at $z\sim 13$ are bright enough to be detected as $\Hdrop$ band dropout with the Roman Space Telescope by comparing its AB magnitudes as a function of redshift in the F158 (H band) and the F184 (H/K band) filters (see Fig.~\ref{fig:H158}). 

  In both panels of Fig.\ref{fig:H158}, the \Lyalpha attenuation  by the IGM starts to take effect at $z\gtrsim 14$ for the $F_{184}$ (H/K) band and at $z\gtrsim 12$ for the $H_{158}$. The dropout criterion is satisfied at $z\sim 12.5$, as shown by the green dashed vertical lines, and the Gunn-Peterson trough will completely suppress the flux in the $H_{158}$ band at $z\gtrsim 14$. In both cases, one will need more than $10^4s$ of total exposure time to capture this effect. In the following sections, we will consider $10^6\Msun$ SMDSs emitting at $z_{\rm emi}\sim 12$ as our ``canonical'' case to contrast against other most likely possible luminous objects observable with RST at the same redshift (i.e. Pop~III/II galaxies). Based on Figs.~\ref{fig:J129} and \ref{fig:H158}, we would expect complete cutoff of the SMDSs fluxes in F129 filter, and the image in F184 filter to be brighter than that in F158 band of the Roman Space Telescope. 
 
\subsubsection{Detection of SMDSs with RST at $z\sim [14,15]$ as $F_{184}$ (H/K) band dropouts} \label{ssec:F184drop}

\begin{figure}[!tb]
\includegraphics[width=\linewidth]{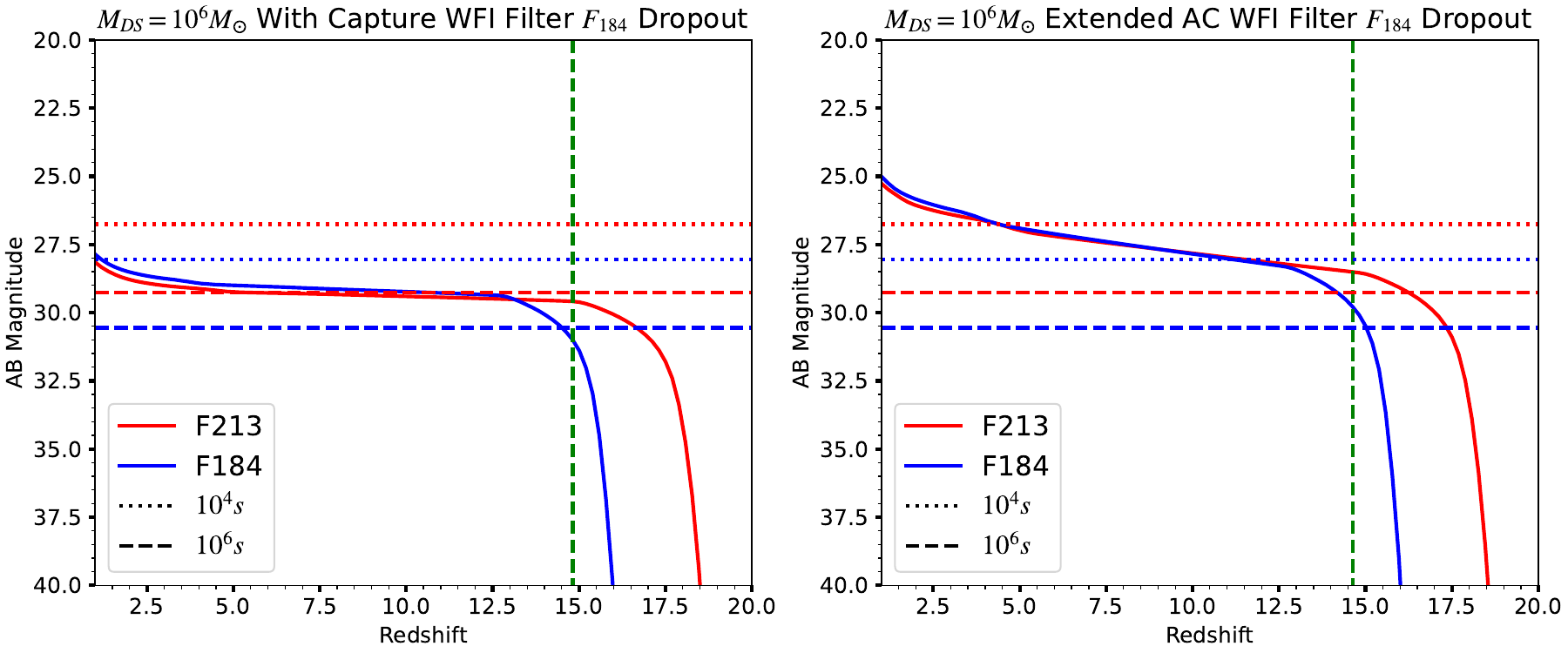}
\caption{$F_{184}$ (H/K) band dropouts: Predicted SDMS magnitude vs. redshift for a $10^6\Msun$ SMDS in the  F213 (K band, red lines) and F184 (H/K band, blue lines) filters of RST. The two panels differ in the different SMDS formation mechanism: DM Capture~(left) and via extended AC~(right). The horizontal dotted lines corresponds to the S/N=5 detection limits with $10^{4}$~s and dashed lines to $10^6$~s of total exposure time. The dropout criterion would be satisfied at around $z_{\rm emi}\sim 14$ for SMDS formed via extended AC for $10^6$~s of observation time (right panel); longer exposure times would be required for the case of SMDS formed via capture (left panel).}
\label{fig:F184}
\end{figure}
In this subsection we consider the possibility to detect SMDSs as photometric dropouts at $z\sim 14$. Specifically, we will demonstrate that SMDSs of mass $M\sim 10^6\Msun$ are bright enough to be detected as F184 (H/K) band dropout ($z_{\rm emi}\sim 14$) with the Roman Space Telescope. In Fig.~\ref{fig:F184} we are comparing the AB magnitudes of $10^6\Msun$ SMDSs formed via either mechanism (DM capture or Extended AC) as a function of redshift in the F184 (H/K band) and the F213 (K band) filters. For SMDSs formed via extended AC (right panel of Fig.~\ref{fig:F184}), the dropout criterion is satisfied at $z\sim 14.2$ (green dashed line) for $10^6$s exposure times. For the case of the dimmer SMDSs formed via DM capture (left panel) longer exposure times would be needed in order to detect them at $z\sim 14$ as $F_{184}$ dropouts. We point out that $z\sim 14$ is the highest redshift SMDSs can be detected (even with ``infinite'' exposure time) as photometric dropouts (F184 H/K droputs) with RST. This is because RST lacks an L band, and, as such, no K band dropout is possible, i.e. no dropout detection at $z_{\rm emi}\gtrsim 15$. 

Having shown that SMDSs are bright enough to be observable as photometric dropouts with RST, at redshifts as high as $z\sim 14$, we move on next to 
a discussion of the objects that may look very similar to SMDS in RST and JWST data:  early galaxies formed of regular, nuclear burning stars (i.e. Pop~III and Pop~II galaxies). 

\section{Population III/II Galaxies}\label{sec:gal}
Early galaxies containing many Population III or Population II stars are the 
competitors to SMDS in observations of high redshift data taken with RST or JWST.
Indeed, according to simulations, large numbers of early galaxies are expected to be found
at $z\gtrsim 10$ and may be hard to differentiate from SMDS.
Before discussing the comparison between SMDS and galaxies in data, in this section we 
describe the relevant parameters and observable properties of Pop~III and Pop~II galaxies.

Whereas in the case of a SMDS the flux is largely given by the SED of a single object, the supermassive dark star in question, the SEDs of galaxies are due to an interplay between the flux coming from all the stars inside the observed galaxy and nebular emissions, which are especially important for young galaxies that actively form stars. For the first type of galaxies those effects have been simulated by~\citet{Zackrisson2011} using the Yggdrasil code. In this paper we use their model grids for the integrated spectra of first galaxies, available at~\url{https://www.astro.uu.se/~ez/yggdrasil/yggdrasil.html}. The relevant input parameters and the possible choices are enumerated below:
\begin{enumerate}
    \item \textit{Initial mass function (IMF)}: The initial distribution of stellar mass population of Pop~III stars for the single stellar populations (SSP) in Yggdrasil. There are three types IMFs we consider, following~\cite{Zackrisson2011} and~\cite{Ilie:2012}, corresponding to three types of Pop~III galaxies: Pop~III.1, Pop~III.2 and Pop~III Kroupa IMF. Pop~III.1 galaxies have a top heavy IMF and a Single Stellar Population (SSP) from~\cite{Schaerer2002AA...382...28S}, and we impose a cutoff on the lower mass end of $50 M_\odot$. Note that ``top heavy" refers to a stellar IMF that, at high mass end, has a less steep slope than the local Salpeter IMF~\citep{Salpeter1955ApJ...121..161S}; this implies a larger fraction of heavy stars than the ``local'' Salpeter distribution\footnote{ For more detailed definitions one can see~\cite{Dave2008MNRAS.385..147D}.}.
  In our case, we model one of the extreme scenarios where all stars are formed within the mass range $50-500 \Msun$; henceforth we call this scenario ``extreme top heavy" where the word ``extreme" refers to the lower mass cutoff rather than to the slope.
    Specifically, for Pop~III.1 galaxies, the simulation adopts a  power-law IMF with a Salpeter slope ($dN/dM \propto M^{-2.35}$) through the  $50-500 M_{\odot}$ stellar mass range. Pop~III.2  galaxies are characterized by a moderately top heavy IMF with a SSP from~\citet{2006ApJ...641....1T,Raiter2010}. Specifically, we model Pop~III.2 galaxies as having a log-normal IMF extending from  $1$ to $500\Msun$ with a characteristic mass $M_c = 10~\Msun$ and distribution width $\sigma=1$. In view of recent simulations
   the mass of Pop~III stars might be lower. Therefore we also include the case of the~\citet{kroupa2001variation} IMF,  usually describing  Pop~II/I galaxies. The stellar masses range in the $0.1-100\Msun$ and the SSP is a rescaled version of the one used in~\citet{Schaerer2002AA...382...28S}. 
    \item \textit{Metallicity ($Z$)}: This index is used to describe the relative abundance of all elements heavier than helium. For reference, the metallicity of sun is $Z_{\odot} \sim 0.02$. The characteristic value of $0.02$ represents typical Pop~I galaxies while $0.0004$ is average for Pop~II galaxies. If one were to include intermediate cases, the following values are available in Yggdrasil model grids for the metallicity of galaxies: $Z=0~(\text{zero-metallicity}), 0.0004, 0.004, 0.008, 0.02$. Since our lowest redshift of interest lies around $z\sim 10$, previous works~\citep[e.g.][]{jaacks2018baseline,2019MNRAS.488.2202J,liu2020did} find via simulations that at such high redshift the metal enrichment process would not be sufficient to make the transition from Pop~III to Pop~II galaxies.  The resulting estimation gives an upperbound of the  mean metallicity: $Z<10^{-2.5} Z_{\odot}\simeq 0.00006$. This value is about an order of magnitude smaller than 0.02 $Z_{\odot}$, which corresponds to the transition to Pop~II galaxies. Therefore, in order to be conservative, we will also include Pop~II galaxies ($Z= 0.0004$) in addition to  Pop.III galaxies ($Z=0$) in our comparisons. 
    \item \textit{Gas covering factor ($f_{\rm cov}$)}:
     This parameter determines the relative contribution of the nebular emissions to the integrated stellar SED. Depending on how compact the ionized hydrogen (HII) region is, the escape fraction ($f_{\rm esc}$) for ionizing radiation (Lyman continuum) from the galaxy into the IGM can vary anywhere from 0-1. Moreover the escape fraction $f_{\rm esc}$ is related to the gas covering factor via: $f_{\rm esc}=1-f_{\rm cov}$. Hence we consider two extreme cases for the gas covering factor: $f_{\rm cov}=1$ (Type A galaxies; maximal nebular contribution and no escape of Lyman continuum photons), and $f_{\rm cov}=0$ (Type C galaxies; no nebular emission).
     \item \textit{Star formation history}: Yggdrasil model grids include instantaneous burst and constant star formation rate (SFR) lasting for $t=$ 10, 30, or 100 Myr. Following~\cite{Ilie:2012, rydberg2015search} we restrict our comparison to the instantaneous burst case, which produces a single age stellar population. 
     We assume that all the stars are formed at $t=1$Myr, as measured from the formation of the galaxy (defined to be at $t=0$), the same as argued in~\cite{Ilie:2012,rydberg2015search}.
    
\end{enumerate}
For future reference we summarize in Tab.~\ref{tab:pop3para} the relevant parameters and the choices made that differentiate between the various Pop~III or Pop~II galaxies we considered in our comparison with supermassive dark stars. 
\begin{table}[!h] 
\centering
 \begin{tabular}{c c c c c c } 
 \hline
Object Name  & $Z_{\rm gas}$ & $Z_{\rm star}$ & Initial Mass Function & Gas Covering Factor &  Star Formation History \\[0.5ex] 
 \hline\hline
   Pop III.1 A & $10^{-7}$& 0 & Extreme Top-heavy IMF & 1 & instantaneous-burst\\Pop III.2 A & $10^{-7}$ & 0 & Middle Top-heavy IMF & 1 & instantaneous-burst\\Pop III.Kroupa A & $10^{-7}$& 0  & Kroupa IMF  & 1 & instantaneous-burst\\  Pop III.1 C & $10^{-7}$& 0  & Extreme Top-heavy IMF & 0 & instantaneous-burst\\Pop III.2 C & $10^{-7}$& 0  & Middle Top-heavy IMF & 0 & instantaneous-burst\\Pop III.Kroupa C & $10^{-7}$& 0  & Kroupa IMF  & 0 & instantaneous-burst\\Pop II.Kroupa A & $0.0004$& $0.0004$  & Kroupa IMF  & 1 & instantaneous-burst\\Pop II.Kroupa C & $0.0004$& $0.0004$  & Kroupa IMF  & 0 & instantaneous-burst\\ 
 \hline
 \end{tabular}
 \caption{Relevant Yggdrasil input parameters for different Pop~III and Pop~II galaxies~\citep[for more details see][]{Zackrisson2011} used in this paper to contrast against supermassive dark stars.}\label{tab:pop3para}
\end{table}

\subsection{Appropriate Choice of which Early Galaxies to compare with SMDS.}
In this section, we will 
imagine a $z\sim 12$ object has been detected with RST as a photometric $\Hdrop$ dropout, and
we will investigate our ability to differentiate whether this object is an SMDS or an early galaxy, using photometry alone.

In previous work on differentiating galaxies vs. SMDSs in JWST, we (two of us)
fixed the stellar masses ($M_{\star}$) of all Pop~II/III galaxies to be equal to the mass of the particular SMDSs considered~\citep{Ilie:2012}.  However, for objects detected  
 with RST as photometric $\Hdrop$ dropouts, if a spectral analysis is not available, the mass and nature of the object in question would still be uncertain. The only available observables, in this scenario, would be the AB magnitudes for this object in the various bands it is detected, and, if resolved, an estimation of its effective radius. 
 
 Hence we address the question: Would it be possible, using only photometry, signal to noise ratios, and image morphology, to differentiate between a SMDSs and a Pop~III/II galaxy as potential candidates for this object? Our approach is to compare SDMSs and Pop~III/II galaxies with same absolute magnitude in the $F_{184}$ band as the SMDSs we compare them against. The reason we chose this band is that it is bridging the gap between the $H_{158}$ and the $K_{213}$ band, which invovle the only three RST filters in which the flux is not completely attenuated by neutral hydrogen in the IGM, for objects emitting at $z\gtrsim 12$. Moreover, for objects at $z\sim 12$, which are our primary targets in this paper, the $F_{184}$ band is not affected by any emission or absorption features in the spectra of galaxies or SMDSs, respectively (See Fig.~\ref{fig:Pop3spectra}). Lastly, for a $\Hdrop$ photometric detection, the object would typically be brightest (have lowest AB magnitude) in the $F_{184}$ filter.
\begin{figure}[!tb]
\includegraphics[width=\linewidth, height=8cm]{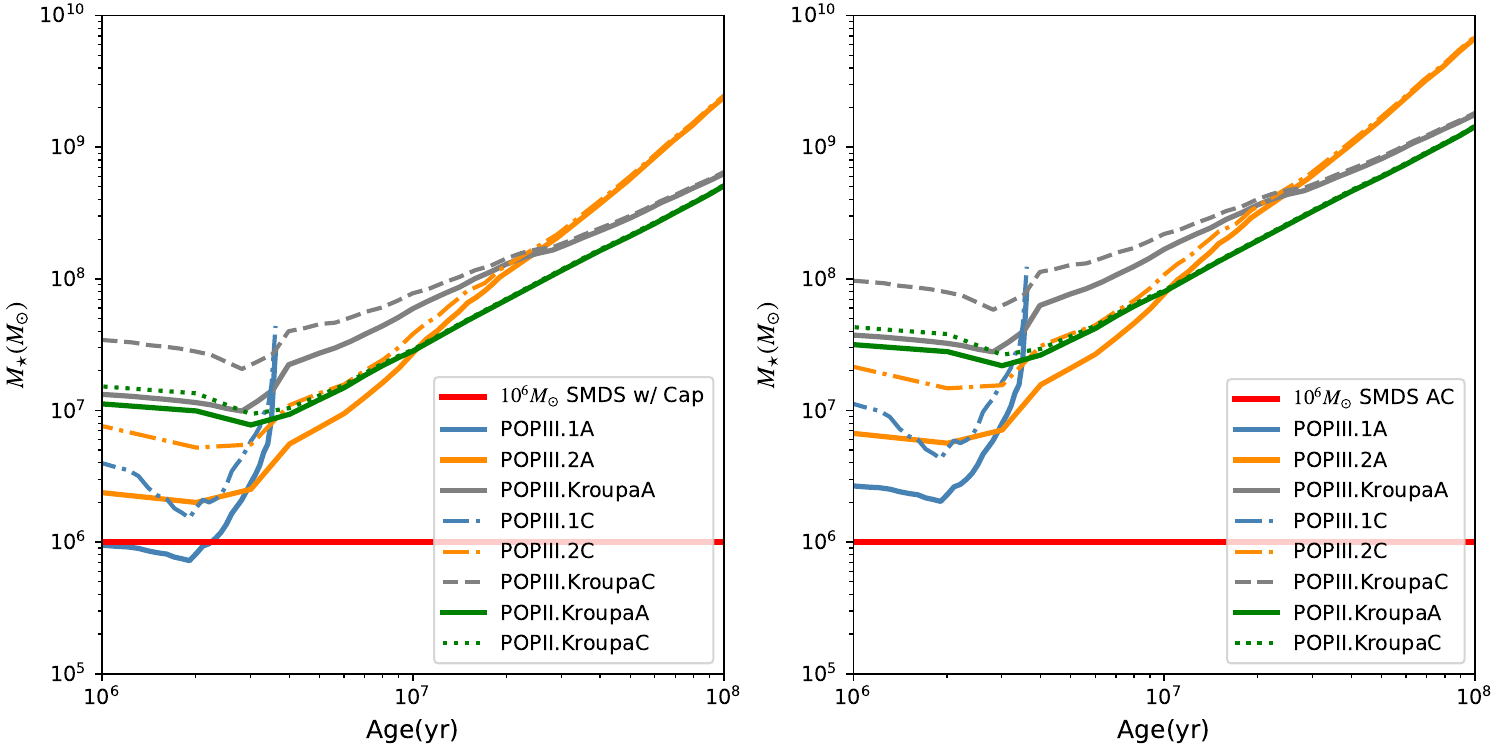}
\caption{ Stellar mass ($M_\star$) of Pop~III and Pop~II galaxies, as a function of age of the galaxy, required to match the flux of a $10^6 M_\odot$ SMDS as shown as a horizontal red line.   The Age in the horizontal-axis is measured from $t=0$, defined as the time the galaxy forms (thus for different galaxies, $t=0$ may correspond to different cosmic times).
We assume an instantaneous burst of star formation at $t=1 $ Myr after galaxy formation.  All objects (at all ages along the curves in the figure) are assumed to produce light at $z_{\rm emi}\simeq 12$. 
For each of the eight types of galaxies considered (see legend and Table~\ref{tab:pop3para} for defining parameters) the stellar mass has been chosen by the requirement that its AB magnitude in the $F_{184}$ filter, at any given time in the plot, matches that of a $10^6 M_{\odot}$ SMDS formed either via the DM capture mechanism (left panel) or via the extended AC mechanism (right panel). 
For the galaxies, the mass necessary to match the  SMDS flux increases with age: the older the galaxy is, some of its stars are already dimmed out. Further, in order to match the F184 brightness of a SMDSs formed via AC (right panel) galaxies have to be more massive than those matching the F184 brightness of a SMDSs formed via DM capture (left panel), since the redshifted fluxes of SMDSs formed via AC are higher for the entire band probed by RST (see Fig.~\ref{fig:spectra_z}). }
\label{fig:Pop3scaled}
\end{figure}
In Fig.~\ref{fig:Pop3scaled} we plot the corresponding stellar mass we found for each of the Pop~III/II galaxies considered in such a way that their AB magnitudes in the $F_{184}$ band match those of a $10^6\Msun$ SMDSs formed via AC (right panel) or via DM capture (left panel), assuming $z_{\rm emi}\sim 12$ for all objects. In order to find the stellar mass ($M_\star$) for each galaxy we assumed, following~\cite{Zackrisson2011}, that the total brightness of the galaxy is proportional to the stellar mass $L \propto M$.\footnote{This assumption starts to break down for galaxies with non-zero gas covering factor whenever the stellar mass is $\gg 10^9\Msun$.}

For galaxies older than $\sim 3$~Myrs, their stellar mass required in order to match the observed flux of a SMDS monotonically  increases with time, as can be seen in Fig.~\ref{fig:Pop3scaled}. This is due to the fact that by $\sim 3$~Myrs after the initial instantaneous burst stars begin to die as they exhaust their nuclear fuel, thus rendering a galaxy dimmer. Since we are requiring a constant flux (AB magnitude) in the $F_{184}$ band, as per our comparison criterion with SMDSs, this dimming of a galaxy of constant mass will be compensated by an increase in the stellar mass of the galaxy.\footnote{Note that the blue lines (for Pop~III.1 galaxies) in Fig.~\ref{fig:Pop3scaled} stop abruptly after $\sim3.5$~Myrs. This is not a numerical artifact. Instead, it represents the fact that those extremely top heavy IMF galaxies cease to emit any significant amount of light once the afterglow of the almost simultaneous SN explosions dissipates.} 

An important takeaway of Fig.~\ref{fig:Pop3scaled} is that, at the same mass, a SMDS is typically much brighter than a galaxy. For this reason SMDSs are a natural solution to the problem posed for the standard cosmological model of $\Lambda$CDM by the JWST data, in terms of too much stellar light coming from high redshift sources, if those are interpreted as giant early galaxies.  

To reiterate, we will compare SMDSs to sets of early galaxies, defined as ``young" ($\sim$ 1 Myr) and ``older" (3.6 Myrs for Pop~III.1 galaxies and $\sim$ 100 Myr for Pop~III.2 or Pop~III.Kroupa galaxies).  We compare SMDS and galaxies that have the same observable magnitude (in the F184 band) and the same redshift.

\section{Signatures of unlensed SMDS vs. the first galaxies with RST}\label{sec:compare}
In the remainder of this paper we present a detailed comparison of the properties of Supermassive Dark Stars vs. those of Population~III/II galaxies, assuming both types of objects could be observed with the upcoming Roman Space Telescope as photometric dropouts at $z\gtrsim 10$. We have shown this to be the case for SMDSs in Secs.~\ref{sec:J-band}-\ref{ssec:F184drop}. For Pop~III galaxies see, for instance,~\cite{Vikaeus:2022MNRAS}, who demonstrate that Pop~III galaxies can be photometrically  detected even in blind surveys with RST, JWST,\footnote{For a JWST specific analysis of the observability of Pop~III galaxies see, for example,~\cite{Zackrisson2011,Pawlik2011ApJ...731...54P,pacucci2015shining}, among others.} and Euclid. Here we perform an analysis regarding the possibility to disambiguate between those two classes of objects, if observed with the RST. In order to do so we use spectral analysis (spectroscopy), color magnitudes analysis (photometry), and image morphology (point source/extended objects). In Sec~\ref{sec:spectroscopy} we discuss the potential to use spectroscopy to tell apart SDMSs from Pop~III or Pop~II galaxies. In Sec.~\ref{ssec:ImgSimul} we present simulated images of SMDSs vs Pop~III galaxies in various RST filters. The potential to disambiguate between Pop~III galaxies and SMDSs via photometry alone is discussed in Sec.~\ref{ssec:Photometry} (AB magnitudes in various RST bands) and Sec.~\ref{ssec:Color} (color-color plots technique). The effects of gravitational lensing are discussed in Sec.~\ref{sec:lensing}.

\subsection{Spectroscopy to Differentiate SMDS vs. Early Galaxies:}
\label{sec:spectroscopy}

\begin{figure}[!htb]
\includegraphics[width=\linewidth, height=15cm]{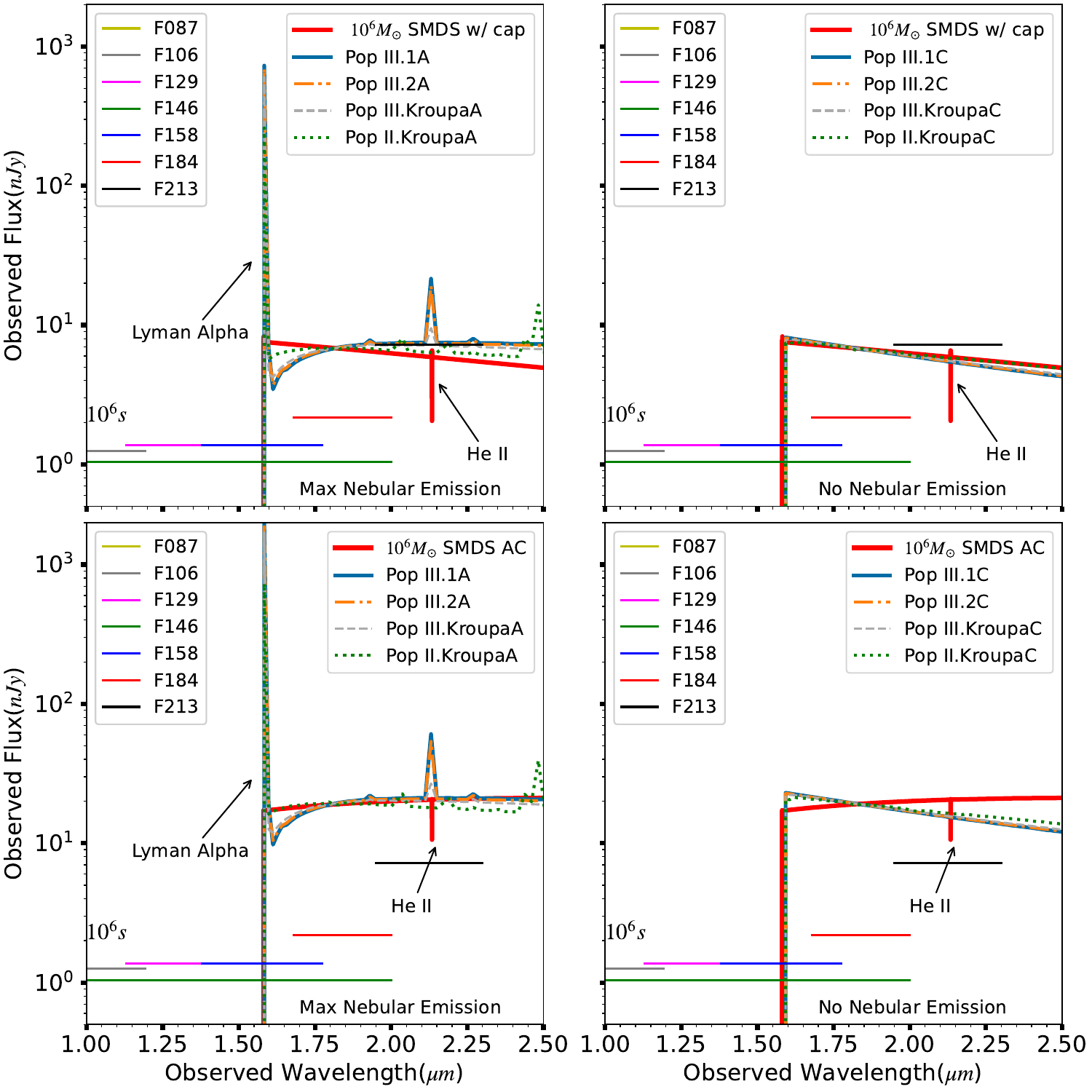}
\caption{Redshifted fluxes of unlensed SMDSs (assuming purely stellar flux) vs various young (1~Myrs old) Pop~III/II starburst galaxies. The SMDSs are assumed, for simplicity, to have a $10^6\Msun$ mass (our fiducial value) with the top two panels for the case of capture and the bottom two panels for the case of extended AC. For the Pop~III/II galaxies the mass is chosen in such a way that it will lead to the same AB magnitude in the $F_{184}$ band, as explained in our discussion of Fig.~\ref{fig:Pop3scaled}. In all four panels, the solid red curves represent the simulated SMDSs spectra at $z_{\rm emi}=12$, formed both by Capture (top two panels) and AC (bottom two panels). In the left two panels we plot the spectra of all Pop~II and Pop~III type A galaxies (maximal nebular emission)  whereas the right two panels are  reserved to Pop~II and Pop~III type C galaxies (no nebular emission).  The spectral lines represents different IMFs in both panels: blue for Pop III.1, orange for Pop III.2, gray for Pop III.Kroupa and green for Pop II.Kroupa. Note that all Pop~III galaxies of the same type (A or C) have spectra that are essentially identical, rendering the corresponding lines difficult to differentiate on the plots. The short horizontal lines are the S/N=5 detection limits of all the Roman WFI filters with respect to $10^6$s of exposure time.}
\label{fig:Pop3spectra}
\end{figure}

\begin{figure}[htb!]
\includegraphics[width=\linewidth,height=15cm]{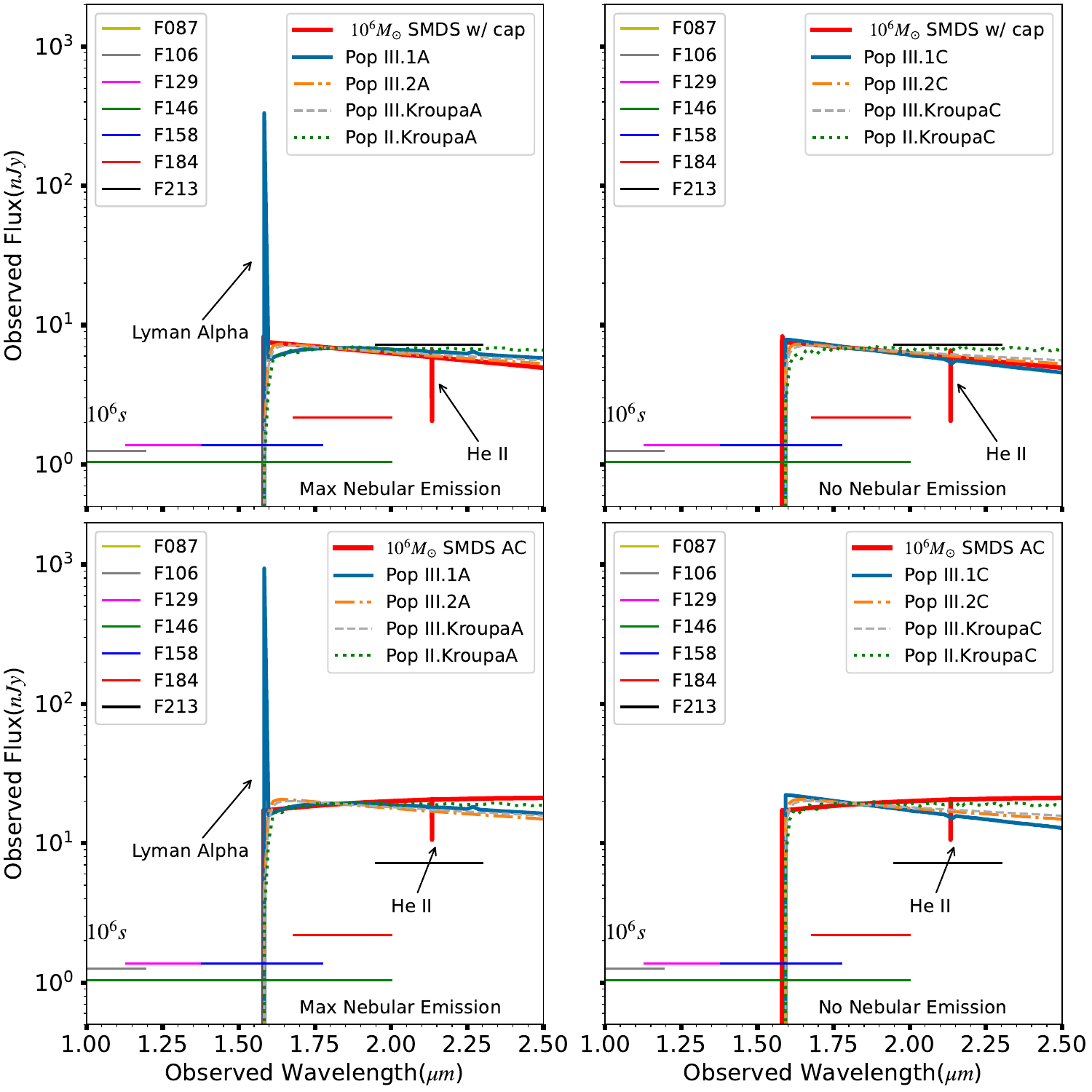}
\caption{Same as Fig.~\ref{fig:Pop3spectra}, but here we have assumed the age of the galaxy to be oldest possible to show the evolutionary trend of galactic SEDs. For Pop III.1A and C, their corresponding ages are $\sim 3.6$ Myr which is the end of their evolutionary stage in Fig.~\ref{fig:Pop3scaled}, and for the rest of galactic candidates, we pick the spectra at the age of $\sim 100$ Myr.}
\label{fig:Pop3spectra100}
\end{figure}

In this subsection we investigate the possiblity to differentiate SMDS vs. early galaxies based on their spectra.  
We will first investigate the difference between SMDS and early galaxy spectra for young (1~Myrs old) Pop~III/II galaxy candidates (see Fig.~\ref{fig:Pop3spectra}) and then turn to the same study for the oldest possible galaxies (see Fig.~\ref{fig:Pop3spectra100}); in both cases the galaxies are assumed to have the same AB magnitude in the $F_{184}$ band as SDMSs counterparts. For ``older" galaxies, the spectra is computed at an age of $\sim 100$ Myrs (measured since the formation of the galaxy), with the exception of the extreme top heavy Pop~III.1 galaxies, which burn through their stars in $\sim 3.6$~Myrs, leading to an abrupt end of their evolutionary stage (see Fig.~\ref{fig:Pop3scaled}). All cases have been chosen such that they could be discovered as $H_{158}$-band dropouts.

We begin with a study of comparison between young galaxies vs. SMDSs, assuming negligible nebular emission for the later.  In Fig.~\ref{fig:Pop3spectra} we compare the redshifted spectra of $10^6~\Msun$ SMDSs formed via the DM capture mechanism (top two panels) or those formed via the extended AC mechanism (lower two panels) to the SEDs of all the young (1~Myrs old) Pop~III/II starburst galaxy candidates considered.  An emission redshift of $z_{\rm emi}\simeq 12$ is assumed, however results and conclusions from this plot will not change significantly even if we took any other emission redshift in our interval of interest: $z_{\rm emi}\in[10-14]$. The most striking finding is that the spectra of young Type C galaxies (no nebular emission) is essentially identical to that of SMDSs formed via the DM capture mechanism, as one can see from the top right panel of Fig.~\ref{fig:Pop3spectra}. The only difference is the absorption feature at $\sim\microm{2.1}$, present only for SMDSs. This feature corresponds to a He~II absorption line, at a restframe wavelength of $\sim\microm{1.6}$, i.e. the HeII $\lambda 1640$ line. However, as shown in~\cite{Vikaeus:2022MNRAS}, the He~II emission/absorption lines are unlikely to be detected in blind surveys, and dedicated spectroscopic followups are necessary. Therefore, in photometric surveys SMDSs formed via DM capture can masquerade for young Pop~III/II galaxies with little to no nebular emission. In the lower right panel of Fig.~\ref{fig:Pop3spectra} we compare redshifted SEDs for Type C Pop~III/II galaxies to those of SMDSs formed via extended AC. In this case SEDs can be differentiated, in principle, based on their distinct slopes of the continuum. The SEDs of young Type A galaxies (maximum nebular emission) are different from those of SMDSs, formed via either formation mechanism, as can be seen from the left two panels of Fig.~\ref{fig:Pop3spectra}. The absence of any emission lines for the SEDs of SMDSs is in stark contrast to the two prominent emission features of the type A galaxies at $\sim\microm{1.6}$ and $\sim\microm{2.1}$. Those two correspond to restframe wavelengths of $\sim \microm{0.12}$ (the \Lyalpha line) and $\sim \microm{0.16}$ (the He~II $\lambda$1640 line). The strength of the \Lyalpha line leads to a higher integrated flux in the F158 band for the galaxies, when contrasted to the SMDSs.\footnote{This will be seen explicitly in Figs.~\ref{fig:PSFCAP} and~\ref{fig:PSFAC}.} We have also considered the possibility that SMDSs and young Pop~III/II galaxies have the same AB magnitude in the F158 band, instead of our choice F184, which was explained in our discussion of Fig.~\ref{fig:Pop3scaled}. In this case the galaxies will look much dimmer than SMDSs in all the other bands, since their flux in the F158 band is primarily driven by the prominent \Lyalpha line, which is absent from purely stellar SMDSs spectra. For this reason, in what follows, our criterion when selecting galaxies to contrast against SMDSs is to require the same integrated flux in the F184 band.   In conclusion, young Type C Pop II/III galaxies without nebular emission (upper right panel) cannot be differentiated from SMDSs on the basis of their spectra;  in all other cases the young galaxies can in fact be differentiated from SMDSs.

In Fig.~\ref{fig:Pop3spectra100} we contrast the SEDs of SMDSs to those of older Pop~III/II galaxies. For all, except the Pop~III.1 galaxies we chose their age to be 100~Myrs. Pop~III.1 galaxies have a top heavy IMF, and, as we mentioned in our discussion of Fig.~\ref{fig:Pop3scaled}, all their stars go dim when the galaxy is roughly 3.5~Myrs old, which is the age we assumed for those type galaxies (Pop~III.1) in Fig.~\ref{fig:Pop3spectra100}. Note how the spectra of all 100 Myrs old galaxies considered contain no distinctive features at the wavelengths of interest. On the other hand, the Pop~III.1A galaxies (age 3.5~Myrs) are distinctive due to the strong \Lyalpha nebular emission line, which should be easily detectable with spectroscopy and also drives up the integrated flux in the F158 to become higher than the corresponding one for the SMDSs.
The main take away from Fig.~\ref{fig:Pop3spectra100} is that old (100 Myr) Pop~III/II galaxies are great SMDSs chameleons. The SEDs of SMDSs formed via DM capture (top row of Fig~\ref{fig:Pop3spectra100}) is nearly identical, up to the He~II $\lambda$1640 absorption line, to the SEDs of all 100 Myrs old galaxies considered (i.e. Pop~III.2 A/C, Pop~III.Kroupa A/C,Pop~II.Kroupa A/C). When contrasting SEDs of SMDSs formed via adiabatic contraction to those of old (100~Myrs) Pop~III/II galaxies (lower row of Fig.~\ref{fig:Pop3spectra100}) we find that the spectra of Pop~II.Kroupa galaxies (dotted green lines) are very similar to those of SMDSs formed via AC. However, the two cases differ in that only the SMDS exhibit a He~II $\lambda1640$ absorption line,  for masses  $M\gtrsim 10^5\Msun$ (see lower right panel of Fig.~\ref{fig:spectra}). In principle one can detect this line with dedicated spectroscopic studies. However, for objects at z$>$10.7, the line is outside of the wavelength range probed by RST and would require followup studies (e.g. with JWST).

We end the discussion of Figs.~\ref{fig:Pop3spectra} and~\ref{fig:Pop3spectra100} by pointing out that the hottest SMDSs ($T_{\rm eff} \gtrsim 5\times10^4$~K, typically reached by SDMSs formed via DM capture) could actually have significant nebular emission, as shown by~\cite{Zackrisson:2011}. This would further complicate the prospects of disambiguation between SMDSs and the first galaxies based on their SEDs.  Namely, for each SMDSs for which nebular emission is significant one could identify, in principle, a galactic counterpart with similar photometric of spectral signatures. Moreover, such objects (SMDSs surrounded by an ionization bounded nebula) could, in principle, be resolved as compact extended objects. We leave the detailed investigation of the possible role of nebular emission from the hottest dark stars on their detectable signatures for a future study. Our expectations are that, whenever SMDSs form a ionization bounded nebula the disambiguation between such an object a compact (i.e. unresolved or poorly resolved) early galaxy would be nearly impossible with RST. For a more in depth discussion see Appendix~\ref{ap:Nebular}.  

In this subsection we have investigated the ability to distinguish SMDSs and early galaxies on the basis of their spectra.
There is one clear signature of an SMDS in the spectra: if a HeII $\lambda$1640 (i.e. restframe wavelength $\microm{0.1640}$) absorption line is seen, then the object is a Dark Star and not a galaxy. However,
the He II emission/absorption lines are unlikely
to be detected in blind surveys, and dedicated spectroscopic followups will be necessary. The Grism spectrometer that will be part of the Wide Field instrument onboard RST is only sensitive to wavelengths as high as $\microm{1.93}$. Note that at $z_{\rm emi}\sim 12$ the only significant spectral feature within the range of the RST Grism is the possible \Lyalpha nebular emission of galaxies  (see Figs.~\ref{fig:Pop3spectra} and~\ref{fig:Pop3spectra100}). 
Thus, until followup spectroscopic data is available (possibly from JWST, which is sensitive to much higher wavelengths), it will be impossible to differentiate $z\sim 12$ SMDS vs old Pop~III/II galaxies (or those with little nebular emission) on the basis of their spectra with RST. In the remainder of the paper we investigate how one could differentiate between SMDSs and Pop~III/II galaxy candidates in the absence of spectroscopic data. 

\subsection{Photometry Alone cannot Differentiate unlensed SMDSs vs. Early Galaxies in RST:}\label{ssec:Photometry}

\begin{figure}[!htb]
\includegraphics[width=\linewidth, height=15cm]{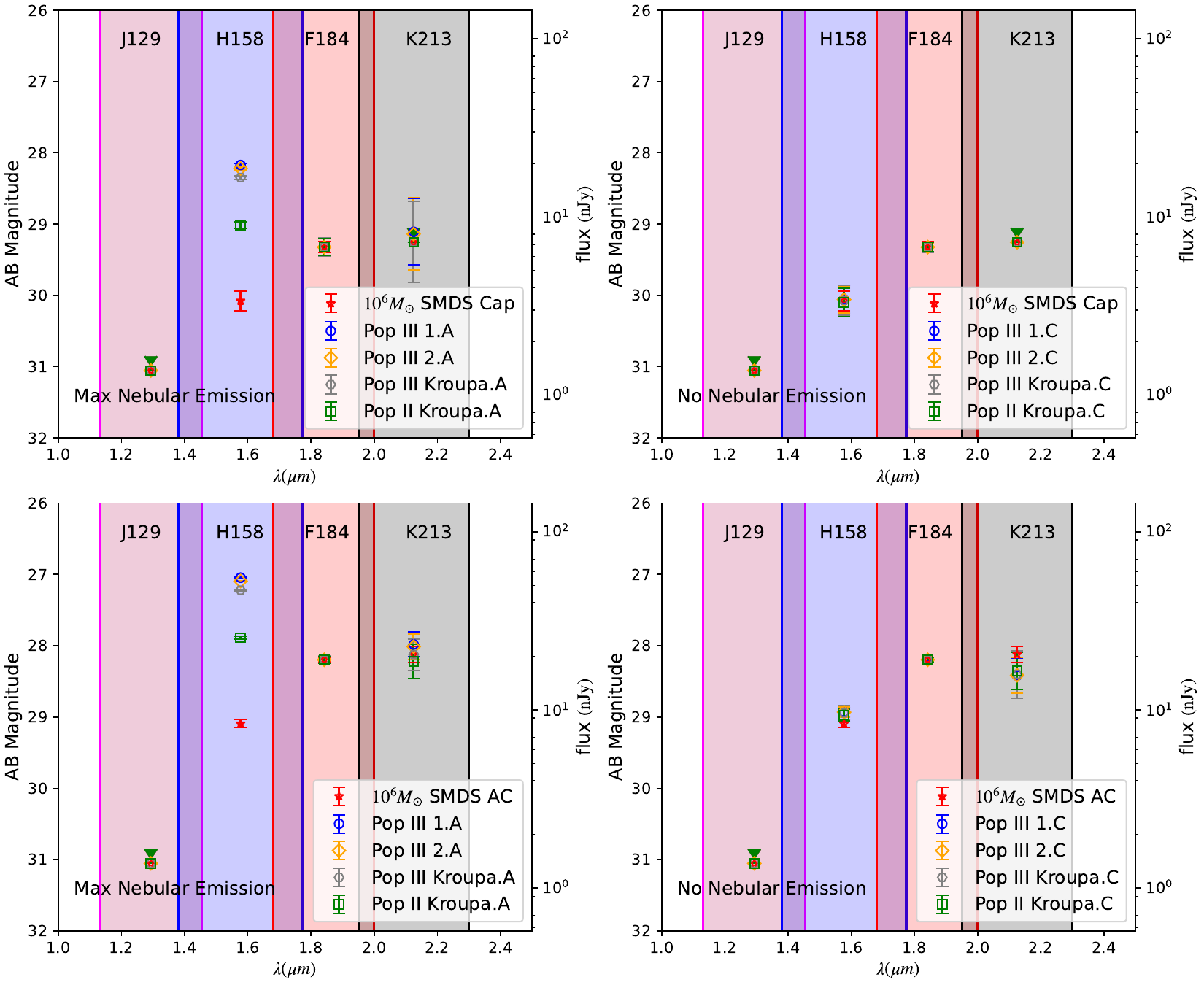}
\caption{AB magnitudes (left vertical axes) and average observed fluxes (right vertical axes) of SMDSs (with purely stellar flux) vs various young (1~Myrs old) Pop~III/II galaxies in the case of no lensing. The SMDSs are assumed, for simplicity, to have a $10^6\Msun$ mass (our fiducial value) with the top two panels for the case of capture and the bottom two panels for the case of extended AC. For the Pop~III/II galaxies the mass is chosen in such a way that it will lead to the same AB magnitude in the $F_{184}$ band, as explained in our discussion of Fig.~\ref{fig:Pop3scaled}, and as one can see from the plots. The SMDSs are represented by red stars. For all galaxies considered the symbols can be found in the legends. On the left two panels we consider the case of galaxies maximal nebular emission (type A), whereas the right two panels are for the case of galaxies without nebular emission (type C). For each object we estimate and plot the 1-sigma uncertainties in the predicted AB magnitudes. In the J129 band, the AB magnitude of both SMDSs and galaxies are predicted to fall below the detection limit because of the Gunn-Peterson trough. Therefore, by convention, we manually assign a value to be equal to the detection threshold (for a $10^6$s exposure) magnitude, and denote that with the upper triangle on top of the marks. Note that with photometry alone both types of SMDSs can be mistaken for galaxies with little to no nebular emission (right panels)}
\label{fig:Pop3mag}
\end{figure}

Early data are likely to be photometric in nature, with detailed spectra only obtained later. Hence in this section we examine
the capability to differentiate SMDS vs early galaxies in photometric data.
We convert the information included in the direct spectroscopic comparison of the SEDs of SMDSs and Pop~III/II galaxies (discussed in Sec.~\ref{sec:spectroscopy}) in a comparison in terms of photometry with RST. We start with Fig.~\ref{fig:Pop3mag}, where we plot the calculated AB magnitudes (left vertical axes) and average fluxes (right vertical axes) of SMDSs and young Pop~III/II galaxies. This figure is the photometric representation of Fig.~\ref{fig:Pop3spectra}, meaning all the parameters are the same as in Fig.~\ref{fig:Pop3spectra}. Specifically, in order to get $m_{AB}$ we convolved, according to Eq.~(\ref{eq:ABmag}), the redshifted SEDs with the RST throughput curves for each of the following bands: J129, H158, F184, and K213. In order to estimate the uncertainties in $m_{\rm AB}$ we used Pandeia to obtain the SNR for each object, assuming $10^6$~s exposures. The same conclusion we drew from Fig.~\ref{fig:Pop3spectra}, in terms of spectra, can be easily justified in terms of photometry from Fig.~\ref{fig:Pop3mag}. Specifically, SMDSs formed via DM capture will have almost identical photometric signatures when compared with Pop~III/II galaxies without nebular emission (top right panel of Fig.~\ref{fig:Pop3mag}). Moreover, now we find that in terms of photometry even the SMDSs formed via AC are very similar with Pop~III/II galaxies without nebular emission (lower right panel of Fig.~\ref{fig:Pop3mag}). 

\begin{figure}[htb!]
\includegraphics[width=\linewidth, height=15cm]{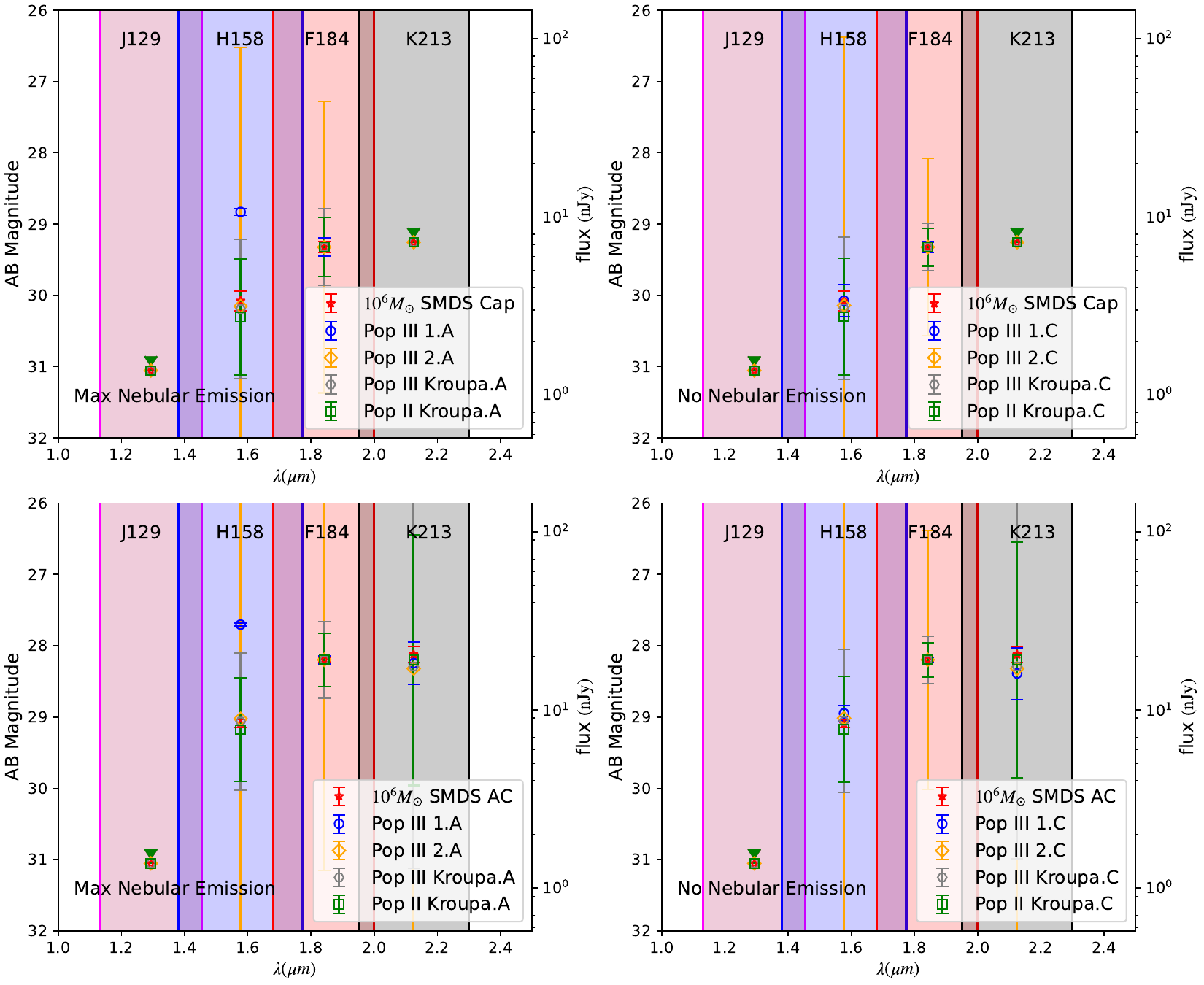}
\caption{Same as Fig.~\ref{fig:Pop3mag} with the 1-sigma uncertainty in the AB magnitude, but here we have assumed the age of the galaxy to be oldest possible. For Pop III.1A and C, their corresponding ages are $\sim 3.6$ Myr which is the end of their evolutionary stage in Fig.~\ref{fig:Pop3scaled}, and for the rest of galactic candidates, we pick the spectra at the age of $\sim 100$ Myr.}
\label{fig:Pop3mag100}
\end{figure}

In Fig.~\ref{fig:Pop3mag100} we repeat the same analysis done in Fig.~\ref{fig:Pop3mag}, considering now old galaxies instead. From the left two panels we can conclude that SMDSs formed via either formation mechanism will have colors (AB magnitudes) that are indiscernible from old Pop~III/II galaxies without nebular emission. And, in fact, most old galaxies should be expected to have little to no nebular emission. The only exception to this are ``old'' Pop~III 1. galaxies, that, just before the end of their evolutionary tracks at $\sim 3.5$~Myrs, can still exhibit quite a lot of nebular emission. This can be explicitly seen in the comparison of left to right panels, where the only galaxies that have different colors in the left (Type A, i.e. maximal nebular emission) to the right (Type C, i.e. no nebular emission) are Pop~III 1 galaxies, i.e. those with the most extreme top heavy IMF. Form them, the strong \Lyalpha nebular emission drives the averages fluxes up in the H158 band (assuming, as we did here $z_{\rm emi}=12$). 

To sum up: We find there is a clear difference between SMDSs formed via either formation mechanism vs. Type A young galaxies based on their vastly different colors in the H158 band (see left panels Fig.~\ref{fig:Pop3mag}), but not vs. Type C young galaxies. Comparing SMDSs with old (100 Myrs old) galaxies, we find that there is no way to tell them apart from Pop~III/II galaxies, using RST photometry (Fig.~\ref{fig:Pop3mag100}).  Since we don't know ab initio what type of galaxy is actually in the sky, we have to conclude that SMDSs cannot be differentiated from all possible galaxies via photometry alone. Thus RST photometric data will not be sufficient to uniquely determine that a $\sim10^6\Msun$ SMDS has been discovered.\footnote{For lower mass SMDS see discussion in Sec.~\ref{ssec:ColorColorMag}, where we show that color color plots can be useful in disambiguating SMDSs formed via AC, if sufficiently lensed.}

\subsection{RST Image Simulations using Pandeia for unlensed SMDS vs. Early Galaxies: Morphology alone cannot distinguish in RST}\label{ssec:ImgSimul}

In this section we address the following question: is it possible, based on image morphology--i.e. distinguishing between point and extended objects--to tell apart unlensed $10^6\Msun$ SMDSs from their counterpart Pop~III/II galaxies (those that have the same observed flux in the F184 RST band). We will assume objects are observed with RST as dropouts with a photometric redshift $z\gtrsim 10$. We start this analysis by estimating the effective angular size of high redshift galaxies as  a function of stellar mass (see Eq.~(\ref{eq:size})). By comparing the effective size of galaxies to the size of a RST pixel we will soon show that unlensed galaxies will barely cover a few pixels. Thus, just as SMDSs, unlensed galaxies (counterparts to SMDSs) will appear unresolved, and as such, are indistinguishable based on image morphology alone, when observed with RST. 

\begin{figure}[!tb]
\includegraphics[width=\linewidth, height=14cm]{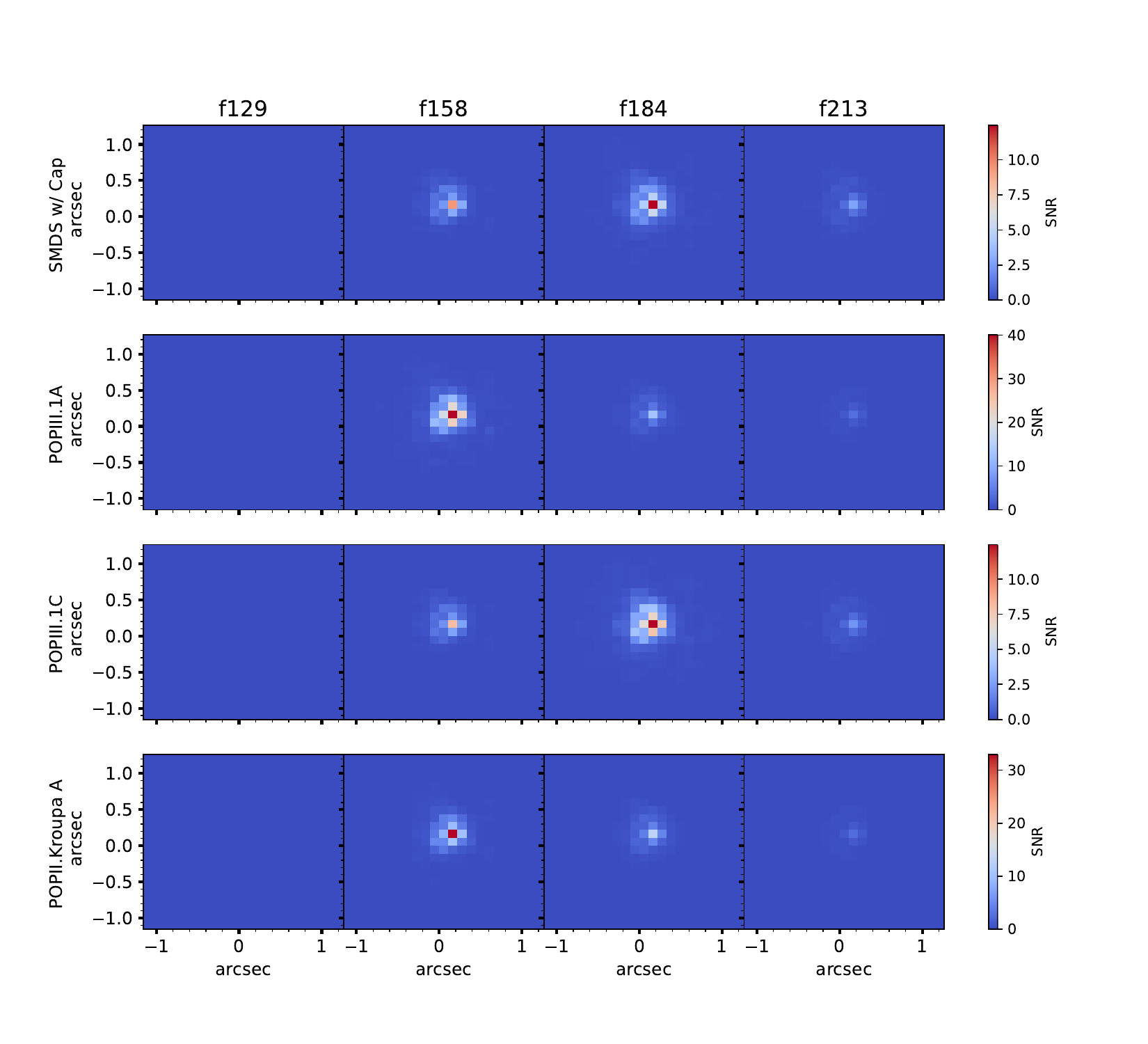}
\caption{$2.2"\times 2.2"$ image stamps of unlensed SMDS formed via Capture vs Young starburst Galaxies:  Simulated images for $10^6 M_{\odot}$ SMDSs formed at $z_{\rm form}=15$ via DM capture are shown in the top horizontal panels. The other three rows are the simulated images for young (1 Myrs) galaxies of type labeled next to each row. All the objects are assumed to emit light at $z_{\rm emi}=12$ and they have the same flux in the F184 band, as per our comparison criterion. All of the images are simulated in the Roman WFI f129, f158, f184 and f213 filters with $10^6$s of exposure. The plot is colored in cool-warm scale by SNR value, and the color bar to the right of each row shows the relative value of SNRs for each of the corresponding horizontal panels.}
    \label{fig:PSFCAP}
\end{figure}

\begin{figure}[!htb]
\includegraphics[width=\linewidth, height=14cm]{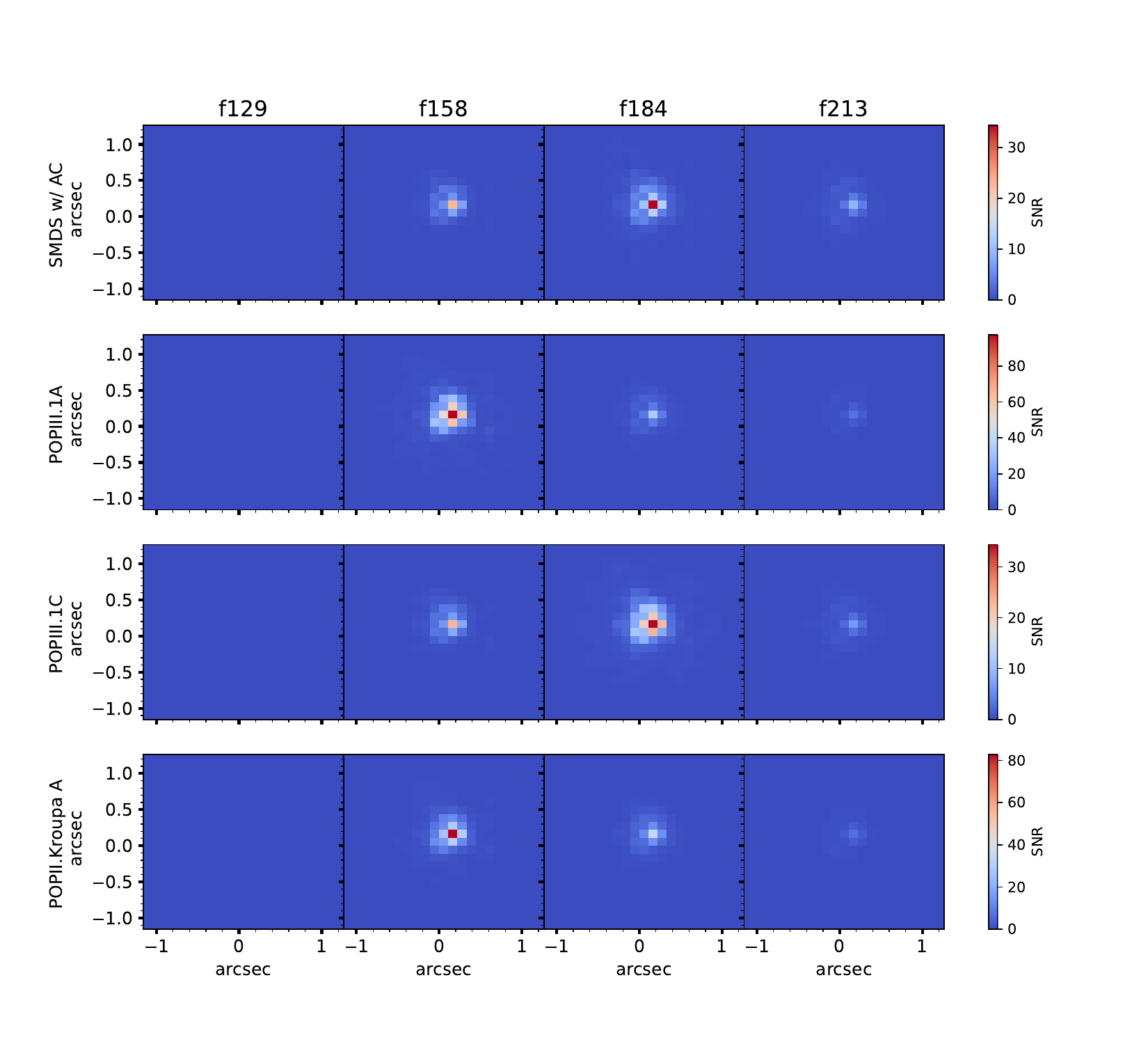}
\caption{Simulated $2.2"\times 2.2"$ image stamps of unlensed SMDS formed via Extended AC vs Young starburst Galaxies: Same as Figure \ref{fig:PSFCAP}, but here we plot the comparison with respect to $10^6M_{\odot}$ SMDS formed via extended AC. }
    \label{fig:PSFAC}
\end{figure}

To reinforce this point, we will also present  RST image simulations of unlensed SMDSs and Pop~III/II galaxies obtained using the Pandeia exposure time calculator (ETC) developed at STScI~\citep{pontoppidan2016pandeia}. Among other things, Pandeia allows the user to simulate the effects of the point spread function (PSF) of different instruments onboard JWST or RST. As such, one can perform high-fidelity modeling of the image(s), as seen with JWST or RST, of any hypothetical source(s) for which the restframe SEDs are known. For our comparison of simulated images for unlensed SDMSs vs galaxies, as viewed in various RST bands, see Figs.~\ref{fig:PSFCAP} and~\ref{fig:PSFAC}.

As stated above, we begin by estimating the effective angular size of galaxies and contrast it against the size of an RST pixel. 
The main ingredient in this calculation is the effective radii ($r_{\rm  eff}$) of those objects. The size of Pop~III/II galaxies are largely uncertain (due to the lack of observational data) and somewhat model dependent. However JWST already offers a glimpse into the size of very luminous galaxies at high redshift. For instance, one of most distant objects found with JWST, Maisie's galaxy, at $z\sim 14$, has an estimated half-light radius of $r_{\rm  eff}=330\pm 30$~pc~\citep{Maisies:2022}. This is in excellent agreement with what one expects from extrapolating results from lower redshift observed galaxies. For instance Fig.~15 of~\citet{2018ApJ...855....4K} presents the redshift evolution of the average size of bright galaxies observed up to 2018, in the redshift range $2-12$. Using their best fit function and extrapolating to $z=14$ we find, remarkably, a value of $r_{\rm 
 eff}=330$~pc, a perfect match to Maisie's galaxy size.

In what follows we will proceed to estimate the galactic stellar mass dependence of the angular size of high redshift galaxies. Recall that for galaxies with a stellar mass not much greater than $10^9\Msun$, the luminosity is directly proportional to the stellar mass. Therefore, by using the size-luminosity relationship from observed data, we can also make an estimate of how the effective size scales with the galactic stellar mass: $r_{\rm eff}\sim L^{\beta} \sim M_{\star}^{\beta} $, where $\beta$ is the scaling coefficient.  
The mass-size relationship for lensed galaxies found with HST at high redshift ($z\sim 6-9$) can be found in Fig~2 and Fig~3 of~\citet{bouwens2021low}.  In what follows we take value of $r_{\rm  eff }\sim 20~\rm pc$ as the typical size of galaxies of $10^6 M_{\odot}$ stellar mass (inferred from their figures) and a value of $\beta=0.5$, as found by~\cite{bouwens2021low}. Thus: 
\begin{equation}\label{eq:reff}
    r_{\rm eff}\approx 20~\unit{pc}\left(\frac{M_{\star}}{10^6M_{\odot}}\right)^{0.5}.
\end{equation}
Since the redshift range considered by~\cite{bouwens2021low} (i.e.$z\in[6-8]$) does not cover our range of interest  ($z\gtrsim 10$), we expect the linear fit given by~\citet{bouwens2021low} to be mildly over-predicting the size of galaxies at higher redshifts, since, at the same stellar mass, the  higher redshift galaxies are typically more compact, according to simulations and recent JWST observations. For Maisie's galaxy $M_{\star}\approx10^{8.5}\Msun$~\citep{Maisies:2022}, leading to a predicted $r_{\rm eff} \approx 355$~pc, which is within the observed $r_{\rm eff} =330\pm 30$~pc, albeit mildly over-predicted, as discussed. 
For  the most massive galaxies we consider (i.e.  $M_{\star}\sim 10^9 M_{\odot}$) we get, using Eq.~(\ref{eq:reff}), an estimated size of $r_{\rm eff} \sim 0.6$~kpc, which agrees very well with the recent results from JWST observations of $z\gtrsim 10$ galaxies~\citep[e.g.][]{GLASSz13,Maisies:2022}. This further validates our method of estimating $r_{\rm eff} $ for a given mass galaxy (Eq.~(\ref{eq:reff})). 
Knowing the effective radius of a galaxy allows us to compute its angular size via the following prescription:  
\begin{equation} \label{eq:size}
    \theta_{\rm eff} \simeq 2\times 20~ \mathrm{pc} \times \frac{(1+z_{\rm emi})^2}{D_{\mathrm{L}}\left(z_{\rm emi}\right)}\left(\frac{M_{\star}}{10^6M_{\odot}}\right)^{0.5}.
\end{equation}
  
 For $z\in[12-14]$ an $r_{\rm eff} $ of 0.6~kpc corresponds to $\theta_{\rm eff} \in[0.16"-0.18"]$.
 The size of the pixel for RST is $\sim 0.11"$. In view of this, even the most massive galaxies we consider ($M_{\star}\sim 10^9\Msun$) have an unlensed size that is below two pixels. However, inherent diffraction always spreads out light over neighbouring pixels, so even a point object (such as a SMDS) will end up covering an area of the detector that is a few pixels across (see top panels of Figs.~\ref{fig:PSFCAP} and~\ref{fig:PSFAC}.)  
 Therefore, it will be difficult to distinguish SMDSs from Pop~III galaxies based on morphology alone, unless the objects of interest are gravitationally lensed, which is a possibility we consider in Sec.~\ref{sec:lensing}. 

Next we investigate, using the Pandeia engine~\citep{pontoppidan2016pandeia}, possible differences between SMDSs and Pop~III/II galaxies based on image morphology and/or SNR values.  Of course, based on the discussion above, we already know that for unlensed objects, the disambiguation based on image morphology between SDMSs and Pop~III/II galaxies will be challenging. The full power of the Pandeia engine will become transparent when we study lensed objects (see Sec~\ref{sec:lensing}), however we use it here as well, as a warm-up exercise, and to confirm our expectations obtained  in the  paragraphs above. The relevant parameters used in our image simulations can be found in Appendix~\ref{Pandeia}. For simplicity, we only assume one object in each scene, although in real scenario one might expect other luminous sources in the adjacent field. Following previous sections, we still use the $10^6 M_{\odot}$ SMDSs and take $10^6$s of total exposure time, as fiducial values. 

{\it{Simulated images of SMDS vs. young (1 Myr) galaxies for the case of no lensing:}} 
In Fig.~\ref{fig:PSFCAP} we present the simulated RST images of a $10^6\Msun$ SMDSs formed via capture (top row) vs. those of select young (1~Myrs) galaxies for the case of no lensing.\footnote{We did not include the other two types of Pop~III galaxies~(Pop~III.2 and Pop~III.Kroupa), since their SEDs are essentially degenerate to those of the Pop~III.1 galaxies we plot here (see Fig.~\ref{fig:Pop3spectra}).} A similar plot for SMDSs formed via extended AC (top panels)  can be found in Fig.~\ref{fig:PSFAC}. The angular diameter of the galaxies is given by Eq.~(\ref{eq:size}), with the mass selected in such a way that the AB magnitude in the F184 band is the same for all objects considered. In order to be conservative, in our simulation galaxies are assigned a spherical shape, such that it would be harder to tell the difference between the extended objects and stellar object based on the image morphology alone. However, RST will be largely insensitive to the degree of symmetry of the object, at the redshifts of interest for us. At $z\sim 12$, even the $10^9\Msun$ galaxies have an effective size that barely covers a few pixels, which renders shape extraction impossible. From the simulation, we see that most of the flux received by the detector is concentrated in the central pixel of our target object. As predicted in the previous sections, we find that all of flux in the $J_{129}$ has been attenuated by the neutral H present in the IGM, leading to detection as photometric dropouts. In the F213 band, due to mostly to the decrease in the capability of the detector all objects appear extremely dim. The strong \Lyalpha emission of young type A galaxies boost their fluxes in the F158 band (as can be seen from Fig.~\ref{fig:Pop3spectra}), which leads to an enhanced SNR in the same band (see second and fourth rows of Fig.~\ref{fig:PSFCAP}).  Note the almost perfect match between the simulated images for Pop~III.1C galaxies  and SMDSs formed via DM capture (see top and third from the top rows of Fig.~\ref{fig:PSFCAP}), thus further demonstrating that SMDSs can pose as young Pop~III type C galaxies (i.e. with little to no nebular emission). This is due primarily to their almost identical spectra (see top right panel of Fig.~\ref{fig:Pop3spectra}) and the fact that at those high redshifts, without gravitational lensing, even the most massive galaxies are not resolved by RST (see Fig.~\ref{fig:Size} for the effect of gravitational lensing). 

In Fig.~\ref{fig:PSFAC} we compare the Pandeia simulated images of SMDSs formed via adiabatic contraction to those of young (1 Myrs) Pop~III/II galaxies, selected following the same prescription followed for Fig~\ref{fig:PSFCAP}, again for the case of no lensing. All objects are assumed to emit at $z_{\rm emi}=12$. All type A young galaxies will have a much higher SNR in the F158 band, when compared to SMDSs with purely stellar spectra. This fact is due, again, to the strong \Lyalpha nebular emission line present in the F158 filter at that redshift.  Note that even if the slope of the UV continuum in the SEDs of SMDSs formed via AC and those of Type C galaxies is different (lower right panel of Fig.~\ref{fig:Pop3spectra}) RST will not be sensitive enough to detect this using photometry alone, as can be seen from comparing the top and third from the top rows of Fig.~\ref{fig:PSFAC}, and as we have explicitly seen in the lower right panel of Fig.~\ref{fig:Pop3mag}. Therefore, young Pop~III galaxies without significant nebular emission can also pose as SMDSs formed via the adiabatic contraction mechanism.    

{\it{Image morphology of SMDS vs. older (3-100 Myr) galaxies for the case of no lensing:}}
Here we discuss the image morphology of the older Pop~III/II galaxies when contrasted to that of SMDSs, as they would be seen by RST. As explained in our discussion of Fig.~\ref{fig:Pop3spectra100}, old (100 Myrs) galaxies no longer produce nebular emission, rendering the distinction between Type C and Type A irrelevant. Old Pop~III/II galaxies have SEDs that are very similar to those of purely stellar SMDSs. However, the older a galaxy is, the more massive it has to be in order for it to match the brightness of a SMDSs of a fixed mass (see Fig.~\ref{fig:Pop3scaled}). Even for the most massive galaxies considered ($M_{\star}\sim 10^9\Msun$), the effective area in RST is below the size of $2\times 2$ pixels, which renders them either impossible to resolve or with SNRs less than 5 in each pixel, if unlensed, as can be seen in Fig.~\ref{fig:Size100}

In summary, we find that unlensed SMDSs can be easily mistaken for unlensed Pop~III.1C galaxies throughout their evolution. For the case of Pop~III.1A (maximum nebular emission) galaxies, the strong \Lyalpha line in the F158 band will be a telltale signature, even in photometry, that can set them apart from SMDSs with purely stellar spectra.   

We end this section with the same cautionary note mentioned at the end of Sec.~\ref{sec:gal} regarding disambiguating between SDMSs and the first galaxies. As pointed out by~\cite{Zackrisson:2011}, the hottest SMDSs (those formed via DM capture) can have a detectable nebular emission contribution to their SEDs. This, in turn would lead them to appear as compact, but extended objects, with SEDs that are similar to those of type A galaxies. In a future publication we plan to analyse this possibility in detail. In this paper we restrict our attention to SMDSs with  purely stellar spectra, which is the most likely scenario, as discussed in Appendix~\ref{ap:Nebular}.


In conclusion, for unlensed objects,  morphology alone will not be enough to clearly identify an observed object as an SMDS vs. early galaxies in RST data. In other words, one cannot differentiate cleanly between an SMDS as a point object vs. galaxies as extended objects. Some types of early galaxies will look different from SMDSs. However, for any image in RST data that matches predictions for an SMDS, there is always some galaxy type that could produce a virtually indistinguishable image. We will address the effects of gravitational lensing on this question in Sec.~\ref{ssec:ImageMorph}.

\subsection{Color-color index comparison for the case of no lensing}\label{ssec:Color}

In this section we investigate the ability to differentiate unlensed SMDSs from  Pop~III/II galaxies using the color-color method, following~\cite{Zackrisson:2010HighZDS,Ilie:2012}. 
In principle different
objects could occupy different locations in a color/color plot, thereby allowing a way to identify the nature of the object in the data.
As before, we focus on $z_{\rm emi}\sim 12$ objects found as $J_{129}$ dropouts. Specifically, we will plot the color-color diagram in $F_{184}-K_{213}$ vs. $H_{158}-F_{184}$ for different objects, as those are the only three RST bands available for objects at $z\in[12-13.5]$. By $z\simeq 13.5$ the the Gunn-Peterson trough will cover the entire F158 band, and as such, for $z_{\rm emi}\gtrsim 13.5$ one can no longer use the color index method, as it requires detection in at least three photometric bands. 

\begin{figure}[!htb]
\includegraphics[width=\linewidth, height=8cm]{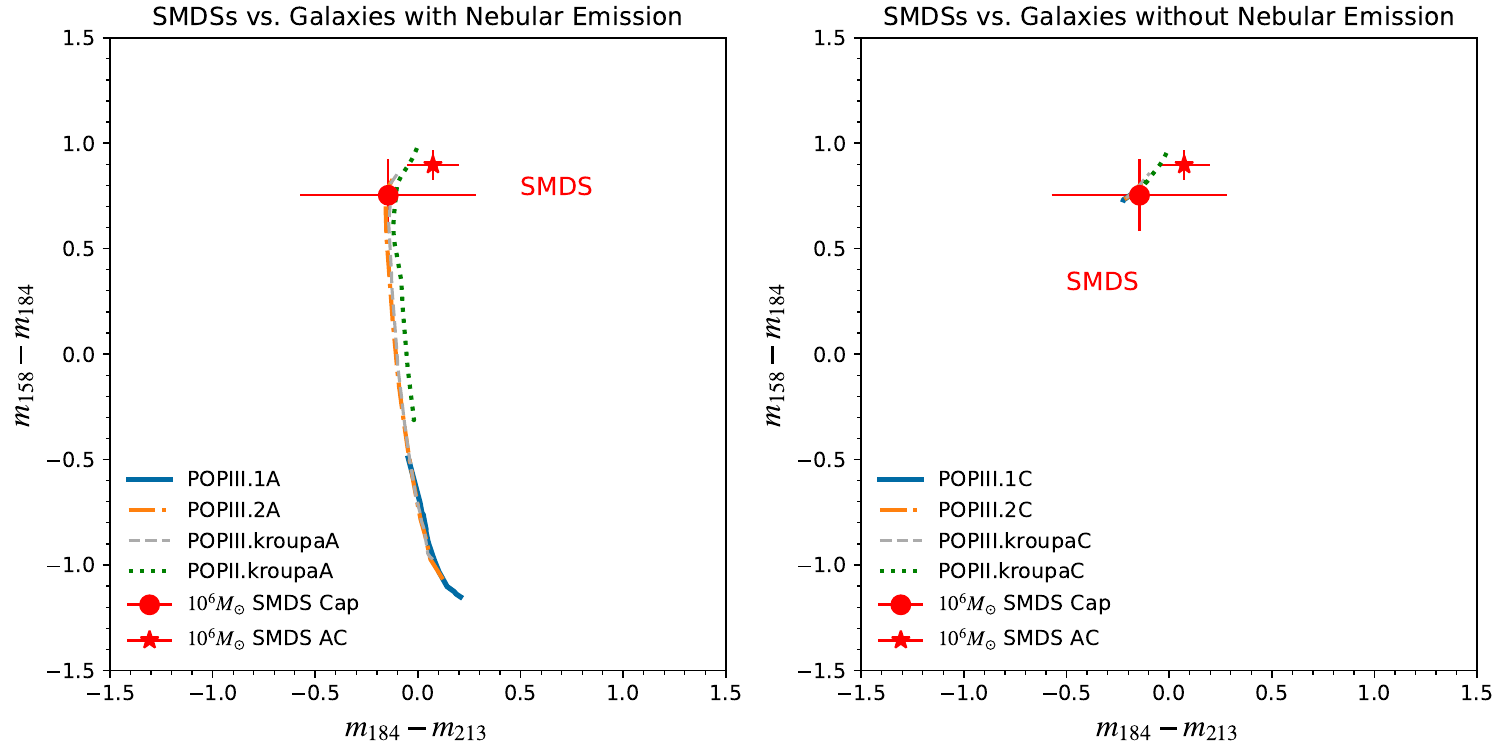}
\caption{Color-color plots of unlensed SMDSs are compared to those of type A (left panel) and Type C (right panel) Pop~III/II galaxies in $m_{184}-m_{213}$ versus $m_{158}-m_{184}$. We assume that all of objects emit light at $z_{\rm emi}= 12$. The emission spectra of Pop~III/II galaxies are adopted from the Yggdrasil model grids of~\cite{Zackrisson2011}. The red circle and stars with 1-sigma error bars represent each individual SMDS formed via Capture and AC respectively of $10^6 M_{\odot}$. The evolutionary tracks for the galaxies are represented by ascending lines, as labelled in the legend, with the youngest galaxies occupying the lowest location in the plot. Note how SMDSs formed via DM capture (red circle) will have the same color index to Type A/C galaxies, at some point in their evolution. Moreover, the SMDSs formed via AC will have very similar (and indistinguishable from within experimental error) color indices when compared to old (100 Myrs) Type A and C galaxies. Hence color-color plots in the case of unlensed objects will not be able to differentiate SMDS from early galaxies.}
\label{fig:Colorcolor}
\end{figure}

In Fig.~\ref{fig:Colorcolor} we present the color-color diagram for unlensed $10^6\Msun$ SMDSs~\footnote{We don't include here lower mass SMDSs since they are not detectable with RST without gravitational lensing, the effects of which are discussed in Sec.~\ref{sec:lensing} (see Figs.~\ref{fig:Colorcolor100} and~\ref{fig:comparecolor100}).} formed via DM capture (red circles) or adiabatic contraction (red stars). On the left/right panels we contrast the colors of SMDSs to those of unlensed Pop~III/II galaxies with maximal/no nebular emission (Type A/C), as they evolve from young (1Myrs; lowest points along the evolutionary tracks) to ``old'' (100 Myrs; highest points along the tracks) galaxies. The error bars of the magnitude difference $\sigma_{m_{1}-m_{2}}$  are calculated by the following formula:
\begin{equation}\label{eq:uncertainty}
    \sigma_{m_{1}-m_{2}}=\sqrt{\sigma_{m_{1}}^{2}+\sigma_{m_{2}}^{2}}=\sqrt{\frac{1}{S N R_{1}^{2}}+\frac{1}{S N R_{2}^{2}}},
\end{equation}
where $\sigma_{m_{1}}$ and $\sigma_{m_{2}}$ represent the standard deviation of the magnitude in the neighboring filters, which is given by the inverse of the SNR value that correspond to the image simulation results calculated by Pandeia. Thus, the color magnitude for SMDS formed via Capture is $m_{158}-m_{184}=0.61\pm 0.17$ and $m_{184}-m_{213}=-0.15\pm 0.43$, and those for SMDS formed via AC are: $m_{158}-m_{184}=0.76\pm 0.07$ and $m_{184}-m_{213}=0.07\pm 0.13$. Note that the colors of unlensed SMDSs are well within one standard deviation of the evolutionary tracks, for Pop~III/II galaxies, with or without nebular emission. 

Therefore it would be impossible to differentiate between unlensed SMDSs and most Pop~III/II galaxies using the color-color plot alone; the inclusion of the error bars for the evolutionary lines of galaxies, which we omitted here for simplicity, would only strengthen this conclusion. The only exception is Pop~III.1A (i.e. top heavy IMF with maximal nebular emission) galaxies (blue solid lines in the left panel of Fig.~\ref{fig:Colorcolor}). Due to them becoming dim at an age of $\sim3.5$ Myrs, their evolutionary tracks never reach the loci occupied in the color color plot by SMDSs. In the next section we will consider the effects of gravitational lensing on both image morphology and color index for the prospects of telling apart SMDSs from the first galaxies.

\subsection{Conclusions for unlensed SMDSs in RST}

Our studies in this Section have shown that, for unlensed objects, there is no way to prove we have found an SMDS in RST data. None of the techniques is good enough: Photometric SEDs, spectroscopy, morphology, and color-color plots are all unable to cleanly find an SMDS that doesn't look like any of the possible galaxy competitors.  Hence, once a candidate SMDS is found via one of these techniques, followup observations e.g. with JWST will be required.  Observation of the He-1640 line would be a smoking gun for a Dark Star. However, for SMDSs at redshifts higher than z$>$10.7, that line is outside the wavelength range of RST. JWST could see it.

\section{Effects of Gravitational Lensing\label{sec:lensing}}

Our discussion in Section 5 above only pertained to unlensed objects, both SMDS and early galaxies.
Now we turn to the case where these objects are gravitationally lensed by foreground material.
The method of strong lensing as a tool to find high redshift objects has been successfully applied and refined in recent years. The detectability of dark stars via lensing was first explored by~\cite{Zackrisson:2010HighZDS}, who found that a magnification factor $\mu\gtrsim 160$ is needed in order to  observe dark stars (at $z_{\rm emi}\simeq10$ ) with low masses $\sim 100 \Msun$ with JWST. For the Roman Space Telescope we find (see Fig.~\ref{fig:ABmagCap} and~\ref{fig:ABmagAC})  that unlensed SMDSs of mass smaller than $\sim 10^6 M_{\odot}$ are too faint to be detected even with exposure times of $\sim 10^6 s$. Therefore, SMDSs of $M\lesssim 10^6\Msun$ can only be detected if they are gravitationally lensed, as we will explicitly show in this section. 

Lensing by massive foreground galactic clusters at $z \sim 0.5$~\citep[as described by][for example]{2009ApJ...707L.102Z} would lead to large magnification factors ($\mu>10$). For example~\cite{2013ApJ...762...32C,2019arXiv190308177L} describe the discovery of high redshift candidate galaxies ($z \sim 11$) through lensing. Perhaps one of the most striking examples of strong gravitational lensing is the recent discovery of the most distant star ($z\sim 6.2$) ever observed: Earendel~\citep{2022Natur.603..815W,EarendelJWST:2022}. The light from the $50-500\Msun$ stellar object  has been emitted when the universe was merely 900 million years old.\footnote{This shatters the previous record held by Icarus, who's light was emitted when the universe was 4.4 billion years old~\citep{2018NatAs...2..334K}.} The lensing effect from the foreground cluster boosts the stellar flux by $1000$ to $40000$ times, providing a great example on how strong gravitational lensing allows the observation of very distant individual stars. Additionally, soon after JWST released its first images, the list of high redshift galaxy candidates has been growing almost on a weekly basis, with the record for the ``most distant candidate galaxy'' being broken multiple times~\citep[e.g.][]{GHz2,GLASSz13,Maisies:2022,z16.CEERS93316:2022,z17.Schrodinger:2022}. If those objects are indeed at the redshifts indicated by photometry, then at least some of them must have been gravitationally lensed in order to have been observed. In a separate publication we plan to analyse if the available data from any of those candidates can be well fit by SMDSs SEDs. Recently we have shown that out of the four $z\gtrsim 10$ spectroscopically confirmed ``galaxies'' found by~\citep{JADES:2022a,JADES:2022b}, the three most distant ones (specifically JADES-GS-z13-0, JADES-GS-z12-0, and JADES-GS-z11-0) are each consistent with a Supermassive Dark Star interpretation~\citep{Ilie:2023JADES}.

In this section we discuss the effects of gravitational lensing on the prospects of telling apart SMDSs from the first galaxies using RST by using image morphology (Sec.~\ref{ssec:ImageMorph}) and color color indices (Sec~\ref{ssec:ColorColorMag}). Below we remind the reader what are some reasonable amounts of lensing we expect based on a non-exhaustive list of previous observations. Besides the extreme case of Earandel, discussed above, we mention other examples of greatly lensed stars: \cite{Kaurov2019ApJ...880...58K} finds a star with $\mu 
\sim 200$ at low redshift $z\sim 0.9$; \cite{Meena2023ApJ...944L...6M}, one star with $\mu \sim 200$ and one star with $\mu \sim 50$ at high redshift $z\sim 4.8$. For galaxies: \cite{Hsiao:2022fzs} finds a triply lensed $z\sim 11$ candidate, with combined $\mu \sim 15$; \cite{Bradley2022arXiv221001777B} find $z\sim 9-13 $ galaxies lensed by $\mu \lesssim 10$; \cite{Adams2023MNRAS.518.4755A}, one lensed by $\mu \sim 6$ with $z\sim 9$.

\subsection{Image Morphology}\label{ssec:ImageMorph}

Gravitational lensing magnification would increase the apparent size of an object, but for nearly point sources (stars), this translates to the magnification in the flux density by a factor $\mu$, called the magnification factor~\citep[for details see][for example]{1992grle.book.....S}. This factor is given by~\citep[e.g.][for a review]{Kilbinger:2014cea}:

\begin{equation}\label{eq:magfac}
    \mu=\frac{1}{\left[(1-\kappa)^{2}-\gamma^{2}\right]}
\end{equation}
where $\kappa$ is determined by the mass field of the lens as a function of the position vector: $\kappa \propto D_{\rm L}D_{\rm LS}/D_{\rm S}$, where $D_{\rm L}$, $D_{\rm S}$ and $D_{\rm LS}$ refer to the angular distance to the lens, the source and their distance in between, respectively. $\gamma$ is the shear factor that determines the distortion of the source object. In a strong-lensing survey out of the catalogue of 12 galaxy clusters from the MAssive Cluster Survey (MACS), 9 of them have a large area ($>2.5~\square^{\prime} $)\footnote{$\square^{\prime}$ is a notation for the unit $\mathrm{arcmin}^2$.} of high lensing magnification ($\mu>10$) for source objects at $z \sim 8$~\citep{2011MNRAS.410.1939Z}. Since our targeted redshift is $z\sim 12$, we find a simultaneous increase in $D_{\rm LS}$ and $D_{\rm S}$, thus a increase in the value of $\kappa$. Therefore, for the same relative position in the sky, a higher redshift corresponds to a higher $\mu$, which means we have a even larger area for high lensing magnification. And since this area is much larger than the size our target sources (order of $\sim \square^{\prime\prime}$),\footnote{$\square^{\prime\prime}$ is a notation for arcsecond$^2$.} it is possible for multiple Pop~III galaxies and/or SMDSs to fit in this magnification area.

It is likely to observe multiple images or giant arcs from strong lensing~\citep[e.g.][]{2022arXiv220703532W} but given the relatively small size of galaxies at very high redshift, they might still appear as spatially unresolved ~\citep[e.g.][]{2013ApJ...762...32C,2019arXiv190308177L}. For simplification, and in order to be the most conservative, we assume a spherical shaped image and that $\gamma =0$ throughout which means the magnification wouldn't distort the shape of the object for each individual image (although this simplification is unlikely in practice  as it requires homogeneous distribution of mass). The reason for this choice is that a distorted image would be a telltale sign of an extended object (galaxy), rather than a point object (SMDS with no nebular emission). As a result of lensing, the surface area is increased by a factor known as the convergence factor. Assuming a uniform convergence, as justified above, this factor simply becomes  $1/(1-\kappa)^2 = \mu$. Furthermore, we consider our sources to be close to the critical curve to a degree such that a high magnification factor ($\mu >10$) could be achieved without generating spatially resolved arc.

\begin{figure}[!htb]
\includegraphics[scale=0.6]{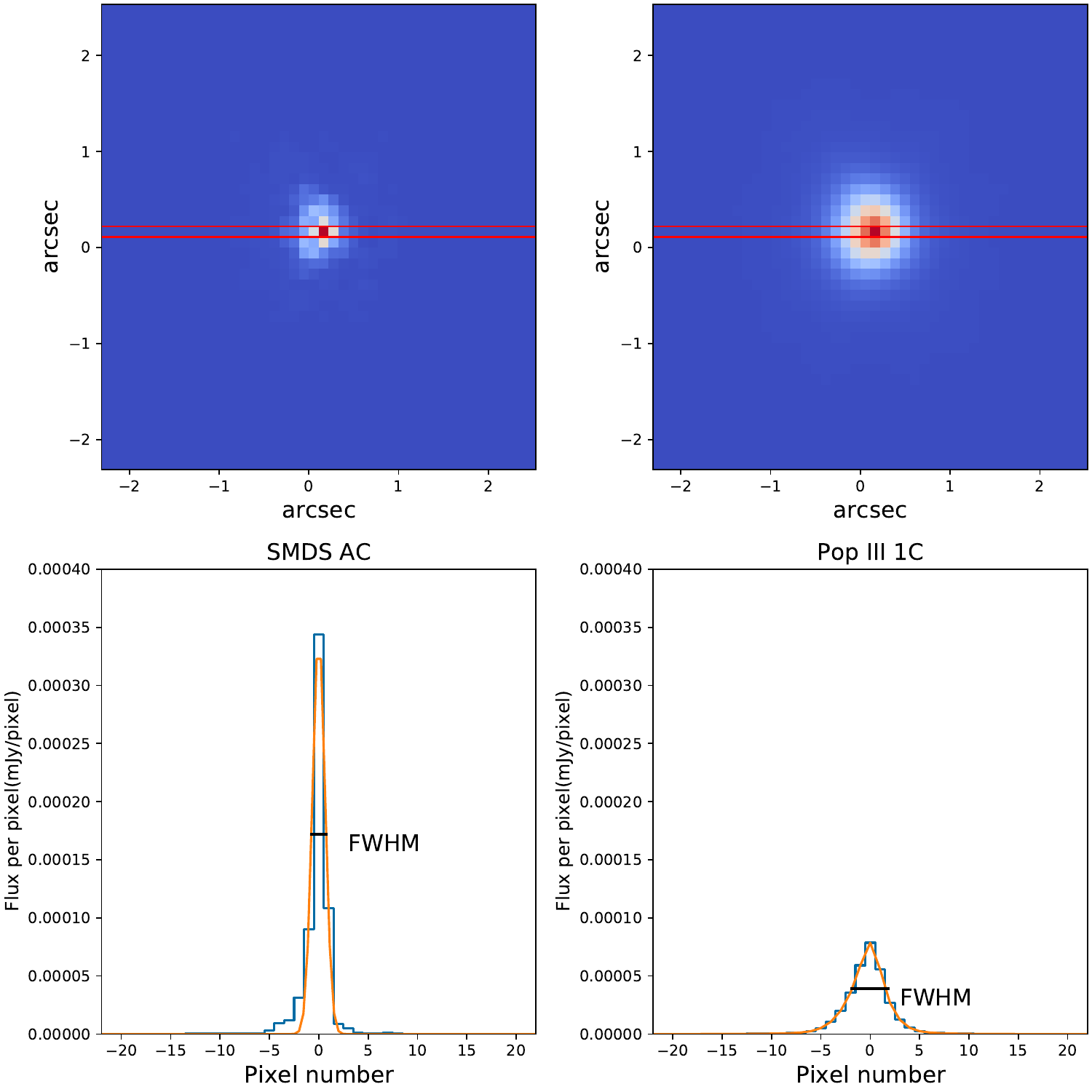}
\caption{In this figure we demonstrate how we estimate the effective angular size in the case of lensed objects as viewed by RST by using  simulated images generated with the Pandeia code. For demonstration purpose we have assumed a large lensing factor of $\mu =100$ and $z_{\rm emi}=12$ for all objects. In the left two panels, we use a $10^6 M_{\odot}$ SMDS formed from AC for the image simulation, whereas the right panels shows a Pop~III.1 type C galaxy at 1Myr of age. The mass, and therefore intrinsic size of the galaxy is chosen in such a way that the magnitude of the galaxy, in the band presented here (F184) matches the magnitude of the SMDSs. In order to estimate the effective size of each object, we take a slice bounded by red lines (top row) to generate the corresponding SNR histogram (bottom row) and estimate the FWHM of the flux as the black solid lines shown in bottom two panels.  The effective size of the lensed Pop~III galaxy ($\approx 0.41"$) is above the resolution limit in this band ($\approx0.19''$). Moreover, the SMDS is still unresolved, as its estimated size is $\lesssim 0.17"$. }
\label{fig:SMDSPSF}
\end{figure}

The main goal of this sub-section is to investigate how gravitational lensing could be used to tell apart first galaxies from supermassive dark stars, as observed with RST. The former are extended objects, with radii from ranging a few tens of parsecs to hundreds of parsecs while the later are puffy stars, with radii of the order of $\sim 10$~AU. As we have seen in Sec.~\ref{ssec:ImgSimul}, RST's angular resolution is not sufficient to tell those two kind of apart conclusively, without gravitational lensing. For each galaxy considered we determined their magnified angular size by using the Pandeia simulator. Our procedure is exemplified in Fig.~\ref{fig:SMDSPSF}, where we extract the effective lensed sizes for two objects of interest magnified by $\mu=100$: a $10^6\Msun$ SMDSs formed via adiabatic contraction (left panel) and a young (1~Myrs) Pop~III.1C galaxy (right panel). Note how now the light from the galaxy covers a significantly larger area of the detector, indicating an extended object. In order to estimate the effective size of each object we use the standard Full Width Half Maximum (FWHM) histogram technique. Namely, we slice the images along a direction. In the figure this is represented by the red horizontal band that goes to the left and the right central pixel of each simulated image. The unit flux values in pixels along this band are then plotted as a histogram (blue lines in the bottom panels of Fig.~\ref{fig:SMDSPSF}). For SMDSs, as they are spherically symmetric stellar objects we apply the conventional Gaussian fit~\citep[e.g.][]{Mighell2005MNRAS.361..861M}. Conversely, for Pop~III galaxies we simply interpolate the flux values at each pixel. The corresponding FWHM values are extracted and plotted as black lines in each of the bottom panels of Fig.~\ref{fig:SMDSPSF}. For instance, for the SMDSs we find that its effective size will be roughly $0.17''$ and for the Pop~III galaxy chosen here the estimated effective size is $\sim0.41''$. For the band considered (F184) the angular resolution, according to the Rayleigh criterion ($1.22\lambda/D$), where $D$ is the diameter of the telescope, leads to a minimum angular size of $\sim 0.19"$ for an object to be resolved. Thus, the SMDSs, even if lensed, is still unresolved, whereas the Pop~III.1C galaxy will show as an extended object. It is important to note here that the estimated effective size for all objects that remain unresolved is merely an upper bound of the actual size of the object.  

\begin{figure}[!htb]
\includegraphics[width=\linewidth]{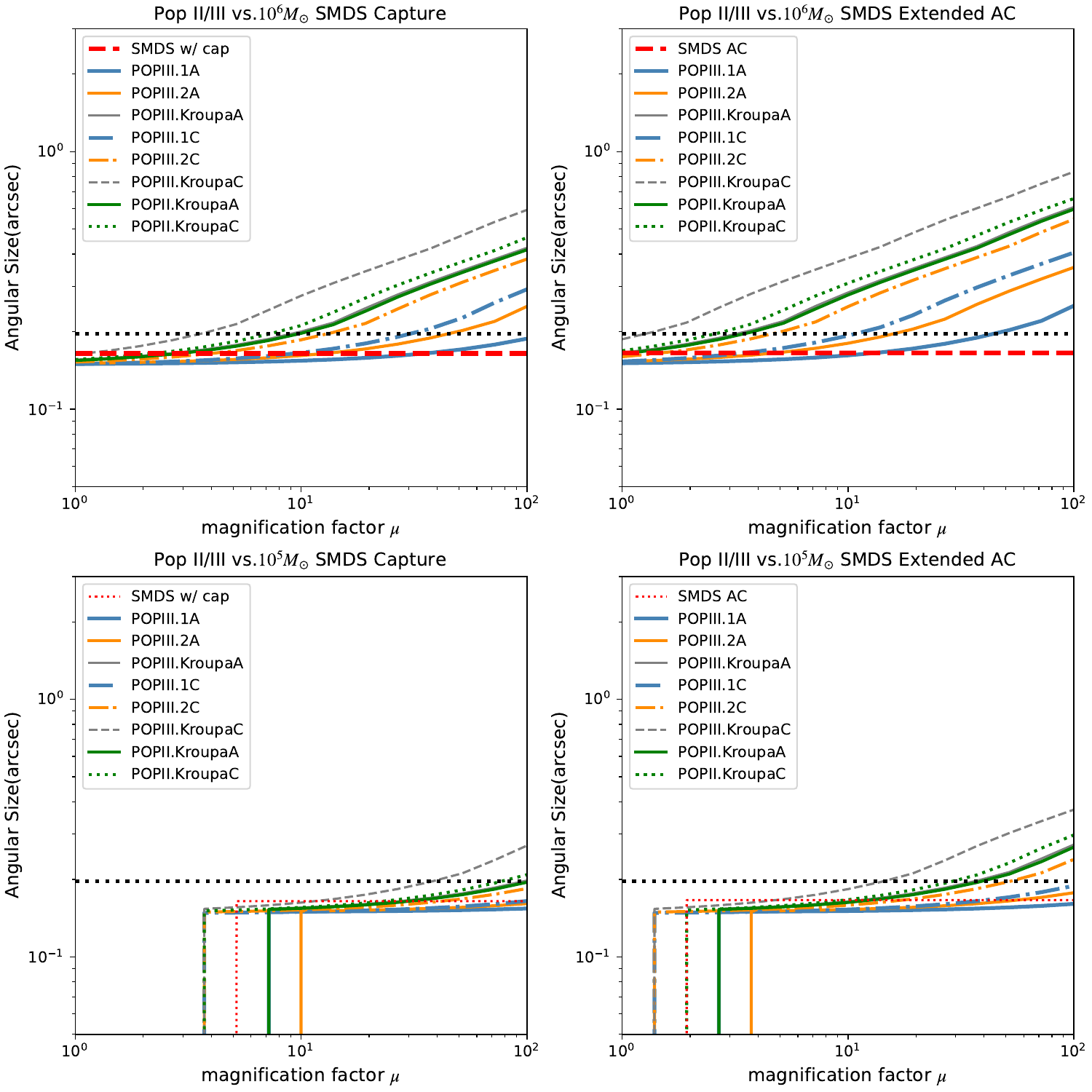}
\caption{Using lensing to differentiate SMDS vs. young Pop III/II galaxies:  Estimated effective angular sizes vs. gravitational magnification factor ($\mu$). For all objects $z_{\rm emi}=12$ is assumed. In all panels, the red  lines represent SMDSs of either $10^6\Msun$ (top row) or $10^5\Msun$ (bottom row) formed via Capture (left column) and AC (right column). The other curves colored according to the legend represent the angular sizes of young (1~Myrs old) Pop~III/II galaxies of various types that are chosen in such a way that they have the same AB magnitude as the SMDSs in the F184 band. The horizontal black dotted lines (the Rayleigh criterion) represent the diffraction limit, above which one can tell an extended object from a point source: i.e. objects with apparent sizes above this horizontal line can be resolved. Note that SMDSs are not large enough to be resolved, even with magnifications as high as $\sim 100$. In contrast, as shown in the top two panels, at $\mu\sim 100$ most of the Pop~III/II galaxies counterparts to $10^6\Msun$ SMDSs will appear as resolved, extended objects. The only exception is the case of PopIII.1A galaxies counterparts of the $10^6\Msun$ SMDS formed via DM capture (see blue solid line in the top left panel) which will still appear as point sources even when $\mu\sim 100$. As shown in the lower two panels, the galaxies matching the lower mass ($10^5\Msun$) SMDSs will be more compact, and many of them appear as point sources even at $\mu\sim 100$. The vertical break in the lines denoting the angular sizes of lower mass SMDS or galaxies (bottom panels) represents the minimum magnification necessary in order to have a detection at the SNR=5 level.}
\label{fig:Size}
\end{figure}

Using the procedure described above, in Fig.~\ref{fig:Size} we estimate the apparent sizes as a function of lensing magnification factor ($\mu$) for young (1~Myrs) Pop~III/II galaxies (upward trending lines, as labeled) that match the integrated F184 flux of supermassive dark stars (top/bottom rows for $10^6/10^5\Msun$ SMDSs) formed via either formation mechanism (left/right columns correspond to DM capture/AC). In the lower panels the evolutionary lines break sharply downward, signifying that below the magnification where this break happens, the SNR in each pixel of the F184 band becomes lower the minimum SNR=5 value required for detection. In each of the four panels the black dashed horizontal line (labelled Rayleigh criterion) is the angular resolution of the RST in the band considered here (F184), estimated to be $\sim 0.19''$. Note that SMDSs at $z_{\rm emi}\sim 12$ are unresolved even at $\mu\gg 100$. 
Thus, for all intents and purposes SMDSs without nebular emission will remain point objects, even with extreme magnifications, and even with telescopes with much better angular resolution than RST.  
The galaxies considered will become resolved once $\mu$ becomes larger than a critical value that can be read off the plot at the intersection of the horizontal dotted line (labeled as Rayleigh criterion) and the upward trending curve corresponding to the galaxy in question. Young Pop~III.Kroupa Type C galaxies will be the easiest to resolve, in view of their large mass needed in order to match the flux of the SMDSs in the F184 band (See Fig.~\ref{fig:Pop3scaled}). In contrast, Pop~III.1 Type A galaxies need the smallest stellar mass to match the flux of the SMDSs, and are more difficult to resolve. When contrasted against a $10^6\Msun$ SMDS formed via DM capture, a Pop~III.1A galaxy will need to be magnified by at least $\mu\sim 125$ in order to have an angular size barely above the resolution limit and, as such, somewhat distinguishable, based on image morphology from the SMDS. For the case of SMDSs formed via AC, the corresponding Pop~III/II galaxies that match their F184 flux are easier to resolve, as it can be seen from contrasting the left and right panels of Fig.~\ref{fig:Size}. 
In the lower two panels of Fig.~\ref{fig:Size} we consider $10^5\Msun$ SMDSs formed via either mechanism. In this case, even at $\mu\sim 100$ there are still many Pop~III/II galactic counterparts that will not be resolved. Thus, for SMDSs with $M\lesssim 10^5\Msun$, differentiation from young ($\sim$ 1 Myrs old) Pop~III/II galactic sources based on image morphology is only possible at $\mu\gg 100$. However, the color-color technique can be very useful in telling apart a $M\gtrsim 10^5\Msun$, formed via AC, from Pop~III/II galaxies (see Figs.~\ref{fig:Colorcolor100} and~\ref{fig:comparecolor100} ).

The main takeaway of Fig.~\ref{fig:Size} is that at SMDSs will appear as point objects, independent of the magnification factor. In contrast,  {\it{all}} young Pop~III/II galactic counterparts of a $10^6\Msun$ SMDS formed via AC would be resolved by RST with $\mu\gtrsim 40$ thus making disambiguation possible based on image morphology whenever $\mu\gtrsim 40$. For $10^6\Msun$ SMDSs formed via Capture, the same would be possible with a higher magnification factor of $\mu\gtrsim 125$ but at $\mu\sim 100$ it will be hard to disambiguate them from young Pop III.1A galaxies. For lower mass SMDSs ($M\lesssim 10^5\Msun$), independent of formation mechanism, one needs $\mu\gg 100$ to be able to use image morphology as a RST differentiating tool from young Pop~III/II galaxies. 
\begin{figure}
\centering
 \includegraphics[width=\linewidth]{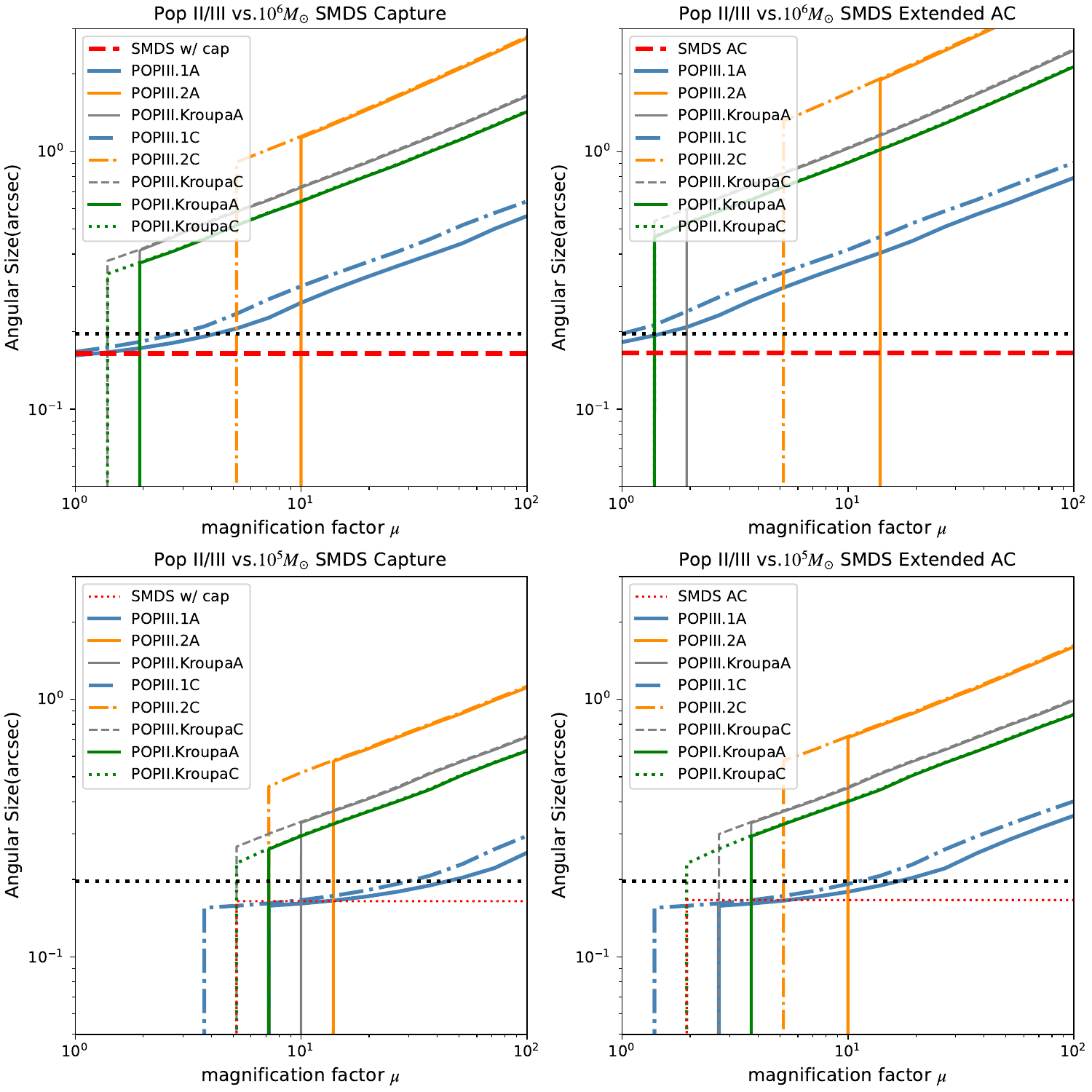}
\caption{Using lensing to differentiate SMDS vs. old galaxies: Same as Fig.~\ref{fig:Size}, but this time considering old galaxies. For all galaxies the age chosen is 100 Myrs old, except for the case of Pop~III.1A/C galaxies, that cease to emit light after about 3.5 Myrs. As in the previous figure, the horizontal black dotted lines (the Rayleigh criterion) represent the diffraction limit, above which one can tell an extended object from a point source: i.e. objects with apparent sizes above this horizontal line can be resolved.
As shown in the top two panels, at $\mu \sim 10$ most
of the Pop III/II galaxies counterparts to $10^6 M_{\odot}$ SMDSs will appear as resolved, extended objects, and can be differentiated from $10^6 M_{\odot}$ SMDSs which appear as  point sources. As shown in the lower two panels, the galaxies matching the lower mass ($10^5 M_{\odot}$ ) SMDSs will be more compact and appear as resolved, extended objects at $\mu \sim 30$ and can be differentiated from SMDS.}
\label{fig:Size100}
\end{figure}

In Fig.~\ref{fig:Size100} we repeat the same analysis done in Fig.~\ref{fig:Size}, this time for older galaxies, with age of 100~Myrs.\footnote{For the case of the top heavy IMF Pop~III.1 galaxies, since most of their stars exhaust the nuclear fuel after $\sim 3.5$ Myrs, we select a time just shy of this value as their ``old'' age.} The top two panels of Fig.~\ref{fig:Size100} demonstrate that most old galaxies that match the flux of a $10^6\Msun$ SMDSs in the F184 band will be above the resolution limit as soon as they are detected at a $S/N=5$ level. The only exception is the Pop~III.1 galaxies (blue lines), which, when contrasted against a $10^6\Msun$ SMDSs formed via DM capture (top left panel)  can be detected as unresolved, as long as it is magnified by $\mu\lesssim 5$. With this in mind we conclude that old galaxies would be, in principle, differentiable from $10^6\Msun$ SMDSs based on image morphology. Regarding the $10^5\Msun$ SMDSs, all of their old Pop~III/II galactic counterparts are large enough to be resolved at $\mu$ ranging from a few to a few tens. In particular, for the case of the $10^5\Msun$ SMDS formed via DM Capture (lower left panel of Fig.~\ref{fig:Size100}) by $\mu\sim 50$ even the more compact Pop~III.1A (blue solid line) galaxies are resolved. When contrasting $10^5\Msun$ SMDSs formed via extended AC (lower right panel) we note that at $\mu\gtrsim 20$ all Pop~III/II galaxies will show as extended objects, in contrast to the SMDS. Thus, differentiation based on image morphology of a $\sim10^5\Msun$ SMDS, formed via either mechanism, from Pop~III/II old galaxies is possible with magnifications $\mu\gtrsim 50$. 

Since the age of an object detected with photometry cannot be estimated exactly, one should view our two extreme cases considered in Figs.~\ref{fig:Size} and~\ref{fig:Size100} as the two limits in between which any observation would lie. Moreover, we point out that the size of the galaxies or other lensed objects will depend on the assumed lensing model and is usually associated with a large error bar, as shown for example in Fig.~8 of~\citet{2019arXiv190308177L}. Thus one should be careful in using the Rayleigh criterion as a disambiguation between galaxies (extended objects) and SMDSs with purely stellar spectra (point objects), especially whenever the objects are barely resolved. 

In summary, in this section we have demonstrated that, with sufficient magnification ($\mu\gtrsim 40$), a $10^6\Msun$ SMDS formed via AC can be differentiated via image morphology from all Pop~III/II galaxies considered.  The galaxies would all appear as extended objects whereas the SMDSs are still unresolved, even for $\mu\gg100$ (see top right panels of Figs.~\ref{fig:Size} and~\ref{fig:Size100}).  In contrast, for all $\mu\lesssim100$, image morphology cannot be used to conclusively differentiate between galaxies and $M \lesssim 10^5 M_\odot$ SMDS (lower panel of Figs.~\ref{fig:Size}) or between galaxies and a $10^6\Msun$ SMDS formed via DM Capture (top left panel of Fig.~\ref{fig:Size}). In the next section we will explore if lensed SMDSs can be differentiated from Pop~III/II galaxies based on their  locations in color-color diagrams.

\subsection{SNR Ratio and Color-color Diagram}\label{ssec:ColorColorMag}

\begin{figure}[!htb]
\centering
\includegraphics[width=0.85\linewidth]{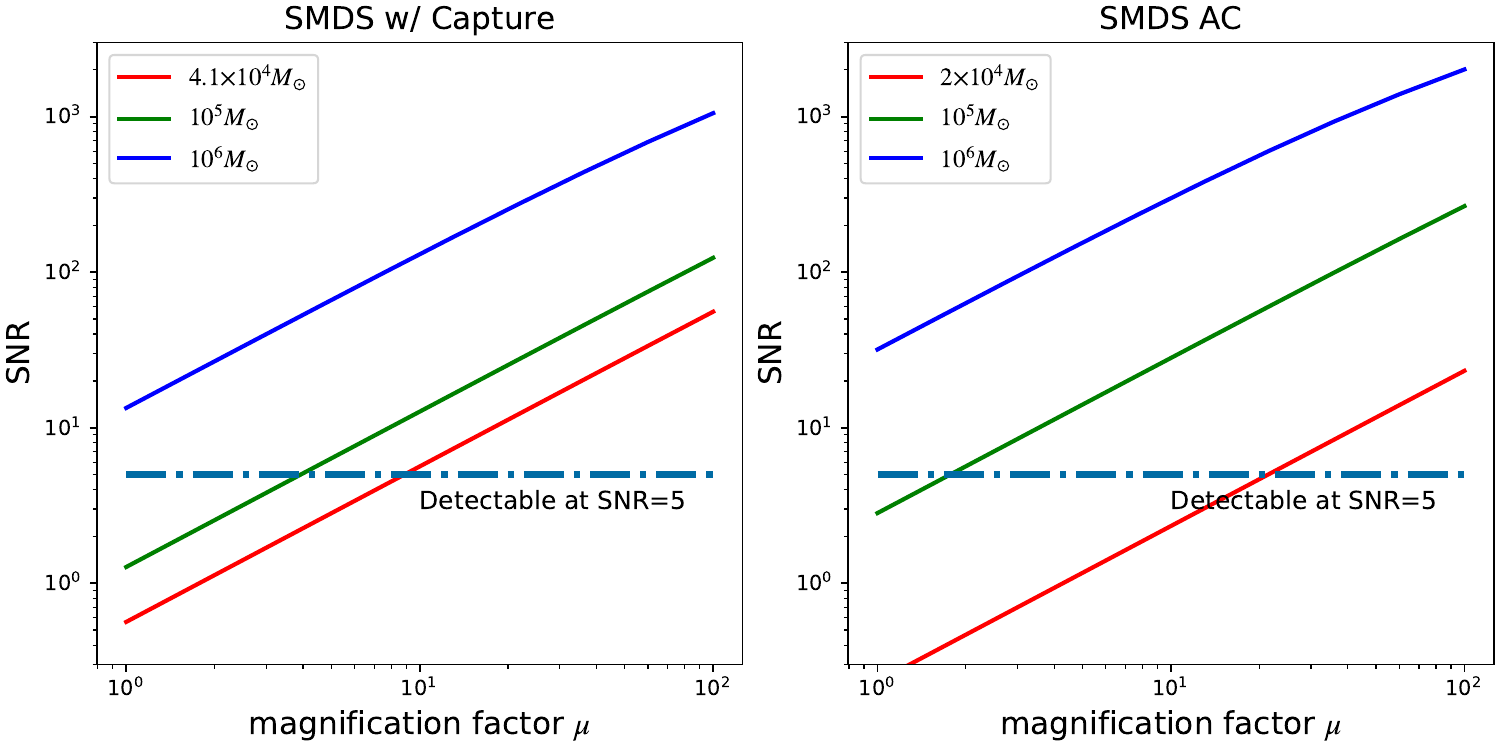}
\caption{SNR values of SMDSs of masses $10^4$--$10^6 M_\odot$ (at $z_{\rm emi}=12$) versus magnification factor ($\mu$) powered by 100 GeV WIMPs. The SNR value was calculated assuming $10^6s$ exposure time and the F184 filter. The left panel represents SMDSs formed via DM Capture and the right panel represents those formed via adiabatic contraction (AC). The horizontal line corresponds to SNR=5, which is usually the minimum value required to claim detection. 
This plot shows that  SMDSs with $M\sim 10^4 M_{\odot}$ could be detectable with moderate amount of lensing ($\mu \sim 10$).}
\label{fig:SMDSvsMu}
\end{figure}

In this subsection we discuss the effects of gravitational lensing on the prospect of detecting smaller mass SMDSs and the possibility to use color-color diagrams to differentiate those SMDSs from Pop~III/II galaxies. We begin our analysis by asking the following question: what is the required magnification factor needed for a SMDS of a given mass in order to be detectable with RST at the level of SNR=5? In order to answer this question we plot in Fig.~\ref{fig:SMDSvsMu} the values of the SNR in the F184 band as a function of magnification ($\mu$) for SMDS of various masses (as labeled) formed either via DM capture (left panel) or adiabatic contraction (right panel).  The SNR values are estimated using Pandia and are selected from the pixel where most of the the photons from the SMDS will generate a signal, in a hypothetical $10^6$~s exposure. We noted before that at the same mass SMDSs formed via AC look brighter that their counterparts formed via DM capture, which can also be seen in the slightly larger SNR values on the curves in the right panel (SMDS AC) of Fig.~\ref{fig:SMDSvsMu}. The main takeaway of Fig.~\ref{fig:SMDSvsMu} is that magnification factors of $\mu\gtrsim 10$ are sufficient to lead to SMDSs of $M\sim10^4\Msun$ (formed via either mechanism) to be detectable with $10^6$~s exposures by RST.

\begin{figure}[!htb]
\centering
\includegraphics[width=0.85\linewidth]{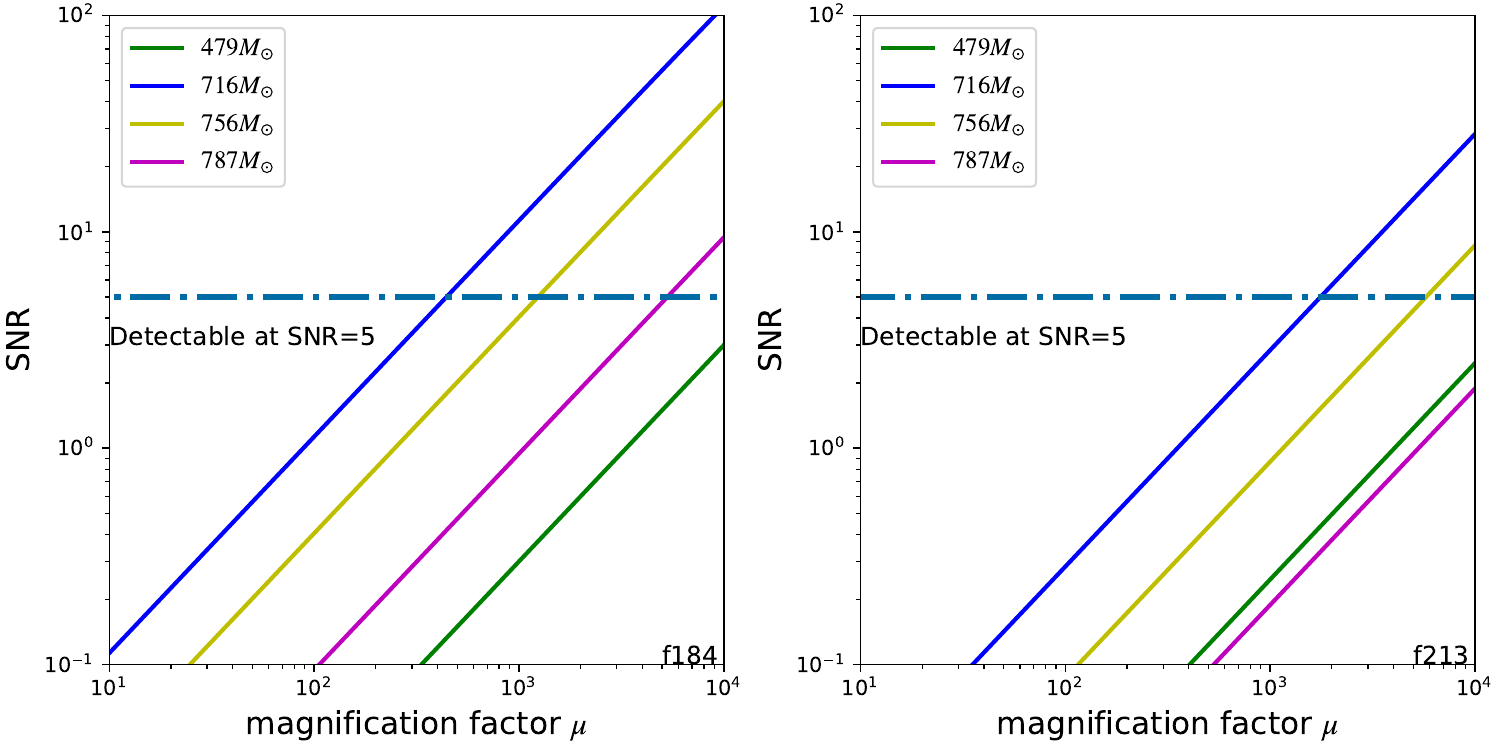}
\caption{SNR values versus magnification factor ($\mu$) for Dark Stars of masses $479$ and $716\Msun$ for the case of extended AC, and $756$ and $787 \Msun$ for the case of captured DM. We take $z_{\rm emi}=12$ and annihilation of 100 GeV WIMPs to be the heat source for the SMDSs. The SNR values are calculated assuming $10^6s$ exposure time and the F184/F213 filters (left/right panels). The horizontal line corresponds to SNR=5, which is usually the minimum value required to claim detection. The parameters for the Dark Star models are taken from~\cite{Spolyar:2009}. Note that for the same object, a higher magnification is needed in order for it to be detected at the same SNR level in the F213 filter vs the F184 filter. This is indicative of the weakened sensitivity at higher wavelengths. The main takeaway of this figure is that Dark Stars with masses as low as $\sim 800\Msun$ can be detected with RST, if magnified by lensing factors $\mu$ of the order of a few hundred.}
\label{fig:DSvsMu}
\end{figure}

In Fig.~\ref{fig:DSvsMu} we repeat the same exercise done in Fig.~\ref{fig:SMDSvsMu}, but this time we consider Dark Stars of lower masses, i.e. with $M\lesssim 10^3\Msun$. We assume $z_{\rm emi}\simeq 12$, and a heat source from 100 GeV WIMPs powering the DSs. The properties of DS models used here (such as $T_{\rm eff} $ and $R$) are taken from Table~2 of~\cite{Spolyar:2009}. In the left panel of Fig~\ref{fig:DSvsMu} we show the SNR values in the F184 RST band, whereas the right panel depicts the SNR values in the F213 RST band. In both cases an exposure time of $10^6$s is assumed. The weakened sensitivity in the F213 filter leads to higher $\mu$ values required for a detection at the same SNR level. Further, we can comment on the difference between the cases of SMDS formed by AC (the 479$~\rm M_\odot$ and 716$~\rm M_\odot$ SMDSs shown in the figure) vs. the DS formed via capture (the 756$M_\odot$ and 787$M_\odot$ DSs shown in the figure).
 SMDSs formed via capture are hotter, leading to a peak in the SED is at shorter wavelengths. Thus, in the two filters considered here, the 479$M_\odot$ and 756$M_\odot$ DSs (formed via capture) are somewhat dimmer and require more magnification (a larger $\mu$) than the 716$M_\odot$ and 787$M_\odot$ DSs formed via AC. 

The main takeaway of Fig.~\ref{fig:DSvsMu} is that Dark Stars as distant as $z_{\rm emi}\simeq12$, with masses as low as $M\simeq 800\Msun$ can be observed with RST, if magnified by $\mu$ of a few hundreds or more. This is consistent with the findings of~\cite{Zackrisson:2010HighZDS}, who shows that for $\mu\simeq 160$ DS at $z_{\rm emi}\simeq12$, with masses as low as $700\Msun$ can be detected at the SNR=5 level in JWST, assuming exposure times of 100 hours. 

\begin{figure}[!htb]
\includegraphics[width=\linewidth]{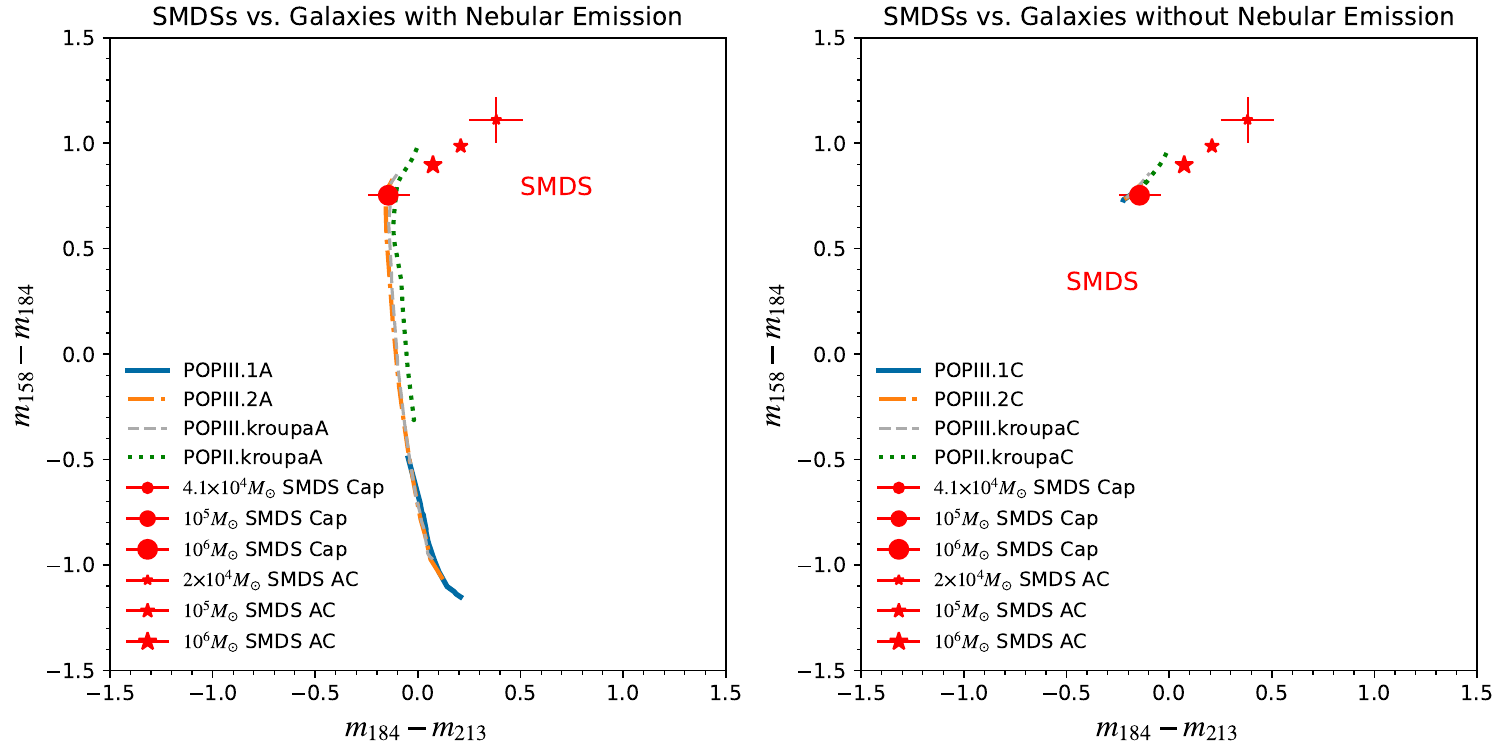}
\caption{Color-color plots as in Fig.\ref{fig:Colorcolor}, but here we have assumed the lensing factor $\mu = 100$. The red circle and stars with 1-sigma error bars represent individual SMDS formed via Capture and AC respectively, this time with masses $2\times 10^4, 10^5, 10^6 M_{\odot}$. Evolutionary tracks are also shown for both type A (left panel) and Type C (right panel) Pop~III/II galaxies. All SMDSs formed via DM capture, regardless of stellar mass, occupy roughly the same spot (red circles), lie along the evolutionary lines of galaxies, and can therefore not be differentiated from galaxies.  On the other hand, SMDSs formed via AC (red stars in both panels) have color-color indices that could differentiate them from galaxies.}
\label{fig:Colorcolor100}
\end{figure}

\begin{figure}[!htb]
\includegraphics[width=\linewidth]{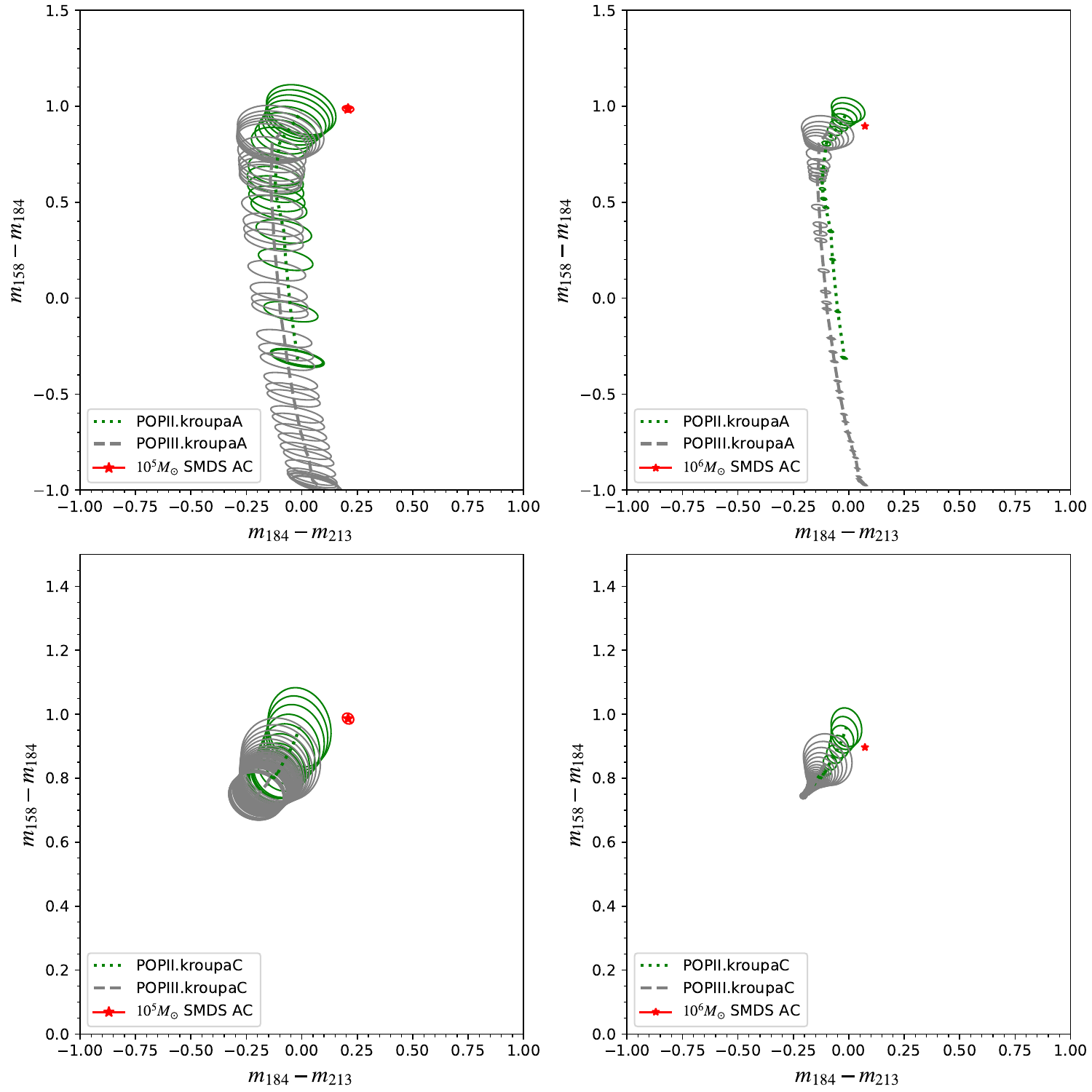}
\caption{Similar to Fig.\ref{fig:Colorcolor100}, we plot the color indices of our $z_{\rm emi}=12$ target objects assuming lensing magnification $\mu = 100$ for the case of SMDS formed by AC of mass  $10^5 M_{\odot}$(left panels) and $10^6 M_{\odot}$(right panels). We include the $2\sigma$ error ellipses both for the SMDS colors and those of the Pop~III/II galaxies along their evolutionary tracks.  Note that we have not plotted lighter SMDS as they are indistinguishable from galaxies at the $2\sigma$ level for this value of magnification.
We only keep Pop~III/II Kroupa galaxies as our galactic candidates since their evolutionary tracks are closest to the chosen SMDSs. The panels on the top row are the comparisons to galaxies with full nebular emission contribution (type A), whereas all the bottom panels are the comparisons to the galaxies with no nebular contribution (type C).
 We find that based on their color indices one can distinguish lensed ($\mu\sim 100$) SMDSs with $M \sim 10^5, 10^6 M_\odot$
 from Pop~III/II galaxies, as observed with RST, at the 2$\sigma$ level;
i.e., SMDSs with $M \sim 10^5, 10^6 M_\odot$ deviate by at least $2\sigma$ in each direction from the evolutionary tracks (green and gray lines) of the color indices of the first generation of galaxies.}
\label{fig:comparecolor100}
\end{figure}

In the reminder of this subsection we focus on the role of gravitational lensing on differentiating SMDSs from Pop~III/II galaxies using the color color technique discussed in Sec.~\ref{ssec:Color}. We start with Fig.~\ref{fig:Colorcolor100}, where the color index analysis in the $m_{184}-m_{213}$ vs. $m_{158}-m_{184}$ plane presented in Fig~\ref{fig:Colorcolor} is redone, assuming now that all objects are magnified with a lensing factor $\mu=100$. 
In the left/right panel we compare the color indices of SMDSs formed via DM capture (red circles) and adiabatic contraction (red star symbols) to the evolutionary tracks of Pop~III/II type A/C. 
For each SMDS we calculate the size of the error bar according to Eq.~(\ref{eq:uncertainty}). The boosted SNRs lead to a significant reduction of the uncertainty in the color indices of SMDSs, as can be seen by comparing Figs.~\ref{fig:Colorcolor} and~\ref{fig:Colorcolor100}. Moreover, since SMDSs formed via DM capture have roughly equal temperature, independent of stellar mass, their SEDs (and therefore colors) do not change significantly as the SMDS grows. For this reason all the color indices of the SMDSs formed via DM capture are stacked upon each other at a point that is very close to the evolutionary tracks of both type galaxies. This reinforces once more that SMDS formed via DM capture can be easily mistaken for early galaxies if neither of those objects is resolved. We don't add here error bars for the points along the evolutionary tracks of galaxies, as those uncertainties will will be dependent on the SNR (via Eq.~(\ref{eq:uncertainty})), and, as such, on the mass of the SMDSs counterpart. However, we revisit this in Fig.~\ref{fig:comparecolor100} where we plot separately each SMDSs formed via AC, which allows for inclusion of uncertainties in color indices of both galaxies and SMDSs.

Fig.~\ref{fig:Colorcolor100} reinforces the point that SMDSs formed via capture are great Pop~III/II galactic chameleons. Independent of their mass, they will occupy the same spot in the $m_{158}-m_{184}$ vs. $m_{184}-m_{213}$ color color plot, a spot that is along the evolutionary lines of both type A (left panel) and type C (right panel) galaxies. 
A natural question arises: can one, in absence of spectral data, tell apart a SMDSs formed via DM capture from the Pop~III/II galaxies counterparts that occupy roughly the same location in the color color plot of Fig.~\ref{fig:Colorcolor100}? First, notice that the color-color technique is very valuable, as it clearly disambiguates between both types SMDSs and Pop~III.1A galaxies. Therefore, considering the case of a $10^6\Msun$ SMDS formed via DM capture, for $\mu\gtrsim 40$, image morphology can break the degeneracy presented in Fig.~\ref{fig:Colorcolor100}, as at those magifications the galactic counterparts (that also occupy the same location in the color color plot) will be all resolved, whereas the SMDSs will still be well under the resolution limit (see Fig.~\ref{fig:Size}). Naturally, for lower mass SMDSs, a larger magnification is required to achieve the same effect.
 
When contrasting SDMSs formed via extended AC (red star symbols in Fig.~\ref{fig:Colorcolor100}) to Pop~III/II galaxies, the situation becomes more promising, as SMDS of smaller and smaller mass deviate more  and more from the evolutionary tracks of both type A and C of galaxies. All SMDSs formed via this mechanism, boosted by $\mu\sim 100$, will be more than $2\sigma$ away from the evolutionary tracks of galaxies. The one caveat here is that Fig.~\ref{fig:Colorcolor100} does not include the uncertainty in the color indices of galaxies, as explained above. We address this point in Fig.~\ref{fig:comparecolor100}, where we present the color index comparison between SMDSs formed via AC of $10^5$ and $10^6 ~\rm M_\odot$ to the evolutionary tracks of Type A/C Pop~III/II galaxies (top/bottom panels) that are selected in such a way that they match the F184 flux of the SMDS, as per our comparison criterion. For all objects a gravitational lensing magnification factor of $\mu=100$ is assumed. 
In order to be conservative, and avoid unnecessarily cluttering the plots, we only keep the Kroupa IMF type galaxies in this figure, since their evolutionary tracks lie closest to the SMDSs.  However, our conclusions are unaffected by this aesthetic simplification. 

For the magnification  $\mu=100$ assumed in  Fig.~\ref{fig:comparecolor100}, we find that $10^5$ and $10^6 ~\rm M_\odot$ SMDS formed via extended AC are separated at the 2$\sigma$ level (in each direction) from the evolutionary tracks of the Pop~III/II galaxies counterparts that have the same F184 flux.
For smaller SMDS, the distinction between SMDS and galaxies is even smaller.  Hence a larger magnification would be required 
to cleanly differentiate the type of object in a color-color plot.  If objects with high enough magnification are found to unambiguously identify a SMDS, such a discovery could indirectly confirm the nature of the DM.

\section{Summary and Conclusions} \label{sec:conclusion}
 Dark Stars (DSs) can form at redshift $z\sim 10-20$. Made almost entirely of hydrogen and helium, they are fueled by Dark Matter heating. 
 As they are relatively cool ($T_{\rm eff} \lesssim 50,000$~K) Dark Stars can grow, via accretion, to become supermassive (SMDS)
 and extremely bright. In this paper we examined the capability of the upcoming Roman Space Telescope (RST) to detect SMDSs.  First we showed that the sensitivity of RST does indeed allow for detection of SMDS, and then we turned to the question of whether or not they can be differentiated in the data from high redshift galaxies.

 Roman Space Telescope will be an excellent observatory in the search for Dark Stars. 
Due to its large field of view, it has great potential for discovery of new objects at redshifts $z< 16$. 
In this paper we considered Supermassive Dark Stars in the mass range $10^4 - 10^6 M_\odot$ formed via two possible mechanisms:  Extended AC and Capture. Here, ``Extended AC" refers to the case in which the DM is accreted gravitationally only; the ``Capture" mechanism relies upon an additional ingredient, capture of further DM via scattering off nuclei in the star. As our fiducial values, we have studied SMDS emitting at $z_{\rm emi} =12$ and DM mass $m_\chi = $100 GeV.

{\it{SMDS detectability in RST:}} \hfil\break
First, we compared predictions of SMDS spectra (simulated with \textsc{TLUSTY} as shown in Fig.~\ref{fig:spectra}) against the sensitivity of RST filters to determine which SMDSs at which redshifts would be detectable by RST, using the standard dropout techniques.
We found that SMDS candidates formed via either the capture or extended AC formation mechanism could indeed be detected by RST using photometry alone.

Figures~\ref{fig:ABmagCap} and~\ref{fig:ABmagAC} show the comparison of AB magnitudes of unlensed SMDSs at different redshifts, in contrast to the detection limit (S/N=5) for each RST filter assuming exposure times of $10^4$s and $10^6$s. 
For SMDSs with masses $\sim 10^6 M_{\odot}$, formed via AC or Capture, we find that at emission redshift $z_{\rm emi}\sim 10-15$ they  would be observable by RST with $\lesssim 10^6$s of exposure time, even if unlensed. With lensing factors of $\mu\gtrsim10$, RST can easily observe SMDSs with masses as low as $M\lesssim10^4\Msun$ (See Fig.~\ref{fig:SMDSvsMu}). SMDSs will appear as J-band dropouts at $z_{\rm emi}\sim 10$ and H-band dropouts at $z_{\rm emi}\sim 13$, and as $F_{184}$ (H/K) dropouts at $z\sim 14$. In order to be able to also include the color index technique, that requires photometric data in at least three bands, we chose $z_{\rm emi}\sim 12$, as our fiducial value throughout this paper.

{\it Differentiating SMDS from high redshift galaxies:} \hfil\break
A key question is whether or not SMDS can be differentiated in the data from high redshift
[zero (Pop~III) or very low metallicity (Pop~II)] galaxies.  For these galaxies, 
the spectral evolution with respect to age, nebular emission, IMF and metallicity were obtained using the YggDrasil~\citep{Zackrisson2011} simulated model grids, and the relevant choice of parameters are given in Table~\ref{tab:para2}.
For the purposes of comparison to SDMS, we chose Pop~III/II galaxies that have the same integrated observed F184 flux as the dark stars in question. 
We used several approaches to see if we can differentiate SDMSs from early galaxies:  SEDs in photometric data, location in color-color plots and image morphology (point vs. extended objects).

Given the resolution of RST, SMDS will be point objects whereas galaxies may be resolved. Without lensing, many galaxies will remain unresolved so that one cannot tell the difference from supermassive stars. The higher the magnification, the more likely that galaxies are resolved whereas SMDS remain point sources. We estimate that magnification $\mu \sim 100$ is sufficient to resolve the all galaxies we have investigated, but upcoming observations will determine this number more accurately. We note that Airy patterns, being specific only for point objects, would definitively identify them as such.
However, very long exposure times $\mathcal{O}$(year) would be needed for their detection, since the first ring is only 1.75\% as bright as the central spot. Therefore, discovery of such a pattern in RST is unlikely for Dark Stars, yet it is a signature of SMDS that future telescopes would be able to see.

For objects only seen with RST  photometry (rather than spectroscopy), SEDs of both SMDS and Pop III/II galaxies can equally well match observations. One approach to differentiate SMDS vs galaxies would be their location in color-color plots. We used the Pandeia engine to simulate RST observations of different candidate objects (e.g. Fig.~\ref{fig:PSFCAP} and~\ref{fig:PSFAC}).\footnote{The simulated SNR values also allow us to estimate the uncertainty in the absolute magnitude, in any RST band in which an object is detected.} Specifically., we use  the F-K vs. H-F color indices for all of the objects of interest (see Figs.~\ref{fig:Colorcolor},~\ref{fig:Colorcolor100} and~\ref{fig:comparecolor100}).

\begin{table}[h!]
\centering
\caption{For the SMDSs of stellar mass and formation mechanism as indicated, the Table summarizes the detectability results based on drop-out, color-color, morphology methods, and presence of the  He~II $\lambda1640$ absorption line for SMDSs. We show results for the case of no lensing as well as with lensing (magnification factors $\mu$ listed in table header). In all cases an exposure time of $10^6$s is assumed.}\label{tab:SMDSdetect}
 \begin{tabular}{lcccccccc}
 \hline
 Formation  &$M_{*} \left(M_{\odot}\right)$  & Drop-out&($SNR > 5$) & Morphology & Morphology & Color-color& Color-color & He~II 1640 \\Mechanism & & No Lensing & $\mu \sim 100$ & No Lensing & $\mu \sim 100$  & No Lensing & $\mu \sim 100$ & Absorp. Line  \\   [0.5ex] 
 \hline\hline
Extended AC & $2.04\times 10^{4}$ & No & Yes & No & No & No & No &No \\
Extended AC & $10^{5}$ &  No & Yes & No & No & No & at 2$\sigma$ &No \\
Extended AC & $10^{6}$ & Yes & Yes & No & Yes & No & at 2$\sigma$ &Yes  \\
Capture & $4.1 \times 10^{4}$ &  No & Yes & No & No & No & No & Yes  \\
 Capture & $10^{5}$ &  No & Yes & No & No & No & No & Yes  \\
 Capture & $10^{6}$ &  Yes & Yes & No & Yes   & No & No &Yes  \\
 \hline
 \end{tabular}
\end{table}

In Table~\ref{tab:SMDSdetect} we summarize the potential to detect SMDSs and then to disambiguate them via image morphology and/or color index techniques from their respective Pop~III/II galactic counterparts. The third and fourth columns take binary (Yes/No) values and address the question of detection at $SNR=5$ level, without lensing (column three) or with $\mu=100$ lensing (column four). Similarly, in the fourth and fifth columns we summarize the answers to the following question: can a SMDSs be differentiated from {\it{all}} possible galactic counterparts based on image morphology, i.e. point source (SMDSs) vs. potentially resolved extended object (galaxy). The last two columns summarize the potential to use the $m_{158}-m_{184}$ vs. $m_{184}-m_{213}$ color diagram to separate SMDSs of a given mass vs. galaxies. 

Differentiating SMDS with masses up to $10^6 M_\odot$ against early galaxies without spectroscopy proves to be difficult. As mentioned above, both types of SMDS as well as early galaxies have SEDs that match RST data $\sim$ equally well.  
For SMDS formed via Extended AC, 
for SMDS masses with $M \sim 10^5, 10^6 M_\odot$,
 the loci occupied by the SMDSs in the color-color plots deviate from the evolutionary lines of Pop~III/II galaxies at the $2\sigma$ level for the case of $\mu\gtrsim 100$ (see Fig.~\ref{fig:comparecolor100}). For smaller SMDS or for higher statistical significance of the difference between
 the objects, even larger magnifications would be required.  For SMDS formed via capture, the locations in color-color plots cannot be differentiated from early galaxies; the only differentiating characteristic would then be that these are point objects whereas galaxies can be resolved at sufficient magnifications. For instance, for the $10^6\Msun$ SMDS via DM Capture, by $\mu \gtrsim100$ all the galactic counterparts are resolved (top left panels of Figs~.\ref{fig:Size} and ~\ref{fig:Size100}).

There is a ``smoking gun" for detection of SMDS once the spectroscopy is available:
a telltale spectroscopic signature, the HeII $\lambda$1640 absorption line. 
While RST does not cover the wavelength band required to find this line (for $z_{\rm emi}\gtrsim 10$), JWST does. Hence the two detectors can be used together in identifying SMDS.
SMDS formed via capture would produce this line; for the cooler SMDS formed via extended AC,
 this feature is present only for extremely massive ($M \gtrsim 10^6\Msun$) SMDSs (See Fig.~\ref{fig:spectra}).  However, there is a caveat: as yet we have not modeled the nebula surrounding the star, which in the case of the hotter SMDS formed via capture could wash out the absorption line and/or produce emission lines. The cooler SMDS formed via extended AC are not likely to be affected.
 In followup work, we plan to use the \textsc{CLOUDY} code to add the effects of the nebula to the stellar spectra produced by TLUSTY. In any case, should a HeII $\lambda$1640 absorption line be found, it would be a strong indicator that an object found in RST is a supermassive star rather than an early galaxy.   

 While the HeII $\lambda$1640 line is a very useful spectroscopic tool, since it is isolated from other neighbouring lines (see Fig.~\ref{fig:spectra}), we note here that SMDSs differ from Pop~III/II galaxies with respect to their spectral signatures in other important ways. First of all, the Balmer break, i.e. the sudden jump in the flux at around $\microm{0.35}$ (restframe), which can be seen in the lower two panels of Fig.~\ref{fig:spectra} as a ``hump-like'' feature. This feature will be more pronounced for Pop~III/II galaxies with nebular emission, but most importantly, the sequence of lines that follows the Balmer break, i.e. the Balmer series, are emission lines for galaxies with nebular emission vs. absorption lines for SMDS with purely stellar spectra. Thus, in addition to the He~II $\lambda$1640 line, as a spectroscopic disambiguation tool, one can also use lines in the Balmer series as well.

{\it On heavier $10^7 M_\odot$ SMDS:}\hfil\break
In this paper we have focused on SMDS with masses up to $10^6 M_\odot$. However, in principle SMDS could grow to become even heavier, e.g. if they formed in $10^8 M_\odot$ minihalos or if smaller minihaloes merged together and the DS remains in a dark matter rich region.  We therefore consider below the detectability of $10^7 M_\odot$ SMDS.  Since total luminosity scales linearly with mass, these objects would be ten times brighter than $10^6 M_\odot$ SMDS and thus easier to detect. We note, however, that differentiation in color-color plots against early galaxies would actually be more difficult (see Figure 21).   On the plus side, galaxies this bright should be easier to resolve. Thus the differentiation of these very heavy SMDS vs. early galaxies should be possible.  Further, the HeII $\lambda$1640 line would be more pronounced (deeper) for these hotter objects; such a line would be detectable in RST only for $z_{\rm emi} < 10$ while JWST could find it even for higher redshift objects.  In summary, $10^7 M_\odot$ SMDS are much brighter, therefore easier to detect as well as differentiate from early galaxies.

{\it{Using RST and JWST together:}}\hfil\break  
The primary advantage of RST is that it will have a much larger field of view than that of JWST. Thus, at the redshifts RST is able to probe, it  will have a larger  probability of successfully finding SMDSs.
Two of us (~\cite{Ilie:2012}) have previously studied SMDS in JWST data (including estimating the numbers of SMDS one could find with JWST). 
As shown in Fig.\ref{fig:telescope}, RST and JWST have comparable detector sensitivity for wavelengths up to $\sim\microm{1.8}$, including the J and H bands relevant for identifying objects as J- and H-band dropouts, corresponding to $z \sim 11$ and $z \sim 13$ respectively.  JWST is far more sensitive at higher wavelengths and thus can find higher redshift objects than RST can, e.g. K-band dropouts at $z\gtrsim 15$. 
JWST can probe, via photometry wavelengths up to $\microm{28}$ (NIRCam up to $\microm{5}$ and MIRI from $\microm{5}$ to $\microm{28}$), whereas the Wide Field Instrument (WFI) onboard RST will be sensitive only up to  $\microm{2.3}$. In terms of spectroscopy the Grism on RST will be probing up to $\microm{1.93}$, whereas NIRSpec onboard JWST is sensitive up to $\microm{5}$. As such RST will be essentially blind with respect to {\it{any}} objects at $z_{\rm emi}\gtrsim 18$. At those high redshifts the Gunn-Peterson trough will cutoff the  entirety of the redshifted flux in the region RST is potentially sensitive to. Moreover, even at the redshift of interest to us in this paper ($z_{\rm emi}\sim 12$) there is quite important information one can gain from possible redshifted spectral features that fall outside of the sensitivity range of RST, such as, for example, the Balmer Break, or other H or He absorption lines present in the SMDS SEDs.  Moreover, with a much larger aperture, JWST has a better angular resolution. Therefore, once a candidate DS object is identified by RST, followup observations with JWST could be used to further verify and validate the Dark Star hypothesis.   As mentioned above, a smoking gun for SMDS is the existence of a He~II $\lambda$1640 line, which would not be expected for the case of early galaxies. While RST wavelengths are insensitive to this line, spectroscopy with JWST may be able to detect it.

We note that the ability of NIRSpec on JWST to obtain clean spectra is remarkable, 
 see e.g. GNz11 in~\cite{GNz11NIRSpec:2023}.  This $z=10.6$ galaxy has a spectrum seen to $m_{\rm AB} \sim  26$ in the H-band, which  would be comparable to the brightness of a $M_\star\simeq 4\times 10^6 M_\odot$ SMDS formed via AC, see Fig.~5 in~\citep{Ilie:2012}. While GNz11 is clearly not a Dark Star, as it is resolved, and moreover shows numerous metal emission lines (see Fig.~1 of~\cite{GNz11NIRSpec:2023}), based on the relatively short total exposure time ($\sim17$ hours) needed to generate its exquisite spectrum with NIRSpec, we would expect spectroscopy with JWST is a viable tool for a SMDSs. In summary, RST will be an excellent instrument for finding SMDS candidates, with followup observations from JWST  potentially providing definitive determination of Dark Star discovery.

\textit{In Conclusion:} \hfil\break
Dark Stars are very bright sources. In fact, a million solar mass dark star can emit as much light as a billion solar mass galaxy so that the two look very similar in RST and JWST observations.  Hence the conundrum posed by the many
observed bright high redshift objects in JWST, in the context of the standard $\Lambda$CDM cosmological model, could be alleviated by the existence of Dark Stars containing much less baryonic mass than would be required of a comparably bright galaxy.

In this paper we showed that SMDSs candidates formed via either the capture or extended AC formation mechanism and emitting at redshifts up to $z_{\rm emi} = 14$ could be found by RST using photometry alone. 
We examined how those candidates could be differentiated from high redshift Pop~III or Pop~II galaxies, and summarized our results in this section.  There are a number of potential signatures, including their location in color-color plots and image morphology.
 SDMSs with purely stellar spectra would appear as point objects, very bright ($L\gtrsim 10^9 L_\odot$), yet relatively cool ($T_{\rm eff} \sim 10^4$~K) sources.  
 The detection of a He~II absorption line at $\microm{0.1640}$ restframe wavelength (the HeII $\lambda$1640 line) with followup spectroscopic observations would be a clear signature of supermassive dark stars, as this feature is absent in any galactic sources. The identification of supermassive dark stars would be truly remarkable, as it would imply the existence of a novel heat source powering stars, such as the DM annihilation we have studied.  Further, such a discovery could  open up the possibility of DM parameter estimations based on observed properties of dark stars.  In contrast, whenever nebular emission becomes significant for  SMDSs~\footnote{See Appendix~\ref{ap:Nebular} for a detailed discussion on what conditions have to be met for this to happen.} the differentiation between those objects and young compact galaxies dominated by zero metallicity stars would become nearly impossible. In turn, this could further help alleviate the ``too many too massive too soon'' mystery posed by the JWST data, as many of those ``galaxies'' could actually be SMDSs disguising under the veil of a nebula.

\begin{acknowledgments}
\section{Acknowledgments} K.F. and S.Z. are grateful for support from the Jeff and Gail Kodosky Endowed Chair in Physics held by K.F. at the Univ. of Texas, Austin. K.F. and S.Z. acknowledge funding from the U.S. Department of Energy, Office of Science, Office of High Energy Physics program under Award Number DE-SC0022021. 
K.F. acknowledges support by the Vetenskapsradet (Swedish Research Council) through contract No. 638- 2013-8993 and the Oskar Klein Centre for Cosmoparticle Physics at Stockholm University. We would acknowledge the useful discussions with James Rhoads, and help from Steve Finkelstein, Volker Bromm through personal communication. We are especially grateful for Jonathan Levine's help with the error ellipses and for him sharing with us the code he uses to generate such ellipses for his research. S.Z. would like to give special thanks to his roommate Changhan Ge who offered help with the Pandeia simulation installation.

\end{acknowledgments}

\appendix

\section{Pandeia Parameters}\label{Pandeia}
In this appendix, we present the key parameters we have used for image simulation, as described in Tables~\ref{tab:para1} to~\ref{tab:para3}. The complete input dictionary can be found on the Pandeia reference paper~\citep{pontoppidan2016pandeia}, as well as at the following link: \url{https://outerspace.stsci.edu/display/PEN/Pandeia+Engine+Input+API}. Below we explain some of the parameters particularly relevant to our work and possible input choices for those. 
\begin{enumerate}
    \item \textbf{Sersic Index}: The morphology of Pop~III galaxies remains largely unknown due to lack of observations. Since the masses of target galaxies are relatively small, we would expect them to have a shape similar to stellar clusters or irregular galaxies. The choice of this index $n=1$ (exponential) reflects that the flux is not overly concentrated at the center. For those galaxies with stellar mass on the order of $M_{\star} \sim 10^9 M_{\odot}$, we use the recent argument on the shape of the first galaxies from~\cite{2022arXiv220211925P}, where they have simulated that galaxies of this mass will have Sersic index $n \lesssim 1.5 $. Also, in the recent results from JWST, ~\citet{GLASSz13} fitted the Sersic index of the observed two galaxies and the results are both $n\lesssim 1$, which is consistent with the simulation. Therefore, we still choose $n = 1$ throughout our paper when we scale up the mass. Although, for smaller galaxies, this might not be accurate description, but it would not significantly change our conclusion since its angular size is relatively small to be resolved. 
    \item \textbf{Group/Integration/Exposure Number}: For simplicity, the readout pattern is chosen as the default for Roman (Deep2). To avoid saturation, the maximal allowed group number($N_{\rm group}$) for this readout pattern is 20, and the integration time ($t_{\rm int}$) is a linear function of the group number. For that case of Deep2, it is given by $t_{\rm int}=t_{\rm frame}\times (20N_{\rm groups}-18)$, where $\mathrm{t}_{\text {frame }}=10.73677 \mathrm{~s}$ as the sample exposure time. Since our desired exposure time would be $\sim 10^6 s$, we choose the maximum value for the group number as $N_{\rm group}=20$.\footnote{Differences in the integration pattern are explained in the fig.~2 of the documentation page:~\url{https://jwst-docs.stsci.edu/jwst-near-infrared-camera/nircam-instrumentation/nircam-detector-overview/nircam-detector-readout-patterns}.} We also have the integration number per exposure ($N_{\rm int}$) and exposure number ($N_{\rm exp}$), that multiply the integration time to give the total exposure time: $t_{\rm exp}=N_{\rm exp}\times N_{\rm int} \times t_{\rm int}$. Since there's no special preference on the choice of these two numbers,\footnote{For a more detailed understanding of the integration process and observation strategies, one can refer to \url{ https://jwst-docs.stsci.edu/understanding-exposure-times}.}~we choose $N_{\rm int}=29$ and $N_{\rm exp}=30$ such that the total exposure time is closest to $10^6s$. 
    \item \textbf{Background Level}: the relevant background properties and spectrum are described at \url{https://jwst-docs.stsci.edu/jwst-general-support/jwst-background-model}. For the range of wavelengts of interest ($\sim[1 - 2]~\mu m $), the background noise contribution from zodiacal lights and galactic dust dominates. To acquire optimal observation results and for simplicity, we choose the minimum zodiacal background as 'minzodi' and set the background level to be its benchmark level. 
\end{enumerate}
In the three tables below we list input parameters specific to SMDSs (Table~\ref{tab:para1}), Pop~III/II galaxies (Table~\ref{tab:para2}), and lastly common to both of those objects (Table~\ref{tab:para3}) In. 
\begin{table}[!htb] 
\centering
\caption{Input dictionary unique to the point objects--SMDSs}\label{tab:para1}
 \begin{tabular}{c c} 
 \hline
 Aperture Size & Sky Annulus \\ [0.5ex] 
 \hline\hline
  0.2" & [0.4",0.6"]  \\ [1ex] 
 \hline
 \end{tabular}
\end{table}
\begin{table}[!htb] 
\centering
\caption{Input dictionary unique to extended objects --Pop~III/II galaxies.}\label{tab:para2}
 \begin{tabular}{c c c c c c c} 
 \hline
 Major & Minor & Geometry & Sersic Index & Norm Method & Aperture Size & Sky Annulus \\ [0.5ex] 
 \hline\hline
  $\mu^{1/2}\times \theta_{\rm eff}$&$\mu^{1/2}\times \theta_{\rm eff}$ & Sersic & 1 & integ infinity & $\mu^{1/2}\times$ 0.1" & $\mu^{1/2}\times$[0.15",0.2"]\\ 
 \hline
 \end{tabular}
\end{table}

\begin{table}[!htb] 
\centering
\caption{Other input parameters used in the Pandeia simulation shared by both SMDSs and Pop III galaxies.}\label{tab:para3}
 \begin{tabular}{c c c c c c c c} 
 \hline
 Background & Background & Background & Aperture &  Group &  Integration & Exposure & Readout \\
  &Level&Subtraction&&Number&Number&Number&Pattern \\[0.5ex] 
 \hline\hline
   minzodi & benchmark & True & imaging & 20 & 29 & 30 & deep2 \\ 
 \hline
 \end{tabular}
\end{table}

\section{Role of Nebular emission for SMDS formed via Capture}\label{ap:Nebular}

For SMDSs formed via DM Capture nebular emission could be significant, and thus potentially render their disambiguation from compact early galaxies nearly impossible. In this appendix we explore the conditions under which such SMDSs will power an ionization bounded nebula. As we shall soon see, the mass of the nebula is {\it inversely} proportional to the number density of H atoms in its region. In turn, since the DM halo contains a finite baryonic mass, there is an upper bound to the mass of the nebula, which translates into a lower bound on the number density of H. When comparing this lower bound to analytic or semi-analytic models of H gas densities within primordial DM halos we can find the maximum SMDS mass for which we expect nebular emission to be significant. 

Under the assumption of equilibrium between ionization and recombination rates, one can estimate the radius of a nebula powered by a hot star as the so called Str\"{o}mgren radius:

\begin{equation}\label{eq:Strom}
    r_{\rm S}=\left(\frac{3Q_*}{4\pi\alpha n_{\rm H}^2}\right)^{1/3},
\end{equation}
where $Q_*$ represents the ionizing flux from the star, $\alpha$ is a temperature dependent recombination coefficient, and $n_{\rm H}$ represents the number density of protons (or electrons) inside the ionized H region. When considering the mass of the HII (i.e. ionized) gas enclosed within the nebula: $M_{\rm HII}=\frac{4\pi}{3}r_{\rm S}^3n_{\rm H}m_{\rm H}$ one can see that indeed $M_{\rm HII}\sim 1/n_{\rm H}$, as alluded to before.

The upperbound to $M_{\rm HII}$ is given by the deducting the total mass of H in star from that in halo:
\begin{equation}\label{eq:Bound}
 M_{\rm HII}\leq f_{\rm H} \left(f_{\rm B} M_{\rm halo} - M_{\rm *}\right),
\end{equation}
where $f_{\rm H}$ represents the H mass fraction, which for primordial DM halos can be assumed to be $75\%$ according to BBN, and $f_{\rm B}$ represents the baryonic mass fraction within the DM halo, which we assumed to be $\sim10\%$.
The inequality saturates whenever one completely ignores the HI and $\rm H_2$ regions around the star, and any other molecular H clouds that could be present within the DM halo. Therefore the upper bound in Eq.~(\ref{eq:Bound}) should be viewed as strict. This, in turn, implies that all the results we below here are to be viewed as conservative.

Using  the Str\"{o}mgren radius from Eq.~(\ref{eq:Strom}) we can convert the upper-bound from Eq.~(\ref{eq:Bound}) into a lower-bound on $n_{\rm H}$:

\begin{equation}\label{eq:nHCrit}
    n_H \geq\frac{Q_*(M_*) m_{\rm H}}{\alpha f_{\rm H}(f_B M_{\rm halo}-M_*)}.
\end{equation}
Throughout $m_{\rm H}$ represents the mass of a proton, and $M_{\rm halo}$ the mass of the DM halo within which the Dark Star is embedded. $Q_*(M_*)$ is the ionization photon flux from a star of mass $M_*$, which we calculate numerically for our SMDSs formed via DM caputre models, from their fluxes ($F_\nu$) via:
\begin{equation}
    Q_*=\int_{h\nu=13.6 \unit{eV}}^{\infty}4\pi R_*^2 \frac{F_\nu(\nu)}{h\nu}d\nu
\end{equation}
Assuming, as typically done, that the equilibrium temperature for the gas within the HII region is $\sim 10^4$~K we can use the following value for the recombination coefficient: $\alpha=2.6\times 10^{13}\percc$~s$^{-1}$. For SMDSs formed via DM capture considered in this paper we tabulate in Table~\ref{tab:SMDSnH} the lower bounds on $n_{\rm H}$, under the assumption that $M_{\rm halo}=10^8\Msun$. If instead we were to assume a $10^7\Msun$ DM halo, which could also in principle host a SMDSs as massive as $10^6\Msun$, the lower-bound values $n_{\rm H}$ would increase by roughly one order of magnitude  for the $\sim 10^4$ and $10^5\Msun$ SMDSs, whereas for the case of the $10^6\Msun$ SMDSs it would approach $\infty$. The physical significance $n_{\rm H,min}\to\infty$ in this context is simple: there is no gas left in the halo, therefore no nebula can be powered.

\begin{table}[h!]
\centering
 \begin{tabular}{lccccc}
 \hline
 Formation Mechanism &$M_{*}$  & $L_{*}$ & $R_{*}$  & $T_{\text {eff }}$ & $n_{\rm H,min}$ \\ &$\left(M_{\odot}\right)$ & $\left(10^{6} L_{\odot}\right)$ & $(\mathrm{AU})$ & $\left(10^{3} \mathrm{~K}\right)$ & ($\percc$) \\ [0.5ex] 
 \hline\hline

Capture & $4.1 \times 10^{4}$ & 774 & 1.8 & 49 & 20 \\
 Capture & $10^{5}$ & $1.75 \times 10^{3}$ & 2.7 & 51 & 60  \\
 Capture & $10^{6}$ & $2.03 \times 10^{4}$ & 8.5 & 51 & 670 \\ \hline
 \end{tabular}
 \caption{For SMDSs formed via DM capture, with parameters described in columns two to five, we tabulate in column six the lower-bound on  $n_H$ obtained according to Eq.~(\ref{eq:nHCrit}). We assumed $M_{\rm halo}=10^8\Msun$.}\label{tab:SMDSnH}
\end{table}

There are several ways to estimate the hydrogen number density in a halo. If one assumes gas with uniform density in a virialized DM halo, then, as shown by~\cite{Tegmark:1997}, the spherical collapse model leads to:
\begin{equation}\label{eq:nHVir}
    n_{\rm H}\simeq 1.01 \left(\frac{1+z}{31}\right)^3\percc
\end{equation}

Note that at the redshifts of interest for us, i.e. $z\in[10,20]$ the estimated values of $n_H$ obtained via Eq.~(\ref{eq:nHVir}) are all much smaller than the lower-bound for $n_{\rm H}$ tabulated in Table~\ref{tab:SMDSnH}. This would indicate that SMDSs formed via DM Capture don't have sufficient material surrounding them to power an ionization bounded nebula. If $n_{\rm H}$ is enhanced to $\sim 100~\percc$ (more than two orders of magnitude when compared to the value predicted by Eq.~\ref{eq:nHVir}) then one expects nebular emission from SMDSs with $M\lesssim 10^5\Msun$, as long as they form in a $10^8\Msun$ DM halo. For the same SMDSs formed inside halos with  $M_{\rm halo}\lesssim 10^7\Msun$ one would require an enhancement to $n_{\rm H}\gtrsim 1000~\percc$ in order to have significant nebular emission. 

The main limitation of the analysis above is the assumption that the gas surrounding the star has uniform density, which is embedded even in our starting point, Eq.~(\ref{eq:Strom}). For hydrogen density profiles that are cored, such as an isothermal sphere, or a truncated isothermal sphere, we expect that sufficient enhancements of $n_{\rm H}$ (with respect to its values predicted by Eq.~(\ref{eq:nHVir})) are possible in the region where an equilibrium between recombination and ionization is attained. We leave the detailed exploration of nebular emission from Dark Stars to for a future dedicated study, where will also include the effects of realistic density profiles for $n_H$. However, the expectation is that, whenever nebular emission becomes significant, SMDSs will essentially look just as compact young galaxies, thus rendering any prospects for disambiguation nearly impossible.        

\section{Differentiating SMDS from other types of SMS}\label{ap:SMDSvsSMSs}
 We present below a brief discussion comparing the SEDs of SMDSs with those of another class of Supermassive Stars (SMSs) considered in the literature: Pop~III Supermassive Stars. Those are precursor to the so called Direct Collapse Black Holes (DCBHs) and could form in atomic cooling DM halos that have been exposed to sufficiently high level of Lyman-Werner (LW) radiation~\citep{Loeb:1994wv,Belgman:2006,Lodato:2006hw,Natarajan:2017}. This radiation can be generated abundantly by the first generation of stars (i.e.  nuclear burning Pop~III stars or Dark Stars). As a consequence of the LW radiation, $\mathrm{H}_2$ is dissociated and stellar formation is inhibited in nearby DM halos, until those reach $M_{\rm halo}\gtrsim 10^8\Msun$, when atomic H cooling becomes efficient. Simulations show that at this stage a catastrophic baryonic runaway collapse could build up a star at rates of up to $\sim 1 \Msun \rm yr^{-1}$. By the time they reach $\sim 10^5\Msun$ those Pop~III Supermassive Stars quickly enter a gravitational instability regime, and collapse to what are typically called Direct Collapse Black Holes (DCBHs). For a review of the DCBH scenario the interested reader can consult~\cite{Inayoshi:2020}. 

 We do note, however, that SMDSs can also form in the case of atomic hydrogen cooling in $10^8 M_\odot$ DM haloes.  This type of SMDS is reviewed in~\cite{Rindler-Daller:2015SMDS} for the case called Large Minihaloes (LMH, see Eqn. (6) in that paper). Indeed it is possible that the heat from dark matter annihilation will  inhibit the SMS formation and instead lead to a Supermassive Dark Star. In the remainder of the Appendix, we compare the observable consequences of Pop III SMS vs. SMDS.

 Pop~III SMSs follow two possible evolutionary tracks: cool (red) tracks at $T_{\rm eff}\lesssim10^4$K~\citep[e.g.][]{Surace:2018ApJ,Vikaeus:2022ApJ} or hot (blue)  tracks at $T_{\rm eff}\gtrsim 10^4$K~\citep[e.g.][]{Surace:2019}. Here we begin a simple comparison between Pop~III SMSs on each of those tracks with SMDSs in terms of observability. Initially we take the case of no nebular emission for either type of object. With this assumption, it should be possible to distinguish SMSs on either track from SMDSs (of either formation via purely gravitational effects or via capture) based on their colors. At the same brightness a SMS on the Red Super Giant Track will be cooler (redder) than a SMDS.
Conversely, at the same luminosity, a SMS on the Blue Super Giant Track will be hotter (bluer) than a SMDSs.

 However, the situation changes when nebular emission is taken into account.
For most Pop~III SMSs, processing of the stellar SEDs by the accretion disk surrounding them is expected, because of their formation from a disk undergoing runaway collapse. In other words, nebular emission is likely for Pop~III SMSs. They exhibit features such as reprocessing of photons and redistribution towards wavelengths longwards of the \Lyalpha line, and the emergence of emission lines (see, for example, bottom panels of Fig.~2 in~\cite{Surace:2019}). As yet we have not included the effects of nebular emissions on Dark Stars and will study these effects in future work using CLOUDY. Hence
we reserve the detailed comparison of Pop~III SMSs SEDs processed by their gas envelope to that of SMDSs with nebular emission (see Appendix~\ref{ap:Nebular}) for a future study. We do expect nebular emission to be important for some but not all SMDS, and that the processing of the stellar SEDs by the gas envelope would lead to making differentiation between Pop~III SMSs and SMDSs in those cases for which nebular emission is important to be quite difficult, if not impossible.  As discussed in Appendix~\ref{ap:Nebular}, we expect the lower mass ($M\lesssim 10^5\Msun$) SMDS formed via capture to experience nebular emission. However, for the most massive ones there is not sufficient H gas to form a ionization bound nebula. Further, we typically expect SMDS formed via AC to avoid nebular emission, due to the large amount of neutral hydrogen in the atmosphere that absorbs most of the ionizing flux, as can be seen in the upper right panel of Figure 2 (see the nose-dive to the left of the Lyman edge in the figure).  Thus we expect SMDS formed via AC as well as the heavier ($M>10^5 M_\odot$) SMDS formed via capture to have unique signatures clearly differentiable from all other objects in JWST.

\bibliography{Main.bib}
\bibliographystyle{aasjournal}

\end{document}